\documentclass[conference]{IEEEtran}
\IEEEoverridecommandlockouts
\usepackage{fourier} 
\usepackage{amsmath,amssymb,amsfonts}
\usepackage{bm}  %bold math fonts

\DeclareMathOperator*{\argmin}{arg\,min}
\usepackage{algorithm}
\usepackage{adjustbox}
\usepackage{siunitx}
\usepackage{comment}
\usepackage{color,soul}
\usepackage{tikz}

\usepackage{pgfplots}
\pgfplotsset{compat=newest}
\usetikzlibrary{plotmarks}
\usetikzlibrary{arrows.meta}
\usepgfplotslibrary{patchplots}
\usepackage{grffile}
\usepackage{helvet}
\usepackage[eulergreek]{sansmath}
 \pgfplotsset{plot coordinates/math parser=false}
\newlength\figureheight
\newlength\figurewidth
\pgfplotsset{
tick label style = {font=\sansmath\sffamily},
every axis label/.append style={font=\sffamily\footnotesize},
}

\pgfplotsset{
  contour/every contour label/.style={
    sloped,
    transform shape,
    inner sep=2pt,
    every node/.style={mapped color!50!black,fill=white,
    font =\sffamily\fontsize{4.5}{5}\selectfont},
    /pgf/number format/relative*={\pgfplotspointmetarangeexponent},
  }
}
\makeatletter
\renewcommand{\fnum@figure}{Figure \thefigure}
\makeatother

\usepackage{multirow}
\usepackage{multicol}
\usepackage{threeparttable}
\usepackage{algpseudocode}
\usepackage{cite}

\usepackage{graphicx}
\usepackage{textcomp}
\usepackage{xcolor}
\usepackage{enumitem}
\usepackage{afterpage}
\usepackage{siunitx}
\usepackage[utf8]{inputenc}
\usepackage{verbatim}
\usepackage{enumitem}
\usepackage{array}
\usepackage{adjustbox}

\def\BibTeX{{\rm B\kern-.05em{\sc i\kern-.025em b}\kern-.08em
    T\kern-.1667em\lower.7ex\hbox{E}\kern-.125emX}}

\definecolor{dgreen}{RGB}{0,130,0}

\algnewcommand\algorithmicswitch{\textbf{switch}}
\algnewcommand\algorithmiccase{\textbf{case}}
\algnewcommand\algorithmicassert{\texttt{assert}}
\algnewcommand\Assert[1]{\State \algorithmicassert(#1)}%
\algdef{SE}[SWITCH]{Switch}{EndSwitch}[1]{\algorithmicswitch\ #1\ \algorithmicdo}{\algorithmicend\ \algorithmicswitch}%
\algdef{SE}[CASE]{Case}{EndCase}[1]{\algorithmiccase\ #1}{\algorithmicend\ \algorithmiccase}%
\algtext*{EndSwitch}%
\algtext*{EndCase}%

\newcommand{\bfmu}{\bm{\mu}}

\begin{document}
\sloppy

\title{Subjective Image Quality Assessment with Boosted Triplet Comparisons }

\author{\IEEEauthorblockN{Hui Men\textsuperscript{1}, Hanhe Lin\textsuperscript{1}, Mohsen Jenadeleh\textsuperscript{1}, Dietmar Saupe\textsuperscript{1}}\\ \vspace{-3pt}
\IEEEauthorblockA{\textsuperscript{1}Department of Computer and Information Science, University of Konstanz, Germany}
Email: \{hui.3.men\,|\,hanhe.lin\,|\,mohsen.jenadeleh\,|\,dietmar.saupe\}@uni-konstanz.de}

\maketitle 
           
\begin{abstract}
In subjective full-reference image quality assessment, a reference image is distorted at increasing distortion levels. 
The differences between perceptual image qualities of the reference image and its distorted versions are evaluated, often using degradation category ratings (DCR). 
However, the DCR has been criticized since differences between rating categories on this ordinal scale might not be perceptually equidistant, and observers may have different understandings of the categories. 
Pair comparisons (PC) of distorted images, followed by Thurstonian reconstruction of scale values, overcomes these problems.  
In addition, PC is more sensitive than DCR, and it can provide scale values in fractional, just noticeable difference (JND) units that express a precise perceptional interpretation. 
Still, the comparison of images of nearly the same quality can be difficult. 
We introduce boosting techniques embedded in more general triplet comparisons (TC) that increase the sensitivity even more. 
Boosting amplifies the artefacts of distorted images, enlarges their visual representation by zooming, increases the visibility of the distortions by a flickering effect, or combines some of the above.
Experimental results show the effectiveness of boosted TC for seven types of distortion (color diffusion, jitter, high sharpen, JPEG\,2000 compression, lens blur, motion blur, multiplicative noise). 
For our study, we crowdsourced over 1.7 million responses to triplet questions. 
We give a detailed analysis of the data in terms of scale reconstructions, accuracy, detection rates, and sensitivity gain.
Generally, boosting increases the discriminatory power and allows to reduce the number of subjective ratings without sacrificing the accuracy of the resulting relative image quality values. 
Our technique paves the way to fine-grained image quality datasets, allowing for more distortion levels, yet with high-quality subjective annotations. 
We also provide the details for Thurstonian scale reconstruction from TC and our annotated dataset, \emph{KonFiG-IQA}, containing 10 source images, processed using 7 distortion types at 12 or even 30 levels, uniformly spaced over a span of 3\,JND units.
\end{abstract}
\vspace{10pt}

%\titlepgskip=-15pt

\maketitle

%------------------------------------------------------------
%------------------------------------------------------------
%------------------------------------------------------------

%cite{liu2015comparison}: [Comparisons between ACR and 3 stimulus-PC, i.e., PC with reference ] 

\section{Introduction}
%------------------------------------------------------------
%Subjective IQA for sequences of distorted image like in TID2013 or KADID comes in different flavors: 1. Double stimulus paired comparisons, 2. double stimulus\,JND assessment, 3. single stimulus ACR, or rather DCR

Full-reference image quality assessment (FR-IQA) quantifies the perceptual image qualities of distorted versions of pristine reference images. In addition, FR-IQA quantifies the trade-off between bitrate and perceived quality in perceptual image compression, which helps optimize encoding parameters. Similarly, the development of other image processing applications such as image restoration and enhancement may profit from knowing the expected perceptual quality of their output images. 
%artificial imagery from generative adversarial networks (a hot topic, but more a case for NR-IQA)

Since it is not feasible to assess perceptual image quality by a subjective study each time in such applications, automated FR-IQA algorithms must be used that estimate the quality from the image data without any human interaction. To develop and train such FR-IQA algorithms, annotated image datasets, derived from subjective studies, are required. In such studies, images are judged by subjects according to their perceived quality, either individually or in comparison with one or more other images. This paper contributes boosting methods for the presentation of the image stimuli in subjective studies that improve the accuracy and sensitivity of the perceptual measurements. 

%------------------------------------------------------------------------------------------
%-----´SUBSECTION --- SUBSECTION --- SUBSECTION --- SUBSECTION --- SUBSECTION -------------
\subsection {Subjective Full-reference Image Quality Assessment}
\label{sec_subFRIQA}
%------------------------------------------------------------------------------------------
In subjective studies, test stimuli may be presented one at a time and rated according to a 5-point absolute category ratings (ACR) scale, i.e., Bad (1), Poor (2), Fair (3), Good (4), and Excellent (5). For each stimulus, the integer values of the ratings from many subjects are averaged, yielding corresponding mean opinion scores (MOS), which serve as scalar perceptual image qualities %\cite{itu1999subjective}
\cite{itu2009methodology}.
ACR is likely to lead to low sensitivity in distinguishing among stimuli of similar qualities. A modified version of ACR, degradation category rating (DCR), provides higher sensitivity \cite{itup800,itup913subjective}.
% [ITU-T P.913: "DCR ratings are minimally influenced by a subject's opinion of the content (e.g., whether the subject likes or dislikes the production quality). Thus, DCR is able to detect color impairments and skipping errors that the ACR method may miss."]
% [ITU-T P.800: "The Absolute Category Rating (ACR) method described in Annex B tends to lead to low sensitivity in distinguishing among good quality circuits. A modified version of the ACR procedure, called the Degradation Category Rating (DCR) [13] procedure, affords higher sensitivity. "
In a DCR test, distorted stimuli are presented with their references, either sequentially or simultaneously side by side. Stimuli are rated according to the 5-point DCR scale, namely Very Annoying (1), Annoying (2), Slightly Annoying (3), Perceptible but not Annoying (4), and Imperceptible (5). Average ratings are called degradation mean opinion scores (DMOS).

The approaches mentioned above, although straightforward, have some limitations. 
\begin{enumerate}
    \item [(1)] Observers may have different understandings of the quality categories \cite{jones1986graphic,chen2009crowdsourceable}, %{hossfeld2016qoe}[removed], 
    leading to large variances of ratings and therefore requiring a large number of ratings to achieve the desired precision of the mean opinion scale of ACR or DCR.
    \item [(2)] There is the danger of a saturation effect. If a subject scores an image in the best (or worst) quality category, another item to be judged may come up with a perceived quality that is even significantly better (or worse). Then this item can only be scored with the same category as before. There is no way to correct previously assigned quality values to accommodate the overall larger than expected dynamic range in quality. 
    \item [(3)] ACR and DCR scales should be regarded as ordinal, not interval scales \cite{jones1986graphic, chen2009crowdsourceable, seufert2021statistical}, even though in practice the categories Bad, Poor, etc.\ are linked to numerical values 1, 2, and so on. This means that pairs of stimuli with an equal difference in MOS, resp.\ DMOS are not generally perceived as having the same perceptual distance. 
    \item [(4)] Given the mean opinion scores $s$ and $s+ \Delta s$ of two images, there is no meaningful interpretation for the difference in perceptual quality based on $s$ and $\Delta s$. It would be desirable to have a scaling property similar to that provided by the peak-signal-to-noise ratio (PSNR) for the case of objective image quality. For example, if two distorted images have a difference of 1\,dB in PSNR, then we know that the mean-square-error in one image is $10^{0.1}\approx 1.259$ as large as in the other one.
\end{enumerate} 

The pair comparison method (PC) is an alternative to ACR and DCR. In the 2-alternative forced-choice (2AFC) setting, observers are presented with pairs of test images and asked to identify the image in each pair with less distortion, i.e., the image with better quality. The PC method is an indirect measurement, and scale values cannot be generated simply by averaging ratings. Instead, an algorithm is required that ``reconstructs'' the latent quality scale values.

%There have been many reconstruction methods proposed. In psychophysics, the field of study that investigates the relationship between physical stimuli and the perceptions they produce, methods based on probabilistic models have become the defacto standard for this purpose.
Many reconstruction methods have been proposed. In psychophysics, the field that studies the relationship between physical stimuli and the perceived experiences they evoke, methods based on probabilistic models have become the de facto standard for this purpose. In Thurstonian models, the perception of each stimulus is modelled quantitatively as a scalar Gaussian random variable. The random variables corresponding to sequences of stimuli under investigation are most commonly taken to have the same variance of 0.5 so that quality differences in PC have a unit variance when the independence of the random variables is assumed. This setting defines the units of measurement. Initially, following Thurstone's pioneering work \cite{thurstone1927law} in 1927, a least-squares approach was taken to solve for the reconstructed scale values based on his model. Nowadays, the method of choice is maximum likelihood estimation (MLE) \cite{perez2017practical}.  

The PC method overcomes all of the above-listed limitations of ACR and DCR. 
\begin{enumerate}
    \item [(1)] Such a comparison does not rely on the particular interpretation of a nominal category of quality. Therefore the task is clear and more natural than ACR and DCR. 
    \item[(2)] By design, the saturation effect is eliminated. 
    \item[(3)] The reconstruction for PC yields quality values on an interval scale. Specifically, according to the Thurstonian model, pairs of stimuli with an equal difference in scale value are perceived having the same perceptual distance in the sense of the old, famous psychological rule of thumb: \textit{Equally often noticed differences are equal, unless always or never noticed}  \cite{thurstone1927equally}. Thus, for a difference of $\Delta s$ on the perceptual quality scale, the proportion of subjects who consider the stimulus with the larger latent scale value to be the one with better quality is a function of $\Delta s$ alone.
    \item[(4)] By appropriately choosing the variance of the Gaussian distribution in the Thurstonian model, we can define the perceptual scale such that one unit difference between the two values of a pair corresponds to a fraction of 75\% of the subjects choosing the correct better quality item among the pair. This corresponds to the usual definition of the just noticeable difference (JND): The JND is the perceptual quality difference for which the probability of detecting the better quality image is 50\%. In PC (2AFC) for this case, therefore, half of the subjects will detect and choose the correct stimulus as the better one, while the other half has to guess and will be correct half of the time, leading to the 75\% ratio.
\end{enumerate} 

%Because of the first point above, forced-choice PCs are easier to decide in subjective studies and therefore take less time than the ACR or DCR categorization task. Moreover, the PC method results in the smallest measurement variance and thus produces the most accurate results \cite{mantiuk2012comparison}.
Based on the first point above, forced-choice PCs are easier to decide in subjective trials and require less time than the ACR or DCR categorization task. In addition, the PC method yields the lowest measurement variance and thus provides the most accurate results~\cite{mantiuk2012comparison}.

In a slight variation of PC, the reference stimulus is placed in the middle of the pair of stimuli to be compared. The task of the observers is to select the one that looks more (or less) similar to the reference \cite{ponomarenko2009tid2008, ponomarenko2015image,sun2017mdid,  men2020subjective}. Similar to DCR, assessing the (dis)similarity of the distorted stimuli to the reference should lead to a more appropriate, informed choice of the subject than a PC without reference. Such an approach can be seen as a special case of triplet comparisons (TC). In TC tests, three stimuli are displayed simultaneously, and the (dis)similarity to the stimulus placed in the middle, called the \emph{pivot}, is asked to be compared. In general, the pivot can be any element of the sequence of distorted stimuli (\textit{general TC}), and the PC with reference corresponds to TC where the pivot is fixed and equal to the reference for all comparisons. In our main experimental study, we applied the latter approach, which we call \textit{baseline TC}. In a secondary experiment, we used general triplets and showed their potential to further increase the performance of FR-IQA compared to DCR and baseline TC.

Like PC, TC avoids the problems discussed above for ACR and DCR. The subject responses to baseline TC can be interpreted as answers to the corresponding 2AFC questions that show only the two distorted stimuli with the reference. Therefore, the same Thurstonian reconstruction may be applied. However, for general triplet comparisons with arbitrary pivot stimuli, this is not possible. 
Triplet comparisons had already been introduced in psychophysics by Torgerson \cite{torgerson1958theory} in 1958, and recently found much interest in vision science and machine learning \cite{haghiri2020estimation}. Many scale reconstruction methods for TC have been proposed. However, with just one exception, none of them are based on the Thurstonian model allowing to give scale values in meaningful JND units as discussed above. Hence, in this paper, we propose a complete method for scale reconstruction from triplet comparisons, based on Thurstone's model and MLE, to produce scale values expressed in perceptual JND units.

%---------------------I/VQA Datasets ----------------
\begin{table*}[t!]
\caption{IQA/VQA/JND Datasets with Artificial Distortions}
\centering
\begin{threeparttable}
    %\renewcommand\TPTminimum{\linewidth}
%\resizebox{1\textwidth}{!}{ 
\footnotesize
\begin{tabular}{l | l r r r r l l r l}
\multirow{2}{*}{\textbf{IQA Datasets}} & Year & SRC\tnote{a}  & DST\tnote{b}  & DST & DST & Method & Scale  & Average & Environment\\
& & & & Types & Levels & & Range\tnote{c} & Responses\tnote{d}\\
\hline
LIVE IQA \cite{sheikh2006statistical} & 2006 & 29 & 779  & 5  & 5--6 & ACR  & $[0,100]$  & 23 & Lab \\
CSIQ IQA\cite{larson2010most} & 2010 & 30 & 866   & 6  & 3--5 & Customized\tnote{e} &  $[0,1]$ & 5--7 & Lab \\
TID2008 \cite{ponomarenko2009tid2008} & 2009 & 25  & 1700   & 17  & 4  & Baseline TC\tnote{f} & $[0,9]$  & 33 & Lab \\
VCL@FER \cite{vclfer} & 2012 & 23 & 552  & 4 & 6 & SS-NS\tnote{g} & [1,100]  & 16--36 & Lab \\
TID2013 \cite{ponomarenko2015image}  & 2013 & 25 &  3000  &  24 &  5 &  Baseline TC\tnote{f}  &  $[0,9]$  &  30  & Lab \\
CID:IQ \cite{liu2014cid} & 2014 & 23 & 690 & 6 & 5 & ACR  & [1,9]  & 17 & Lab \\
MDID \cite{sun2017mdid} &  2016 & 20 & 1600\tnote{h}  & 5 & 4 & Baseline TC\tnote{f}  & $[0,8]$ & 33--35  & Lab \\
KADID-10k \cite{lin2019kadid} & 2019 & 81  & \num{10125}   & 25 & 5 & DCR & $[1,5]$ & 30 & Crowdsourcing\\
\hline
KonFiG-IQA (Part A)& 2021 & 10 & 840 & 7 & 12 & Baseline TC & $[0,\infty)$ & 97& Crowdsourcing \\
KonFiG-IQA (Part B)& 2021 & 10 & 300 & 1 & 30 & General TC & $[0,\infty)$ & 582 & Crowdsourcing \\
\hline
\multirow{2}{*}{\textbf{VQA Datasets}} \\
\\ \hline
EPFL-PoliMI \cite{5496296}& 2009 & 6 & 72\tnote{i} & 2 &  6  & SS-CS\tnote{j} & [0,5] & 17--23 & Lab\\
LIVE VQA \cite{seshadrinathan2010study} & 2010 & 10 & 150  & 4 & 3--4  & ACR & $[0,100]$  & 38 & Lab\\
IVP \cite{zhang2011ivp} & 2011 & 10 & 128   & 4 & 4--5  & ACR & $[1,5]$  & 42 & Lab\\
CSIQ VQA \cite{CSIQ}  & 2014 & 12 & 216  & 6&  3 & SS-NS\tnote{g} & $[0, 100]$ & 35 & Lab\\
MCL-V \cite{lin2015mcl} & 2015 & 12 & 96   & 2 &  4  & PC & $[0,8]$  & 32 & Lab\\
NFLX \cite{netflix} \cite{vmaftech}& 2016 &  9 & 70    & 2 & 7--11  &  DCR & $[0,100] $  & N/A & Lab \\
\hline
\multirow{2}{*}{\textbf{JND Datasets}} & &  & &  &  &   & JND  &  &  \\
 & &  & &  &  &   & Range  &  &  \\
\hline
MCL-JCI \cite{jin2016statistical} & 2016 & 50 &  5000 & 1 & 100 & PC\tnote{k} & $\{1,...,100\}$ & 30 & Lab \\
VVC-JND \cite{shen2020jnd}& 2020 & 202 &  7878 & 1 & 39 &  PC\tnote{k} & $\{13,...,51\}$ & 20 & Lab\\%,shen2020just
MCL-JCV \cite{wang2016mcl} & 2016 & 30 & 1530 & 2 & 51 & PC\tnote{l}  & $\{1,...,51\}$ & 50 & Lab\\
VideoSet \cite{wang2017videoset} & 2017 & 220 & \num{44800}\tnote{m} & 1 & 51 & PC\tnote{k}  & $\{1,...,51\}$ & 30 & Lab\\
SIAT-JSSI~\cite{fan2019interactive} &2019& 10&3510&2&51/300 & PC\tnote{k} & $\{1,...,51\}$,$\{1,...,300\}$\tnote{n} &36&Lab\\
JND-Pano~\cite{liu2018jnd}&2018&40 & 4000& 1& 100&  PC\tnote{o}& $\{1,...,100\}$&  25 & Lab\\
QAD-HEVC \cite{huang2017measure}& 2017 & 40& 2040& 1& 51& PC\tnote{k}& $\{1,...,51\}$ & 30& Lab %Huang~et~al.
\end{tabular}
%}
\begin{tablenotes}
\scriptsize%
\item[a] SRC: number of source images/videos
\item[b] DST: number of distorted images/videos
\item[c] Range of MOS/DMOS (possibly scaled to a larger interval), reconstructed scale from comparisons, or JNDs
\item[d] Total number of quality ratings, resp.\ comparisons, divided by the number of images or videos. For JND datasets, it is the number of observers per source stimulus.
\item[e] All distorted images of one sequence were displayed simultaneously. Observers were asked to place the images in relation to each other according to the overall quality they perceived. 
\item[f] Baseline TC: PC with the reference image provided.
\item[g] Single stimulus with numerical scales (SS-NS), observers are asked to assign each stimulus an integer from a given range. 
\item[h] For each source 20 source images, 80 distorted images with combined distortions were generated by concatenating 4 types of distortions, in the order of Gaussian blur, contrast change, compression (JPEG or JPEG\,2000), and Gaussian noise. Each distortion was randomly distorted at one of the four levels.
\item[i] Each of the 6 source videos was distorted by a simulation of packet loss distortion at 6 different packet loss rates with 2 channel realizations, resulting in a total of 12 distorted versions.
\item[j] Single stimulus with continuous scale (SS-CS), subjective quality scores of the videos were obtained using the single stimulus method with a 5-point continuous scale.
\item[k] Observers were asked to compare two images displayed side by side and determine whether the differences between them are noticeable. 
\item[l] Observers were asked to determine whether the differences between the two video clips displayed one after another are noticeable. 
\item[m] Four different resolutions were generated from each source video. The whole process of encoding and JND evaluation was carried out for each resolution. Therefore, the total number of distorted/encoded sequences in the VideoSet is $4 \times 220 \times 51$.
\item[n] Reference stereoscopic images were compressed using H.265 intra coding with the quantization parameter ranging from 1 to 51 and JPEG\,2000 with the compression ratio ranging from 1 to 300. %(Matlab function “imwrite”)
\item[o] Subjects wore a head-mounted display device and were free to control the field of view to compare two panoramic images whether they could notice a difference.
\end{tablenotes}
\end{threeparttable}
\label{tb:ivqadatabase}
\end{table*}

% -------------------------------------------------
% --------------NOTE for LIVE-IQA -----------------
%- Scores: for distorted images, [1,100] (DMOS--> zsocred --> map to [1,100]); ref. images are of score 0
% Dist. levels: 5 distortion types for four distortion types, for JPEG2000, it's various (not counted yet)

%-------------------------------------------------
%--------------NOTE for CSIQ IQA -----------------
%- customized subjective method: all of the distorted versions of the reference image were displayed simultaneously and placed in relation to each other based on the overall quality by the observers.
%- Distortion levles; Except for contrast, all other 5 distortion types were distorted at 5 levels. For contrast, images were distorted at 3-4 levels.

%---------------NOTE for MDID---------------------
%* Multiplicative distortions between each other (except for JPEG multiplied with JPEG 2000) were involved, resulting in 1600 distorted images in total.

%------------------------------------------------------------------------------------------
%-----´SUBSECTION --- SUBSECTION --- SUBSECTION --- SUBSECTION --- SUBSECTION -------------
\subsection {Current Visual Quality Datasets}
%------------------------------------------------------------------------------------------

%Besides FR-IQA, there are other cases where a visual reference stimulus is distorted at different levels of severity and compared. In full reference video quality assessment (FR-VQA) short image sequences of a few seconds are considered and rated according to visual quality.
In addition to FR-IQA, there are other applications in which a visual reference stimulus is distorted to various levels of severity and compared. For example, in full-reference video quality assessment (FR-VQA), short image sequences of a few seconds are viewed and evaluated for visual quality. Since video data transmission is the dominant load on the Internet traffic, FR-VQA for compressed video streaming is the most relevant application of visual quality assessment methods. Recently, it has been proposed to replace the quality assessment scale based on MOS or DMOS from subjective categorical or nominal ratings by the just noticeable  difference \cite{lin2015experimental}. In JND assessments, the distorted reference images or videos are also compared to the reference, and the minimal distortion level that leads to a perceivable difference of the stimulus is reported. In all of these cases of visual quality assessment, the boosting techniques that we have developed can be applied to increase the sensitivity, allowing for a more fine-grained visual analysis of the range of distortions.

In Table\,\ref{tb:ivqadatabase} we present an overview of the currently available datasets for subjectively assessed visual quality for FR-IQA, FR-VQA, and JND. In this paper, we contribute a new dataset, KonFiG-IQA (Konstanz Fine-Grained IQA), which is also listed in the table. Several points are noteworthy:
\begin{enumerate}
          \item[(1)]  Quality and JND assessment techniques of the current datasets are dominated by the classical approaches such as ACR, DCR, and their variations where a discrete or continuous ratio scale replaces the categorical scale.
        %\textcolor{red}{and SSCQE \emph{(should it be single stimulus with numerical or continuous scale?)}. [The latter refers to video only where viewers dynamically rate the quality of an impaired video sequence using a slider mechanism with an associated quality scale.]}
        For TID2008, TID2013, and MDID, baseline triplet comparisons were carried out. For the scale reconstruction, the Swiss-system tournament style point scoring resp.\ a ranking procedure based on insertion sort was used.
        %in contrast to the state-of-the-art Thurstonian MLE-based approach. 
        The only dataset for which a probabilistic MLE-based reconstruction from comparisons was carried out is MCL-V, without final conversion into JND units. Here, the Bradley-Terry model was used, which is very similar to the Gaussian Thurstonian model.
        Thus, in all current IQA and VQA datasets, possibly except for MCL-V, artefacts due to nonlinear scaling of perceptual quality may be present.
    %A Swiss-system tournament is a non-eliminating tournament format that features a fixed number of rounds of competition, but considerably fewer than for a round-robin tournament; thus each competitor (team or individual) does not play all the other competitors. Competitors meet one-on-one in each round and are paired using a set of rules designed to ensure that each competitor plays opponents with a similar running score, but does not play the same opponent more than once. The winner is the competitor with the highest aggregate points earned in all rounds. All competitors play in each round unless there is an odd number of them. [Wikipedia]
    \item[(2)] In all current IQA and VQA datasets, only a small number of distortion levels were applied, up to 6 for images and up to 11 for video. This choice corresponds to the small number of only 5 nominal quality values available in ACR and DCR. It would be desirable to introduce IQA and VQA datasets with a larger number of distortion levels, especially at the high end of quality. This would allow for training machine learning techniques for objective quality assessment aimed at applications delivering high quality imagery and streaming video over the internet at a  minimal but sufficient bitrate.
    \item[(3)] Two recent trends can be observed. In 2019, the first crowdsourced FR-IQA dataset was introduced (KADID-10k), and more are likely to come, like the two sets included in this paper. Moreover, since 2016 the first datasets for JND were established both for images and video, and more of them can be expected, including crowdsourced JND datasets. 
\end{enumerate}
The new dataset from our study, KonFiG-IQA, stands out from the rest of the FR-IQA datasets in the following aspects: 
\begin{enumerate}
    \item[(1)]  The number of distortion levels is larger and designed by perceptual consideration, namely 12, resp.\ 30  distortion levels, perceptually equally spaced over a range of 3\,JND.
    \item[(2)] The number of ratings, resp.\ triplet comparisons, averaged per distorted image, is much larger, 97 per image in Part A and up to 875 in Part B. This allowed for an extensive analysis of reliability and convergence of the resulting scale values.
    \item[(3)] The scale values are derived from the probabilistic Thurstonian MLE process and converted to give perceptually linear quality scale values in meaningful JND units. 
\end{enumerate}

%------------------------------------------------------------------------------------------
%-----´SUBSECTION --- SUBSECTION --- SUBSECTION --- SUBSECTION --- SUBSECTION -------------
\subsection {Boosting for Visual Quality Assessment: Motivation}
%------------------------------------------------------------------------------------------

When designing visual quality datasets, the creators usually try to cover the complete range of visual quality with only a few samples taken from a very large pool of images or videos. In the case of FR-IQA or FR-VQA, for each source stimulus, a large range of values for the distortion parameters of the chosen distortion types may be applied, yielding stimuli that are very close to the original as well as others with severe distortions. It may be hard to reliably assess the resulting small and large quality differences in subjective quality assessment experiments. Let us illustrate this, assuming the Thurstionian model for visual quality impairment. 

Consider a sequence of $M+1$ images $I_0,\ldots, I_{M}$, where $I_0$ is a pristine source image and $I_1,\ldots, I_{M}$ are increasingly distorted versions of the source. Let their image qualities on the impairment scale be modelled by random variables having Gaussian distributions with means $\mu_0, \ldots, \mu_{M}$ and standard deviations equal to $\sigma = \sqrt{0.5}/\Phi^{-1}(0.75) \approx 1.0484$, where $\Phi$ is the normal cumulative distribution function (CDF). This particular choice of the variance scales the quality values to be expressed in convenient JND units as pointed out in Subsection\,\ref{sec_subFRIQA}. 

When comparing two such stimuli that are close to each other in their mean values, the corresponding effect size~$\Theta$ determines the difficulty of assessing their difference. A common way to define the effect size is the standardized mean difference. In our case, 
$$
    \Theta = \frac{|\mu_i-\mu_j|}{\sigma} \approx 0.9539\,\Delta\mu_{i,j}.
$$ 
Smaller effect sizes indicate the necessity of larger sample sizes. Effect sizes $\Theta \in [0.2,0.5), [0.5,0.8), [0.8,1.3) $ and $\Theta \ge 1.3$ are called small, medium, large, and very large, respectively. For example, at $\Delta \mu = 1$\,JND, we have $\Theta = 0.9539 \in [0.8,1.3)$, and thus, a large effect size. This is in line with the detection rate of 50\% at 1\,JND quality difference. However, for differences $\Delta \mu < 0.2097$\,JND we have $\Theta < 0.2$ and therefore a very small effect size. 

Typically, image quality datasets have hundreds of images with perceptual qualities ranging over just a few JND. If one were to compare these images to each other, such small effect sizes would become relevant. Therefore, quality assessment techniques that enlarge the effect size would be beneficial, allowing to distinguish image qualities with small differences with a smaller number of samples. For this purpose, we propose and study our boosting methods in this contribution.

It is quite natural that small visual differences are difficult to assess. It has been observed, in addition, that large quality differences are hard to quantify: Stimulus differences larger than about 1.5\,JND cannot be reliably assessed by the human visual system, presumably due to a kind of saturation effect by overwhelming noise \cite{keelan2003iso}. 

In addition to the problem of subjectively quantifying large distortions reliably, there is a numerical problem to reconstruct such large quality differences from paired comparisons with subsequent Thurstonian scale reconstruction.
%even if one assumes that large differences can be judged by human observers just as well as smaller distortions. 
In the Thurstonian model, a quality difference is given by a normal random variable with unit variance and the mean equal to the quality difference on the perceptual quality scale. Thus, a fraction $p \in (0,1)$ of observations that correctly identify the better quality image in the given pair gives rise to the reconstructed quality difference $\Phi^{-1}(p)$, where $\Phi$ again denotes the normal CDF. However, when the quality difference is large, most observers ($k$ out of $n$) will agree on which image is of better quality. Therefore, the fraction $p = k/n$ is close to 1 and $\Phi^{-1}(p)$ depends very sensitively on $p$ when $p$ is near 1 or 0. For $k=n$, we even have $p=1$ and $\Phi^{-1}(1) = \infty$. So, if just one observer would change his/her response, the reconstructed quality difference between the stimuli would drastically change. 
%\begin{figure}[t!]
%\centering{
%\includegraphics[width=0.98\columnwidth]{images/rmse.png}
%}
%\caption{In this simulation we consider reconstruction of a quality difference from a set of $n$ pair comparisons of two stimuli with a given difference of mean observed qualities, shown on the horizontal axis. On the vertical axis the resulting root-mean-square error of the  reconstruction of the difference is shown. This simulation confirms that it is not advisable to apply paired comparisons, when the difference to be assessed is so strong such that nearly all observers agree about which of the two stimuli should be chosen as the better (or stronger) one.}
%\label{rmse_mean_difference}
%\end{figure}

We analyse this effect by simulating subjective PC using the probabilistic model to compute $E_{\text{rms}}$, the root of the expected square error of the reconstruction as follows. We assume a quality difference on $\Delta \mu$ and collect $n$ responses for the corresponding pair comparison. Then we make use of the binomial distribution with probability $\Phi\left(\Delta \mu\right)$ to get the result.\footnote{
    The straightforward implementation of this formula will lead to the so-called zero-frequency problem when $k=0$ or $k=n$. In these cases, $\Phi^{-1}\left(\frac{k}{n}\right)$ will be $\pm \infty$, rendering a reconstruction infeasible. To avoid this problem, it is common practice to install a `prior' by adding half a vote to either option, thus computing  $\Phi^{-1}\left(\frac{k+0.5}{n+1}\right)$, which is what we have done here as well.
}

\begin{equation*}
%\begin{split}
   {E_{\text{rms}}(\Delta \mu, n) =}
\end{equation*}
\vspace{-15pt}
\begin{equation*}
{\textstyle \left[ 
    \sum_{k=0}^n 
    {n \choose k} \Phi\left(\Delta \mu\right)^k \left(1 - \Phi\left(\Delta \mu\right)\right)^{n-k}
    \left( \Phi^{-1}\left(\frac{k}{n}\right) - \Delta \mu \right)^2
\right]^{1/2}}
%\end{split}
\end{equation*}

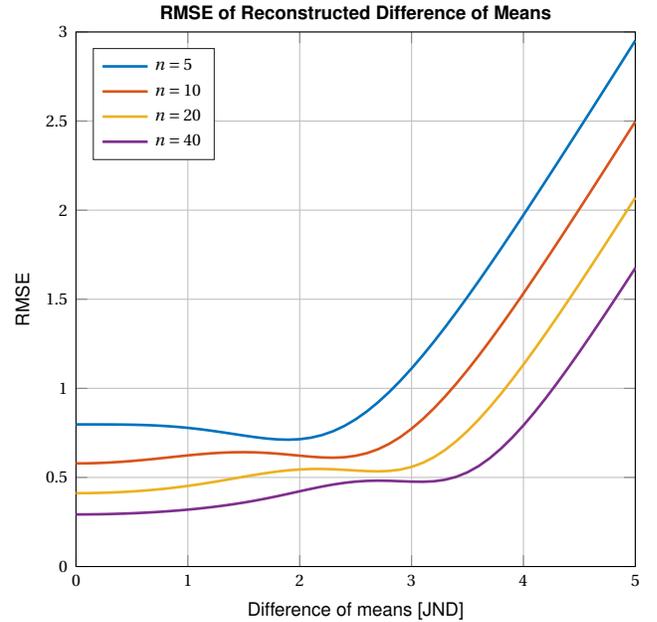
\begin{figure}[t!]
\centering{
\definecolor{mycolor1}{rgb}{0.00000,0.44700,0.74100}%
\definecolor{mycolor2}{rgb}{0.85000,0.32500,0.09800}%
\definecolor{mycolor3}{rgb}{0.92900,0.69400,0.12500}%
\definecolor{mycolor4}{rgb}{0.49400,0.18400,0.55600}%
\begin{tikzpicture}

\begin{axis}[%
width=0.41\textwidth,
height=2.8in,
at={(0in,0in)},
scale only axis,
xmin=0,
xmax=5,
label style={font=\sffamily},
xticklabel style={font=\sffamily\scriptsize},
yticklabel style={font=\sffamily\scriptsize},
xlabel style={font=\sffamily\scriptsize},ylabel style={font=\sffamily\scriptsize},
title style={yshift=-1ex,font=\bfseries\sffamily\scriptsize},
xlabel={Difference of means [JND]},
ymin=0,
ymax=3,
ylabel={RMSE},
axis background/.style={fill=white},
title={RMSE of Reconstructed Difference of Means},
xmajorgrids,
ymajorgrids,
legend style={at={(0.03,0.97)}, anchor=north west, legend cell align=left, align=left, draw=white!15!black, font = \sffamily\scriptsize}
]
\addplot [color=mycolor1, line width=1.0pt]
  table[row sep=crcr]{%
0	0.79755834655254\\
0.1	0.797544641877022\\
0.2	0.797474453924602\\
0.3	0.797263103139056\\
0.4	0.796777740655601\\
0.5	0.795849143280305\\
0.6	0.794287082734938\\
0.7	0.791898365028754\\
0.8	0.788506644262075\\
0.9	0.783973220856261\\
1	0.778218186240038\\
1.1	0.771241412778177\\
1.2	0.763142940791409\\
1.3	0.754142207101491\\
1.4	0.744595204718338\\
1.5	0.735007974460021\\
1.6	0.726043764222455\\
1.7	0.718519857183403\\
1.8	0.713388896991108\\
1.9	0.711699450211577\\
2	0.714532879204812\\
2.1	0.722919375145468\\
2.2	0.737744490495025\\
2.3	0.759665099105736\\
2.4	0.78905511240694\\
2.5	0.825993631886905\\
2.6	0.870294621135682\\
2.7	0.921565057102022\\
2.8	0.979273707028966\\
2.9	1.04281546416603\\
3	1.11156271575545\\
3.1	1.18490143839073\\
3.2	1.26225363119679\\
3.3	1.34308927958518\\
3.4	1.42693111719015\\
3.5	1.51335485244994\\
3.6	1.60198677873252\\
3.7	1.69250003717056\\
3.8	1.7846103187152\\
3.9	1.87807146503451\\
4	1.9726712200868\\
4.1	2.0682272584356\\
4.2	2.16458354349508\\
4.3	2.26160702851473\\
4.4	2.35918469193303\\
4.5	2.45722088848697\\
4.6	2.55563499321047\\
4.7	2.65435931432893\\
4.8	2.75333725146163\\
4.9	2.85252167662283\\
5	2.95187351683449\\
};
\addlegendentry{$n=5$}

\addplot [color=mycolor2, line width=1.0pt]
  table[row sep=crcr]{%
0	0.578594518037086\\
0.1	0.579159974184384\\
0.2	0.580846216283203\\
0.3	0.583621200251917\\
0.4	0.587426306950109\\
0.5	0.592169667636458\\
0.6	0.597718964878803\\
0.7	0.603895514279479\\
0.8	0.610471468664699\\
0.9	0.617171719313539\\
1	0.623681540199729\\
1.1	0.629660340782572\\
1.2	0.634761207392164\\
1.3	0.638655359362669\\
1.4	0.641060308368198\\
1.5	0.641770398790816\\
1.6	0.640688459379023\\
1.7	0.637857381110505\\
1.8	0.633490361104958\\
1.9	0.627998067942521\\
2	0.622009808474803\\
2.1	0.616383710620788\\
2.2	0.612198167437171\\
2.3	0.610714441016956\\
2.4	0.613300993744947\\
2.5	0.621317346976686\\
2.6	0.635970595622444\\
2.7	0.658175927924533\\
2.8	0.688460979731979\\
2.9	0.726941975137078\\
3	0.773371717476633\\
3.1	0.827233818921098\\
3.2	0.887849080857797\\
3.3	0.954468195882211\\
3.4	1.02633930028042\\
3.5	1.10275023157518\\
3.6	1.18305083565581\\
3.7	1.26666172913819\\
3.8	1.35307485990444\\
3.9	1.44184962009097\\
4	1.53260688047864\\
4.1	1.62502232480916\\
4.2	1.71881982810261\\
4.3	1.81376524532184\\
4.4	1.90966076435307\\
4.5	2.00633986416123\\
4.6	2.10366286250184\\
4.7	2.20151301204603\\
4.8	2.29979309444513\\
4.9	2.3984224605623\\
5	2.49733446747667\\
};
\addlegendentry{$n=10$}

\addplot [color=mycolor3, line width=1.0pt]
  table[row sep=crcr]{%
0	0.411611608294449\\
0.1	0.41198858118118\\
0.2	0.413123605065175\\
0.3	0.415029048275095\\
0.4	0.417725602665513\\
0.5	0.421242040309948\\
0.6	0.425614101437998\\
0.7	0.430881683642719\\
0.8	0.437083325059152\\
0.9	0.444247048449733\\
1	0.452377133766208\\
1.1	0.46143739504927\\
1.2	0.471332924960771\\
1.3	0.481893651834623\\
1.4	0.492863889595946\\
1.5	0.503901899827565\\
1.6	0.514592228314409\\
1.7	0.524471572675681\\
1.8	0.533066821942927\\
1.9	0.539942305569538\\
2	0.544752538547974\\
2.1	0.547296812767293\\
2.2	0.547572499929652\\
2.3	0.54582431157158\\
2.4	0.542586268820314\\
2.5	0.538710973316176\\
2.6	0.535376436642245\\
2.7	0.53405499584801\\
2.8	0.536425524734383\\
2.9	0.544217043226359\\
3	0.558996435956091\\
3.1	0.581950149431349\\
3.2	0.613732951189286\\
3.3	0.654437153429529\\
3.4	0.703679548475808\\
3.5	0.760754087035161\\
3.6	0.824788989773768\\
3.7	0.894869899451673\\
3.8	0.970118878664796\\
3.9	1.04973609369201\\
4	1.13301610548253\\
4.1	1.21934929481166\\
4.2	1.30821571038061\\
4.3	1.39917572858405\\
4.4	1.49185991105129\\
4.5	1.5859592359277\\
4.6	1.68121620838632\\
4.7	1.7774170094582\\
4.8	1.8743846760326\\
4.9	1.97197323325569\\
5	2.07006267608494\\
};
\addlegendentry{$n=20$}

\addplot [color=mycolor4, line width=1.0pt]
  table[row sep=crcr]{%
0	0.292319433340024\\
0.1	0.292571538272745\\
0.2	0.293329811759806\\
0.3	0.294600173627407\\
0.4	0.296392639871496\\
0.5	0.298721561037576\\
0.6	0.301605986682631\\
0.7	0.305070192186881\\
0.8	0.309144416735806\\
0.9	0.313865856524344\\
1	0.319279893415562\\
1.1	0.325441339878861\\
1.2	0.332415051522665\\
1.3	0.340274560137379\\
1.4	0.349096565205111\\
1.5	0.358948657783471\\
1.6	0.369868267483025\\
1.7	0.381833129448237\\
1.8	0.394727415629405\\
1.9	0.408311736876828\\
2	0.422207388707259\\
2.1	0.435903859089458\\
2.2	0.448793864120456\\
2.3	0.460233990870299\\
2.4	0.469623963496241\\
2.5	0.476495236710163\\
2.6	0.480600084784253\\
2.7	0.48199432166088\\
2.8	0.481108323645659\\
2.9	0.47879998349428\\
3	0.476377709672319\\
3.1	0.475571470511889\\
3.2	0.47842058147419\\
3.3	0.487053559068049\\
3.4	0.503376614118188\\
3.5	0.528756835211122\\
3.6	0.563827022246073\\
3.7	0.608490557551016\\
3.8	0.662094307517678\\
3.9	0.723664774261831\\
4	0.792114353602143\\
4.1	0.866378728799048\\
4.2	0.945489379485145\\
4.3	1.0286013804375\\
4.4	1.11499586041967\\
4.5	1.20407036593776\\
4.6	1.29532471646276\\
4.7	1.38834618064037\\
4.8	1.48279567353303\\
4.9	1.57839558490347\\
5	1.67491933713337\\
};
\addlegendentry{$n=40$}

\end{axis}
\end{tikzpicture}%
}
\vspace{-10pt}
\caption{In this simulation, we consider the reconstruction of a quality difference from a set of $n$ pair comparisons of two stimuli with a given difference of mean observed qualities, shown on the horizontal axis. On the vertical axis, the resulting RMSE of the reconstruction of the difference is shown. This simulation confirms that it is not advisable to apply paired comparisons when the difference to be assessed is so strong that nearly all observers agree on which of the two stimuli should be chosen as the better (or stronger) one.}
\label{rmse_mean_difference}
\end{figure}

Figure\,\ref{rmse_mean_difference} illustrates this root-mean-square error (RMSE) as a function of $\Delta \mu$ for $n=5,10,20,40$. For increasing quality differences, we see that $E_{\text{rms}}$ is stable and nearly constant until about 2 or 3\,JND. From then on, the error increases approximately linearly. This gives another reason to restrict paired comparisons (resp.\ triplet comparisons) to cases where image quality differences are not too large, i.e., up to about 2 or 3\,JND. 

With our boosting methods implemented to enlarge the distortions applied to source images, this effect of decreased psychovisual sensitivity and increased reconstruction noise at large distortion levels can be overcome partially, as our experiments will show. However, by boosting image differences also between distorted images, we will show that also for large distortions, fine-grained quality scaling can be achieved reliably.

%\begin{figure}[t!]
%\centering{
%\includegraphics[width=0.24\textwidth]{images/SRC28_grd.png}$~$\includegraphics[width=0.24\textwidth]{images/SRC28_grdzoom.png}  \\\vspace{3pt}
%\includegraphics[width=0.24\textwidth]{images/SRC28_JPEG2000_47.336416.png}$~$\includegraphics[width=0.24\textwidth]{images/Ampli_SRC28_jpeg2000_47.336416-small.png}}
%\caption{{\textcolor{red}[add caption]}This figure shows an example of boosting by artefact amplification and zooming. The top left image is an original, undistorted source image. On the right a zoomed crop of is pictured. The source image was compressed by JPEG\,2000 with a compression ratio of 1:47.34, resulting in the lower left image.
% 47.336416 this was the original number put. at level 2
%The distortions due to JPEG\,2000 compression are barely visible. However on the lower right, the same compressed image is shown after artefact amplification and zooming. Now, by this boosting technique, the distortions in the flower petals and stigma became emphasized and clearly visible.}
%\label{BoostingExample}
%\end{figure}

\begin{figure}[t!]
\centering{

\begin{tikzpicture}

\begin{axis}[%
width=1.65in,
height=2.2in,
at={(0in,0in)},
scale only axis,
axis on top,
xmin=0.5,
xmax=384.5,
tick align=outside,
y dir=reverse,
ymin=0.5,
ymax=512.5,
axis line style={draw=none},
ticks=none,
title style={font=\bfseries\sffamily\scriptsize, yshift=-1ex},
title={(b) Compressed by JPEG 2000}
]
\addplot [forget plot] graphics [xmin=0.5, xmax=384.5, ymin=0.5, ymax=512.5] {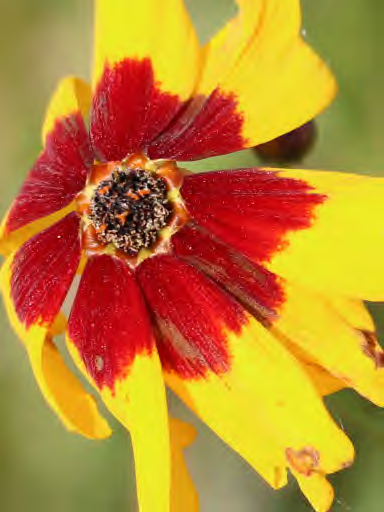};
\end{axis}

\begin{axis}[%
width=1.65in,
height=2.2in,
at={(0in,2.45in)},
scale only axis,
axis on top,
xmin=0.5,
xmax=384.5,
tick align=outside,
y dir=reverse,
ymin=0.5,
ymax=512.5,
axis line style={draw=none},
ticks=none,
title style={font=\bfseries\sffamily\scriptsize, yshift=-1ex},
title={(a) Reference Image}
]
\addplot [forget plot] graphics [xmin=0.5, xmax=384.5, ymin=0.5, ymax=512.5] {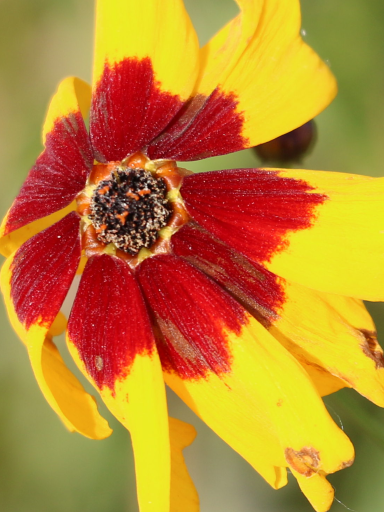};
\end{axis}

\begin{axis}[%
width=1.65in,
height=2.2in,
at={(1.67in,0in)},
scale only axis,
axis on top,
xmin=0.5,
xmax=384.5,
tick align=outside,
y dir=reverse,
ymin=0.5,
ymax=512.5,
axis line style={draw=none},
ticks=none,
title style={font=\bfseries\sffamily\scriptsize, yshift=-1ex},
title={Zoomed Version of (b)}
]
\addplot [forget plot] graphics [xmin=0.5, xmax=384.5, ymin=0.5, ymax=512.5] {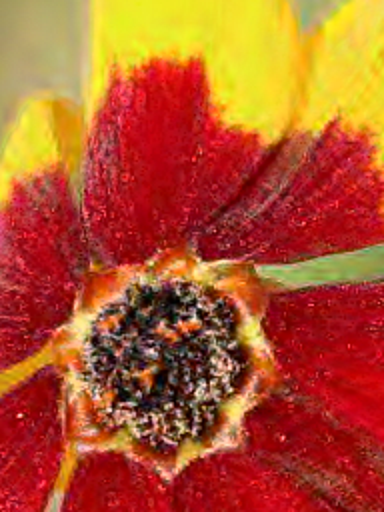};
\end{axis}

\begin{axis}[%
width=1.65in,
height=2.2in,
at={(1.67in,2.45in)},
scale only axis,
axis on top,
xmin=0.5,
xmax=384.5,
tick align=outside,
y dir=reverse,
ymin=0.5,
ymax=512.5,
axis line style={draw=none},
ticks=none,
title style={font=\bfseries\sffamily\scriptsize, yshift=-1ex},
title={Zoomed Version of (a)}
]
\addplot [forget plot] graphics [xmin=0.5, xmax=384.5, ymin=0.5, ymax=512.5] {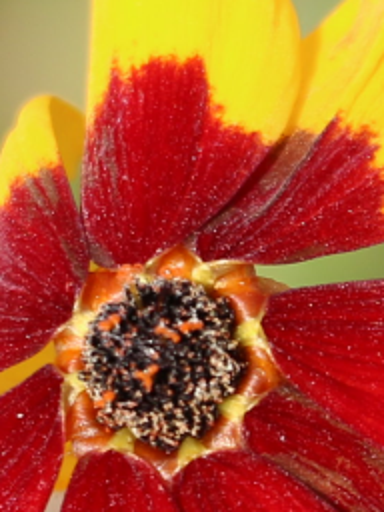};
\end{axis}

\end{tikzpicture}%
}
\caption{This figure shows an example of boosting by artefact amplification and zooming. The top left image is an original, undistorted source image. On the right, a zoomed crop of it is pictured. The source image was compressed by JPEG\,2000 with a compression ratio of 1:47.34, resulting in the lower-left image.
% 47.336416 this was the original number put. at level 2
The distortions due to JPEG\,2000 compression are barely visible. However, on the lower right, the same compressed image is shown after artefact amplification and zooming. Now, by this boosting technique, the distortions in the flower petals and stigma became emphasized and clearly visible.}
\label{BoostingExample}
\end{figure}

%------------------------------------------------------------------------------------------
%-----´SUBSECTION --- SUBSECTION --- SUBSECTION --- SUBSECTION --- SUBSECTION -------------
\subsection {Boosting by Artefact Amplification and Zooming}
%------------------------------------------------------------------------------------------

The approach of boosting for visual quality assessment is to enlarge the differences between stimuli artificially. In the basic setting, using baseline triplets for comparison, we are asking subjects to identify the distorted image of the presented pair that most closely resembles the reference image, which is equal to the source image for generating the distorted versions. Here it is the distortions in each derived image that need to be enlarged for the boosting effect. Distortions typically occur in two main aspects, in terms of greyscale intensity, respectively color, or spatially. Therefore, the first two boosting techniques are as follows:
\begin{enumerate}
    \item[(1)] Artefact amplification (A): In a triplet comparison, the similarity of two distorted images with respect to the pivot image has to be judged. Artefact amplification scales the pixel-wise differences of each distorted image in the three color channels linearly. 
    \item[(2)] Spatial zooming (Z): A linear scaling of the size of the image enlarges the visual representation of image differences. Due to limitations on available screen size, spatial zooming may imply the necessity to crop the distorted and zoomed image.
\end{enumerate}
In Figure\,\ref{BoostingExample} we show an example of how this boosting reveals otherwise invisible or hard to detect distortions caused by JPEG\,2000 compression. 

%------------------------------------------------------------------------------------------
%-----´SUBSECTION --- SUBSECTION --- SUBSECTION --- SUBSECTION --- SUBSECTION -------------
\subsection {Boosting by the Image Flicker Method}
\label{SubsectionFlicker}
%------------------------------------------------------------------------------------------

In PC, the image stimuli are usually displayed on a screen side by side. To detect a small detail that differs between two images, the observer must search over both images and memorize the last examined detail of one image when the eye fixation point moves to the corresponding location in the other image. This task can be difficult. It is at the core of the popular fun game for kids, where differences between two seemingly identical comic drawings are to be found. 

The change detection task can be simplified by displaying the two images in the same screen space, one in the foreground and the other one invisible in the background. The observer can toggle the view between the two images using keystrokes or mouse clicks. In this setting, the eye needs to scan only half of the screen space, and the saccadic eye movements between the two images are replaced by key clicks. Moreover, from an evolutionary perspective, it is plausible that the fundamental ability of human perception to detect a change in the visual environment has developed to a high level. Thus, small changes in the visual field should be easier to detect than differences of two objects (or images) next to each other. 

The visual sensitivity to contrast change has been researched for a long time~\cite{watson1986temporal}. For simple test scenes, contrast is defined as the ratio of target intensity to background intensity. When the contrast changes periodically, e.g., in a sinusoidal fashion, the change becomes visible when its amplitude surpasses a certain threshold, called the contrast threshold. Its inverse is the contrast sensitivity, and its variation as a function of temporal frequency can be described by the temporal contrast sensitivity function (TCSF). For sufficiently high luminances, the contrast sensitivity reaches a maximum of about 200 near a frequency of 8\,Hz.

In 2014, the above concepts were applied for the first time in an image flicker viewing method for subjective assessment of barely visible image compression artefacts \cite{hoffman2014new}. Observers were presented with an original reference image, temporally interleaved with a test image, which was reconstructed  from the compressed reference. The flicker frequency was chosen as 7.5\,Hz, close to the maximum of the TCSF. This method was expected to make even subtle artefacts visible that would be undetectable in a side-by-side comparison. The paper did not compare the performance with that achievable by the side-by-side display, however.

In our work, we provide such studies by including the flicker viewing method as our third option of boosting techniques:
 \begin{enumerate}
    \item[(3)] Image flicker (F): Two images to be compared are displayed temporally, interleaved at a frequency of \mbox{8\,Hz}.
\end{enumerate}
The application of the flicker viewing technique in a triplet comparison requires adapting the visual appearance of the displayed scene. The triplet $(i,j,k)$ has the pivot image $I_j$, and we are asking the observer to answer the question of whether the perceived differences in the left image pair $(I_i,I_j)$ are larger or smaller than those in the right pair, $(I_k,I_j)$. Therefore, with the flicker viewing technique, we show two flickering images side by side: On the left, image $I_i$ alternates with the pivot, and on the right, it is $I_k$ that alternates with the pivot.

\begin{figure*}[t]
\centering{
\includegraphics[width=0.99\textwidth]{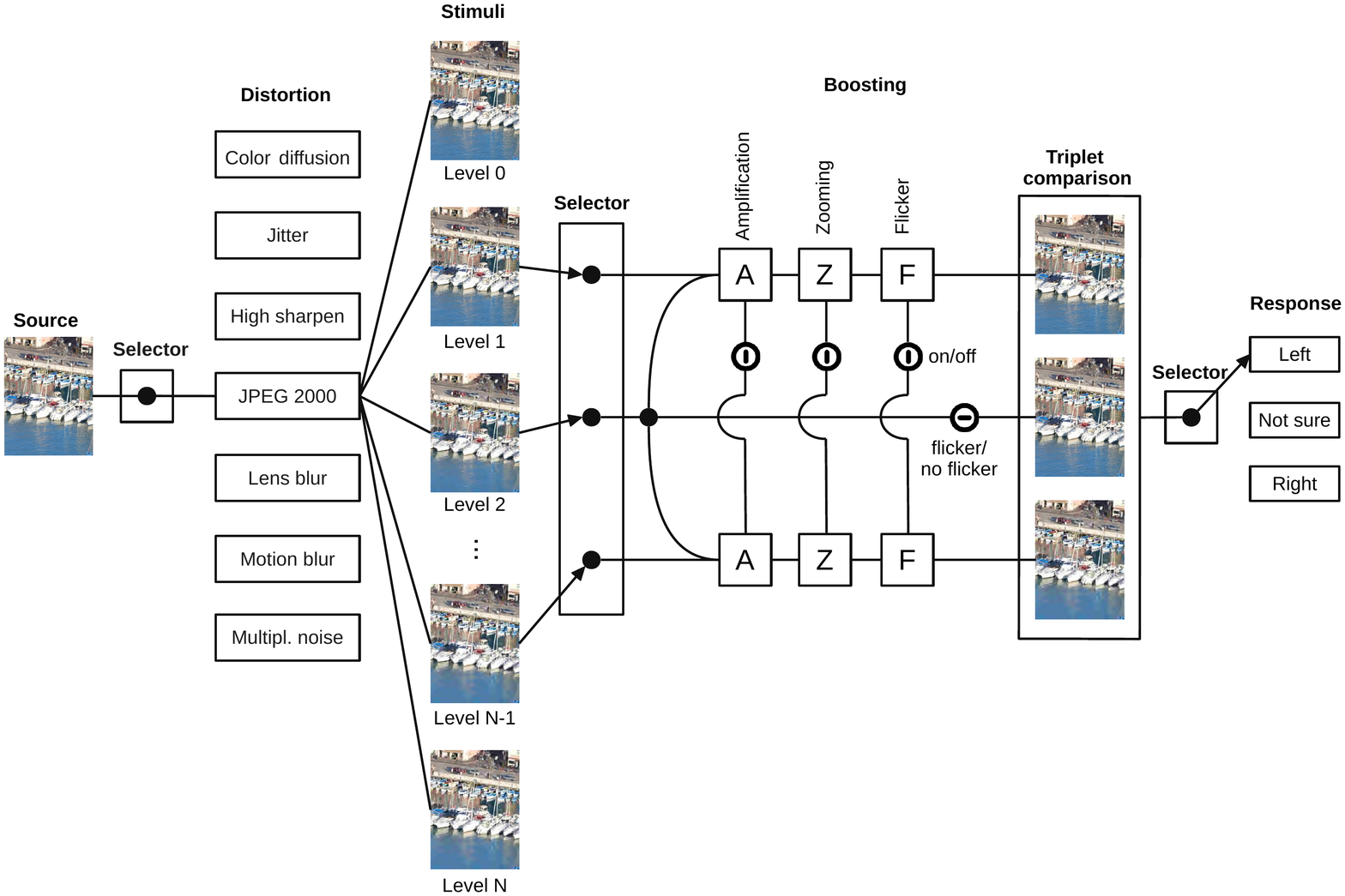}}
\caption{Flowchart of the proposed boosted triplet comparison method for subjective IQA.  Each pristine source image is distorted by seven types of distortion in $N$ levels, respectively. For a given distortion type, samples of three distorted images are drawn, and for each triplet, subjective assessments are made as to whether the left or the right image is perceptually closer to the centre, pivot image. %To enhance the perceptual sensitivity of comparisons between distorted images, three boosting strategies (amplification, zoom, and flicker) are deployed individually or in combination.
Three boosting strategies (amplification, zoom, and flicker) are deployed individually or in combination to enhance the perceptual sensitivity of comparisons between distorted images.
With the flicker option turned on, only two (flickering) images are shown side by side. The yielded triplet comparison results are to be used to estimate image quality impairment scales by Thurstonian reconstruction. 
}
\label{fig:btcflowchart1}
\end{figure*}

%------------------------------------------------------------------------------------------
%-----´SUBSECTION --- SUBSECTION --- SUBSECTION --- SUBSECTION --- SUBSECTION -------------
\subsection {Contributions}
%------------------------------------------------------------------------------------------
We expect (and we will show) that the boosting methods, outlined in the previous subsections, increase the measurement sensitivity on the visual impairment scale, enabling the detection of more subtle artefacts. In our experiments, we investigated the performances by measuring sensitivities with respect to the magnitude of the applied distortion. We selected ten source images from the MCL-JCI dataset \cite{jin2016statistical}, each of which is distorted by seven types of distortion: color diffusion, jitter, high sharpen, JPEG\,2000 compression, lens blur, motion blur, and multiplicative noise. Figure\,\ref{fig:btcflowchart1} shows the flowchart of our boosted triplet comparison method for subjective IQA.

Along with the boosting of sensitivity, however, we have to accept that the absolute values of impairment, given in JND units, will be different and typically larger than those obtained using plain pair or triplet comparison or by using the DCR method. 
For example, if a particular distortion produces an impairment of 1.5\,JND, measured by plain comparison, we may obtain a much larger impairment of perhaps as much as 3\,JND when using one of the boosting methods. Therefore, we have also devised a method to adaptively transform the boosted impairment quality values back so that they approximately match the range of impairment scales as measured by plain comparison, however, without sacrificing the better discrimination ability.

To summarize, in this paper, we present the first study on the potential of perceptual boosting techniques in the context of subjective image quality assessment. The main contributions are:
\begin{enumerate}
    \item We propose three boosting strategies (artefact amplification, zooming, and flicker) that can enlarge the sensitivity of pair and triplet comparisons, as well as increase the accuracy of visual quality assessment.
    \item We propose a method based on Thurstone's model and MLE to reconstruct the perceptual qualities of images from triplet comparisons.
    \item We generate an IQA dataset of 1140 images with distortions from 10 source images. We provide the responses to all triplet comparisons from a large subjective crowdsourcing campaign together with the reconstructed quality values from plain triplet comparison and seven types of boosted triplet comparison. This dataset will be made available after publication.
    \item We provide an extensive performance analysis of boosted triplet comparisons for image quality assessment, including measures of true positive responses, detection rates, sensitivity, effect size, convergence, correlation, and time complexity.
\end{enumerate}

%------------------------------------------------------------------------------------------
%-----´SUBSECTION --- SUBSECTION --- SUBSECTION --- SUBSECTION --- SUBSECTION -------------
\subsection {Glossary}
%------------------------------------------------------------------------------------------

\begin{description}
\item[Sequence.] A sequence is a list of images $(I_0,I_1,\ldots,I_K)$ where $I_0$ is a pristine \textit{\textbf{reference image}} and the others are increasingly distorted versions of it. The index $k$ of $I_k$ is called the \textit{\textbf{distortion level}} of $I_k$.
%\item[Reference.] xxx
%\item[Distortion level.] xxx
\item[Triplet.] A triplet is a list $(i,j,k)$ of three non-negative indices referring three stimuli $ (I_i,I_j,I_k)$ of an image sequence. In a \textit{\textbf{triplet comparison}}, the observer is asked to determine whether the left or the right image ($I_i$ or $I_k$) is perceptually closer to the so-called \textit{\textbf{pivot image}} $I_j$ in the middle.
\item[Baseline triplet.] A triplet of the form $(i,0,k)$, where the pivot is the pristine reference image $I_0$. 
\item[General triplet.] A triplet of the form $(i,j,k)$, where the pivot could be any image (could be the pristine reference image or a distorted one). 
\item[Pivot.] The stimulus that is placed in the middle of a triplet. 
\item[JND.] The natural unit on the perceptual scale. The perceptual difference between the two compared images is 1~JND (just noticeable difference) when the difference is perceived by a random observer of the population of observers with a probability of 0.5. 
\item[HIT.] A human intelligence task is a self-contained, virtual task that a worker can work on. A HIT may contain several questions to be answered.
\item[Assignment.] A completed HIT that is submitted by a unique worker. 
\item[Response.] Equals vote or answer for a pair or triplet comparison.
\item[Rating.] Selected image quality or degree of impairment in an ACR or DCR assessment. 
\item[Plain.] A plain triplet comparison is a conventional one, where the images to be compared are not processed (boosted).
\item[A, Z, F.] Abbreviations for triplet comparisons with artefact amplification, zooming, and flicker, respectively. 
\item[AZ, AF, ZF, AZF.] Abbreviations for triplet comparisons of combinations of A and Z, A and F, Z and F, and A, Z, and F.
\end{description}

%------------------------------------------------------------------------------------------
%----- SECTION --- SECTION --- SECTION --- SECTION --- SECTION --- SECTION ----------------
%----- SECTION --- SECTION --- SECTION --- SECTION --- SECTION --- SECTION ----------------
%----- SECTION --- SECTION --- SECTION --- SECTION --- SECTION --- SECTION ----------------
\section{Related Work}
%------------------------------------------------------------------------------------------

%------------------------------------------------------------------------------------------
%-----´SUBSECTION --- SUBSECTION --- SUBSECTION --- SUBSECTION --- SUBSECTION -------------
\subsection{Boosting  by  Artefact  Amplification  and  Zooming}
%------------------------------------------------------------------------------------------
\label{subsec:boosting}

Amplification and zooming are common techniques applied in image and video processing, for the purpose of highlighting details of an image or for overall image enhancement. For instance, histogram equalization enhances contrast by enlarging small intensity differences. Image sharpening enhances the appearance of edges, e.g., by adding to the input image a signal that is a scaled high-pass filtered version of the original image. Motion and color magnification reveals subtle variations in video sequences that cannot be perceived by human senses, such as heartbeats showing in faces \cite{elgharib2015video}, and can even reconstruct sound from videos of objects subtly vibrating in response to sound \cite{davis2014visual}. Exaggeration also is one of Disney's twelve basic principles of animation \cite{johnston1981illusion}, applied in particular to motion. The cartoon characters were designed to maintain the illusion that they follow the laws of physics, however, in a wilder, more extreme form.

In spite of widespread applications of amplification and zooming in multimedia applications, we have not become aware of any previous work on adapting, applying, and validating these techniques for the purpose of subjective visual quality assessment.

The only exception is our earlier work \cite{men2019visual}, in which we applied image distortion amplification and zooming to properly cropped frames to compare interpolated video frames with the corresponding ground-truth frames.
However, this technique was a side issue in that contribution, and there was no  systematic analysis of the performance and potential of amplification and zooming. 

%------------------------------------------------------------------------------------------
%-----´SUBSECTION --- SUBSECTION --- SUBSECTION --- SUBSECTION --- SUBSECTION -------------
\subsection{Boosting  by  the  Image  Flicker  Method}
%------------------------------------------------------------------------------------------

In \cite{hoffman2014new}, flicker tests were proposed for the task of determining the JND of distorted images. In order to test for noticeable distortions in compressed images, two images were placed side by side. One of them was a still reference image. 
The other was an animation in which a reference image alternated with the distorted image at the same spatial position. The display frame rate was 30\,fps, and the alternation occurred every four frames, yielding a flicker frequency of 7.5\,Hz.
Observers were asked to identify which of the two images was the still (or non-flickering) one (two-alternative forced-choice). 
%(buffer toggles per second). 

Shortly later, the ISO/IEC standard 29170-2:2015(E)  was set up with recommendations for the subjective quality evaluation of single frames from JPEG\,XS encoded videos \cite{isoflicker}. The focus of the standard was on visually lossless, low-latency, and lightweight video compression schemes. Therefore, subjective tests were prescribed for image JND assessment rather than conventional MOS from ACR or DCR procedures. The standard directly follows the ideas in \cite{hoffman2014new}, with one notable exception: The flicker rate was recommended to be 10\,Hz instead of 7.5\,Hz.

The first study, based on the new standard, was published in 2017. In a large lab experiment with 120 subjects, the flicker test was used for video frames, produced by the Video Electronics Standards Association (VESA) Display Stream Compression, which is a lightweight codec designed for visually lossless compression \cite{allison201775}.
An in-depth discussion of this application of the flicker test for the criterion of perceptually lossless compression as prescribed in the ISO/IEC standard was presented in \cite{allison2018perspectives}. 
The authors' conclusion was that ``if the goal is to conservatively evaluate the possibility that a compression artefact might be visible under any situation, then the flicker paradigm is a viable approach as it highlights differences between images regardless of whether they are noticeable in the absence of a reference.'' However, a quantitative comparison of the sensitivity of the flicker test protocol versus the traditional side-by-side presentation was not included. 

%``In this procedure, the reference is presented sequentially in the same location as the processed image; the image differences should be extremely salient due to sensitive motion and change detectors in the visual system.''
%``The results of our evaluation of VESA DSC1.2 using the ISO/IEC 29170-2015 flicker protocol show that this forced choice paradigm is a highly effective means of evaluating sensitivity to image differences. The design of this test protocol and the visually lossless criterion applied is extremely sensitive,''

In 2016, a JPEG\,XS Call for Proposals with subjective quality evaluations based on the flicker test of the ISO/IEC standard was issued \cite{jpegxscfp}. The results of the submissions to the call were summarized in \cite{willeme2018overview}. Different from the ISO/IEC standard recommendations, a flicker frequency of 8\,Hz was applied, and subjects were given a third option for their response, namely to cast a no-decision vote. This was deemed to alleviate subject fatigue.

In other recent work, the flicker test was applied to a palette of different image modalities: High dynamic range (HDR) images \cite{sudhama201885}, foveated images in head-mounted displays (HMD) \cite{thirumalai2020p}, and stereoscopic imagery \cite{mohona2021subjective}.

In our pre-study \cite{lin2020subjective}, presented at the ICME 2020 Workshop on Data-driven Just Noticeable Difference for Multimedia Communication, we have provided the first experiment to compare the performance of the flicker test with conventional side-by-side comparisons. The purpose of the tests was to assess the JND for JPEG image compression. As a result, we reported that the flicker test was about twice as sensitive as the classical side-by-side comparisons with forced choice. However, this experimental study was small, and the focus was rather on a new adjustment method for JND detection using a slider-based design. Moreover, the flicker tests were done in a lab situation, while the classical 2AFC pair comparison ran on a crowdsourcing platform. So the result of the comparison regarding sensitivity is only a preliminary. Our contribution here provides a much more elaborate study targeted specifically at the validation of the performance of several boosting techniques, flicker being one of them.

The previous works on the image flicker method mentioned above have applied flicker in a single stimulus or in a double stimulus method where one of the two stimuli was a still image. We extend these procedures by comparing two flickering stimuli in the context of a triplet comparison. Moreover, in past approaches, the flickering was between the undistorted reference image and a test image. We will show the advantages of considering flicker images in comparisons where the flicker is between two test images.
\subsection{Reconstruction of Scale Values from Triplet Comparisons}
%------------------------------------------------------------------------------------------

Direct quality assessment proceeds by collecting and averaging quality ratings from a sufficiently large set of observers. Absolute category rating is the most common technique in visual quality assessment. Scale value reconstruction is an indirect procedure, deriving scales of latent variables from pair comparisons of the perceptual quality or from the comparison of quality differences in triplets or quadruplets. Other approaches are possible, like reconstruction from rankings of images in subsets of stimuli. For the application of boosting methods in subjective visual quality assessment, indirect methods seem more appropriate because boosting enhances the perception of differences between a test stimulus and its corresponding reference. 

One of the first indirect approaches, based on scaling of perceived distances of stimuli, attained from triplet comparisons, was proposed in 1952 by Torgerson \cite{torgerson1952multidimensional} and named the \textit{method of triads}. Although the goal was multi-dimensional scaling, it is clear that the method can also be used to derive scalar values of a latent variable. In a nutshell, for the 1D case, the reconstruction is based on a model of random variables $X_i$, $i = 0,\ldots,M$,  for the latent stimuli qualities with the assumption that their pairwise distances 
$$
    D_{i,j} = |X_i-X_j|
$$ 
 are random variables with a normal distribution of unit variance. Then scale values of these distances can be reconstructed from the pairwise comparison of distances arising from the triplet comparisons. This step is analogous to the Thurstonian scale reconstruction from pair comparison of stimulus values, where instead of scales for the random variables $X_i$, scales for the distances $D_{i,j}$ are considered.

However, since the triplet comparisons $(i,j,k)$ give information only about differences in distances
(namely whether $D_{i,j} < D_{k,j}$), the reconstruction of the distances can be determined only up to an additive constant. In \cite{torgerson1952multidimensional}, the least squares solution to solve the problem of the additive constant was proposed. 

At the end, a square matrix of distances $\left(d_{i,j}\right), i,j=0,\ldots,M$ is yielded, from which a one-dimensional embedding can be generated. One can solve the optimization problem, where estimates for the latent variables are found as a minimizer of a cost function, for example,
$$
    {\displaystyle \left(\hat{\mu}_{0},\ldots ,\hat{\mu}_{M}\right)} = \arg {\displaystyle \min _{\mu_{0},\ldots ,\mu_{M}}\sum _{i<j}\left(|\mu_{i}-\mu_{j}|-d_{i,j}\right)^{2}.\,}
$$

The method of triads has been criticized as being ad hoc \cite{luce1963psychophysical}, as it does not directly follow the basic setup of Thurstonian models, where the latent variables of the stimuli themselves are normally distributed. Moreover, distances are non-negative and cannot be modelled accurately by normal distributions.

In our contribution, we propose a complete solution to the reconstruction of scale values from triplet questions that strictly adheres to the considerations of Thurstonian models. Let us assume such a model of a set of normally distributed random variables $X_i$, $i = 0,\ldots,M$ of equal variance, for the visual qualities of a corresponding set of stimuli. We will make use of a formula for the probabilities $\Pr\left(D_{i,j} < D_{k,j}\right)$ for the outcome of a given triplet comparison $(i,j,k)$, that was derived by Ennis et al.\ \cite{ennis1988variants} in 1988. 

%Let $Q_{i,j,k}$ denotes the empirical estimation of $\Pr\left(D_{i,j} < D_{k,j}\right)$, 
% -----

Let $Q_{i,j,k}$ denotes the empirical estimation of\linebreak $\Pr\left(D_{i,j} < D_{k,j}\right)$ given by the fraction of responses to the comparison for the triplet $(i,j,k)$ that indicate that the left stimulus $I_i$ is closer to the pivot $I_j$ than the right one, $I_k$. Then the task to be solved is to reconstruct the mean values of the model such that the model predictions for the TC outcomes match the empirical data:
$$
   \Pr(D_{i,j} < D_{k,j}) \approx Q_{i,j,k}.
$$
For this purpose, in \cite{ennis1988variants} the method of least squares was proposed, 
$$
    {\displaystyle \min _{\mu_{0},\ldots,\mu_{M}} \sum _{i<k, j\ne i,k} \left(Q_{i,j,k}-\Pr\left(D_{i,j} < D_{k,j}\right)\right)^{2}.\,}
$$
This is equivalent to the MLE in which the prediction errors $  \Pr(D_{i,j} < D_{k,j}) - Q_{i,j,k}$ are modelled as independent normal random variables with equal variance \cite{press1992least}. This assumption generally cannot hold since for small or large probabilities $ \Pr(D_{i,j} < D_{k,j})$ near 0, resp.\ 1, the error distribution necessarily must be skewed. Therefore we favor the general MLE method, i.e., to maximize the model likelihood of the set of observations $Q_{i,j,k}$. 

While this choice follows the common approach taken in psychometrics \cite{wichmann2001psychometric}, the most widely used one-dimensional scale reconstruction method for vision science applications is probably maximum likelihood difference scaling (MLDS). 
It solves the difference scaling problem of quadruplet questions $(i,j,k,l)$, where the perceptual distance of the first pair of stimuli, $(I_i,I_j)$, is compared to that of the second pair, $(I_k,I_l)$, in a 2AFC setting. In MLDS, the decision variable employed by an observer of such quadruplet questions is modelled as
$$
    Z = |x_j-x_i| - |x_l-x_k| + N_{\sigma},
$$
where $x_i, i=0,\ldots,M$ are the (crisp) unknown qualities and $N_{\sigma}$ is a zero-mean Gaussian noise term with variance $\sigma^2$. The unknown variance characterizes the difficulty of the particular set of quadruplet questions together with the uncertainty of the subjects.

It is worth noting that MLDS presents an approach of a fundamentally different type than Thurstonian models. In Thurstonian models, the decision variable is $Z = |X_j-X_i| - |X_l-X_k|$ and deterministic, but the qualities on the perceptual scales are uncertain, given by normally distributed random variables $X_i,i=0,\ldots,M$.
%Any uncertainties arising from subject variability are not separately modelled but lumped together into these random variables. 
In MLDS, it is just the opposite. The quality values are crisp, while there is uncertainty in the decision variable. 

The number of free parameters for the $M+2$ unknowns in MLDS is $M$, and one may set the range of scale values to $[x_0,x_{M}] = [0,1]$ and solve for the variables $x_1,\ldots,x_{M-1}$ and $\sigma$, using MLE. Alternatively, one can set $x_0 = 0$, the variance $\sigma^2$ to a fixed value, and then solve for the scales $x_1,\ldots,x_{M}$. 

In this paper, we contribute a method for the selection of the variance $\sigma^2$ of the noise term such that the resulting reconstruction yields scale values in approximate JND units.

A recent survey discusses the MLDS method, its variations, and a very large number of applications in different fields \cite{maloney2020measuring}. Two contributions, most closely related to our work for visual quality assessment, are \cite{charrier2007maximum} and \cite{charrier2012optimizing} where quadruplet comparisons for image sequences with distortions due to compression were undertaken and analysed by MLDS.

Although MLDS was designed for scale reconstruction from quadruplet comparisons, it is clear that it can also be applied to triplet comparisons $(i,j,k)$, simply by restricting to quadruplets of the form $(i,j,j,k)$. Then the decision variable is $Z = |x_j-x_i| - |x_k-x_j| + N_{\sigma}$. In practice, it can be expected that its normal distribution is very similar to that for the decision variable 
$Z = |X_j-X_i| - |X_k-X_j|$ which arises from the Thurstonian model ($X_i,X_j,X_k$ normally distributed with variance 1/2). However, for our particular applications in visual quality assessment (FR-IQA), we prefer a reconstruction method that can produce scale values in JND units. MLDS was not designed for that purpose. 

There are several other methods for scale reconstruction from triplet comparisons, some of which have recently originated from the machine learning community \cite{haghiri2020estimation}. Usually, these methods are for multi-dimensional scaling. Some of them can be restricted to the one-dimensional case. In Section\,\ref{sec:tripletrecons}, we compare our results with those computed by MLDS and stochastic triplet embedding (STE) \cite{van2012stochastic}. In terms of correlation with ground truth, all methods showed excellent performance. However, as for MLDS, also STE cannot be expected to yield estimates on JND scales. %However, we show how to calibrate MLDS and STE by tweaking a parameter such that the correspondence to JND scales improves by much.

Let us finally remark that there also is an ISO standard that proposed triplet comparison \cite{keelan2003iso}. Observers rate each image in a triplet using a 5-point ACR scale, and from that, all three pair comparisons in the triplet are deduced. The selections of the triplets in an experiment is prescribed and rather restricted. This setting would not support the boosting strategies discussed in this paper.

%Ordinal embedding based on STE \cite{van2012stochastic} assumes that the probability that an observer gives a correct answer for a triplet question whether $i$ is closer to $j$ or to $k$ is
%\begin{equation}
%    p_{ijk} = \frac{\exp (-|| y_i - y_j ||^2) }{\exp (-|| y_i - y_j ||^2) + \exp (-|| y_i - y_k ||^2)}.
%\end{equation}
%Similar to MLDS, the qualities of the stimuli were estimated by maximizing the likelihood of the given answers to the triplets. A modified version of STE, t-STE, was also proposed in \cite{van2012stochastic}, where the Gaussian functions were replaced by Student-t functions with a heavier tail kernel to make the statistic more robust.

%Non-metric multidimensional scaling (NMDS) \cite{shepard1962analysis} only took the rank order of the dissimilarities between the stimuli into consideration. It aimed at finding a multidimensional Euclidean representation of stimuli such that the pairwise distances of the estimation are consistent with a monotonic transform of the dissimilarities between the stimuli. Although NMDS is easy to apply because of its non-parametric feature, it has a limitation. It needs the full dissimilarity matrix, indicating that all possible triplets need to be rated.

%-----------------------------------------------------------------------------------
%-----------------------------------------------------------------------------------
%-----------------------------------------------------------------------------------
\section{Boosting Strategies}
%-----------------------------------------------------------------------------------

We apply three ways to boost the perceptual sensitivity of comparisons between distorted images: 1) \emph{Artefact amplification} amplifies the artefacts of distorted images relative to their references, 2) \emph{Zooming} enlarges the visual representation of the images, and 3) \emph{Flicker} increases the perceptual sensitivity to the distortions by rapidly alternating between distorted images and their corresponding reference images. 

\begin{figure*}[t]
\centering{
\begin{tikzpicture}

\begin{axis}[%
width=1.37in,
height=2in,
at={(0in,2.05in)},
scale only axis,
axis on top,
xmin=0.5,
xmax=384.5,
y dir=reverse,
ymin=0.5,
ymax=512.5,
axis line style={draw=none},
ticks=none,
title style={font=\bfseries\sffamily\scriptsize, yshift=-1ex},
title={Reference}
]
\addplot [forget plot] graphics [xmin=0.5, xmax=384.5, ymin=0.5, ymax=512.5] {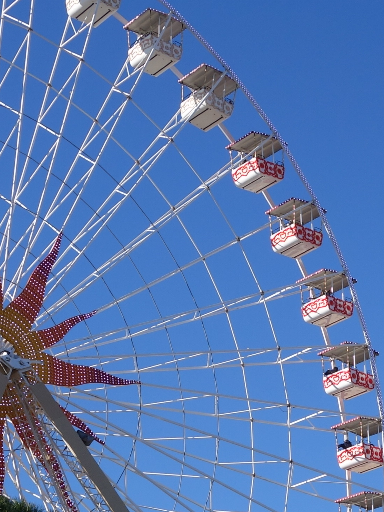};
\end{axis}

\begin{axis}[%
width=1.37in,
height=2in,
at={(1.39in,2.05in)},
scale only axis,
axis on top,
xmin=0.5,
xmax=384.5,
y dir=reverse,
ymin=0.5,
ymax=512.5,
axis line style={draw=none},
ticks=none,
title style={font=\bfseries\sffamily\scriptsize, yshift=-1ex},
title={Distorted}
]
\addplot [forget plot] graphics [xmin=0.5, xmax=384.5, ymin=0.5, ymax=512.5] {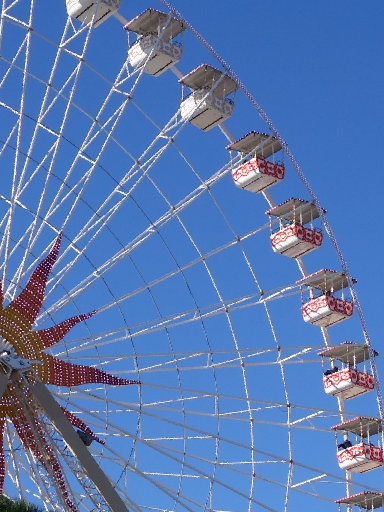};
\end{axis}

\begin{axis}[%
width=0.02in,
height=4.6in,
at={(2.77in,-0.45in)},
scale only axis,
xmin=-1,
xmax=1,
ymin=0,
ymax=10,
axis line style={draw=none},
ticks=none
]
\addplot [color=black, dashdotted, forget plot]
  table[row sep=crcr]{%
0	1\\
0	1.5\\
0	2\\
0	2.5\\
0	3\\
0	3.5\\
0	4\\
0	4.5\\
0	5\\
0	5.5\\
0	6\\
0	6.5\\
0	7\\
0	7.5\\
0	8\\
0	8.5\\
0	9\\
0	9.5\\
0	10\\
};
\end{axis}

\begin{axis}[%
width=1.37in,
height=2in,
at={(2.80in,2.05in)},
scale only axis,
axis on top,
xmin=0.5,
xmax=384.5,
y dir=reverse,
ymin=0.5,
ymax=512.5,
axis line style={draw=none},
ticks=none,
title style={font=\bfseries\sffamily\scriptsize, yshift=-1ex},
title={$\bm{\alpha}$ = $\bm{1.5}$}
]
\addplot [forget plot] graphics [xmin=0.5, xmax=384.5, ymin=0.5, ymax=512.5] {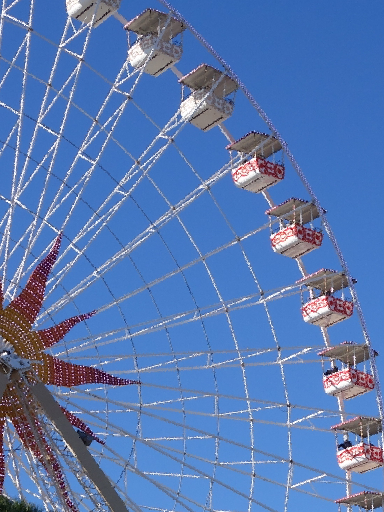};
\end{axis}

\begin{axis}[%
width=1.37in,
height=2in,
at={(4.19in,2.05in)},
scale only axis,
axis on top,
xmin=0.5,
xmax=384.5,
y dir=reverse,
ymin=0.5,
ymax=512.5,
axis line style={draw=none},
ticks=none,
title style={font=\bfseries\sffamily\scriptsize, yshift=-1ex},
title={$\bm{\alpha}$ = $\bm{2}$}
]
\addplot [forget plot] graphics [xmin=0.5, xmax=384.5, ymin=0.5, ymax=512.5] {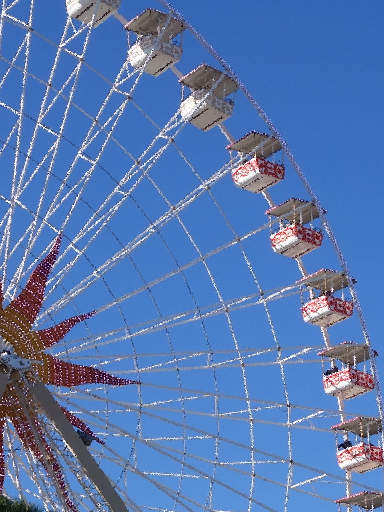};
\end{axis}

\begin{axis}[%
width=1.37in,
height=2in,
at={(5.58in,2.05in)},
scale only axis,
axis on top,
xmin=0.5,
xmax=384.5,
y dir=reverse,
ymin=0.5,
ymax=512.5,
axis line style={draw=none},
ticks=none,
title style={font=\bfseries\sffamily\scriptsize, yshift=-1ex},
title={$\bm{\alpha}$ = $\bm{3}$}
]
\addplot [forget plot] graphics [xmin=0.5, xmax=384.5, ymin=0.5, ymax=512.5] {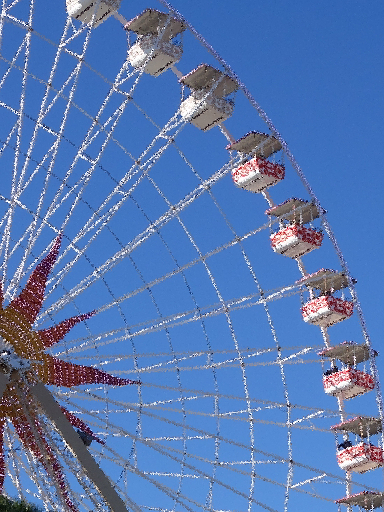};
\end{axis}

\begin{axis}[%
width=1.37in,
height=2in,
at={(0in,0in)},
scale only axis,
axis on top,
xmin=0.5,
xmax=384.5,
y dir=reverse,
ymin=0.5,
ymax=512.5,
axis line style={draw=none},
ticks=none
]
\addplot [forget plot] graphics [xmin=0.5, xmax=384.5, ymin=0.5, ymax=512.5] {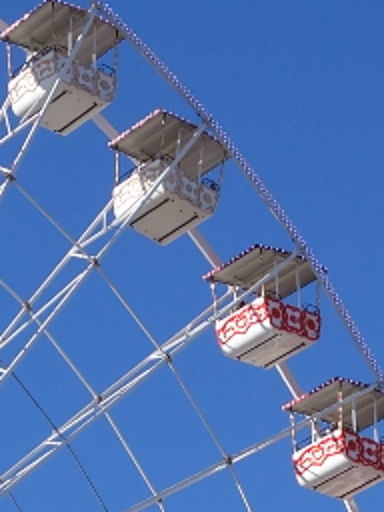};
\end{axis}

\begin{axis}[%
width=1.37in,
height=2in,
at={(1.39in,0in)},
scale only axis,
axis on top,
xmin=0.5,
xmax=384.5,
y dir=reverse,
ymin=0.5,
ymax=512.5,
axis line style={draw=none},
ticks=none
]
\addplot [forget plot] graphics [xmin=0.5, xmax=384.5, ymin=0.5, ymax=512.5] {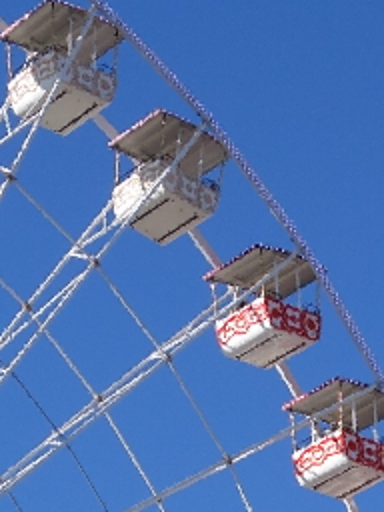};
\end{axis}

%\begin{axis}[%
%width=1.4in,
%height=2in,
%at={(2.8in,0in)},
%scale only axis,
%xmin=-1,
%xmax=1,
%ymin=0,
%ymax=10,
%axis line style={draw=none},
%ticks=none
%]
%\addplot [color=black, dashdotted, forget plot]
%  table[row sep=crcr]{%
%0	1\\
%0	1.5\\
%0	2\\
%0	2.5\\
%0	3\\
%0	3.5\\
%0	4\\
%0	4.5\\
%0	5\\
%0	5.5\\
%0	6\\
%0	6.5\\
%0	7\\
%0	7.5\\
%0	8\\
%0	8.5\\
%0	9\\
%0	9.5\\
%0	10\\
%};
%\end{axis}

\begin{axis}[%
width=1.37in,
height=2in,
at={(2.8in,0in)},
scale only axis,
axis on top,
xmin=0.5,
xmax=384.5,
y dir=reverse,
ymin=0.5,
ymax=512.5,
axis line style={draw=none},
ticks=none
]
\addplot [forget plot] graphics [xmin=0.5, xmax=384.5, ymin=0.5, ymax=512.5] {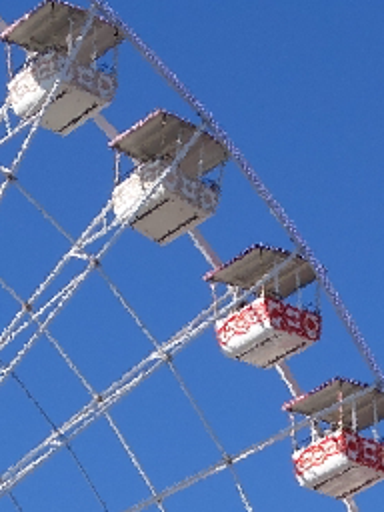};
\end{axis}

\begin{axis}[%
width=1.37in,
height=2in,
at={(4.19in,0in)},
scale only axis,
axis on top,
xmin=0.5,
xmax=384.5,
y dir=reverse,
ymin=0.5,
ymax=512.5,
axis line style={draw=none},
ticks=none
]
\addplot [forget plot] graphics [xmin=0.5, xmax=384.5, ymin=0.5, ymax=512.5] {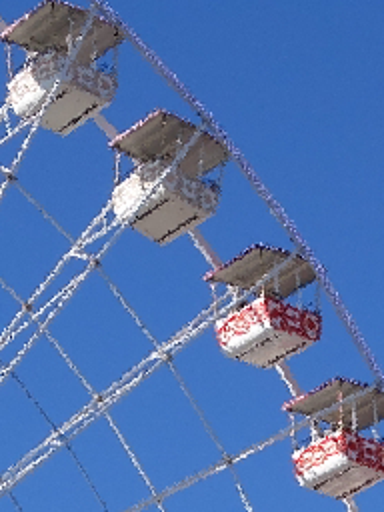};
\end{axis}

\begin{axis}[%
width=1.37in,
height=2in,
at={(5.58in,0in)},
scale only axis,
axis on top,
xmin=0.5,
xmax=384.5,
y dir=reverse,
ymin=0.5,
ymax=512.5,
axis line style={draw=none},
ticks=none
]
\addplot [forget plot] graphics [xmin=0.5, xmax=384.5, ymin=0.5, ymax=512.5] {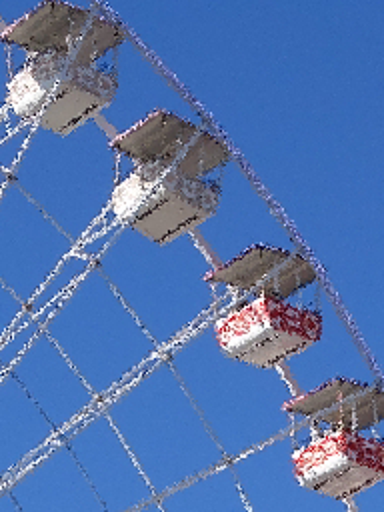};
\end{axis}
\end{tikzpicture}%
}
\vspace{-25pt}
\caption{%\textcolor{green}{[temporal Figure, to be refined]}
%vis_ampli_new.png
Illustration of artefact amplification and zooming. The upper row shows an original source image and its distorted version when jitter is applied, at a level corresponding to 0.5\,JND. The artefacts are  amplified with factors $\alpha=1.5$, $2$, and $3$ in the three top right images. Differences between the reference and the distorted image are barely visible, but with increasing amplification, they become better noticeable. The bottom row presents a zoomed version of the upper row. The visibility of the distortions is further enhanced by zooming.
}
\label{vis_ampli}
\end{figure*}

%-----------------------------------------------------------------------------------
\subsection{Artefact Amplification (A)}\label{sec:ampli}
%-----------------------------------------------------------------------------------
Many ways can be conceived to amplify artefacts due to distortions in images. For this study, we consider one of the simplest kinds, namely linear pixel-wise scaling of RGB color differences between the distorted and the reference image. Let $v, \hat v$ be the RGB pixel values of a pixel in the reference and a distorted image, respectively. Then we replace $\hat v$ by ${\hat v}'=v + \alpha (\hat v-v)$. 

The multiplication by the factor $\alpha > 1$ ensures consistency with Fechner's law \cite{fechner1966elements}, which states that the subjective sensation is proportional to the logarithm of the stimulus intensity.  In our context, this means that equal relative increments of distortion, i.e., the same factor $\alpha$ applied in artefact amplification, should correspond to equal increments of perceived impairment in these images.

\begin{algorithm}[t!]
\caption{Pixel-wise artefact amplification}
\label{alg:boostedpcalg1}
\begin{algorithmic}[1]
\State $\alpha \gets$ 2 \Comment{default amplification factor}
\State $v\gets(v_\text{r},v_\text{g},v_\text{b})$ \Comment{ground truth  pixel}
\State $\hat v\gets(\hat{v}_\text{r},\hat{v}_\text{g},\hat{v}_\text{b})$ \Comment{distorted  pixel}
%\For{$r$ component}
% \If{$\hat{v}_\text{r}-v_\text{r} > 0$}
% \State $\alpha_{\text{r},\max}\gets(255-v_\text{r})/(\hat{v}_\text{r}-v_\text{r})$
% \ElsIf{$\hat{v}_\text{r}-v_\text{r} < 0$}
% \State $\alpha_{\text{r},\max}\gets-v_r/(\hat{v}_r-{v}_\text{r})$
% \Else
%  \State $\alpha_{\text{r},\max} \gets \alpha$
% \EndIf
% \EndFor
% \For{$\text{g},\text{b}$ components}
% \State same as above for r component;
% \EndFor
%  \State $\alpha \gets \min (\alpha, \alpha_{\text{r},\max} ,\alpha_{\text{g},\max}, \alpha_{\text{b},\max})$
\For{c $\in \{\text{r,g,b}\}$ }
\If{$\hat{v}_\text{c}-v_\text{c} > 0$}
\State $\alpha_{\text{c},\max}\gets(255-v_\text{c})/(\hat{v}_\text{c}-v_\text{c})$
\ElsIf{$\hat{v}_\text{c}-v_\text{c} < 0$}
\State $\alpha_{\text{c},\max}\gets-v_\text{c}/(\hat{v}_\text{c}-{v}_\text{c})$
\Else
 \State $\alpha_{\text{c},\max} \gets \alpha$
\EndIf
\EndFor
%\For{$\text{g},\text{b}$ components}
%\State same as above for r component;
%\EndFor
 \State $\alpha \gets \min (\alpha, \alpha_{\text{r},\max}, \alpha_{\text{g},\max}, \alpha_{\text{b},\max})$
 \State $\hat{v}' \gets v + \alpha (\hat{v} -v)$ \Comment{amplified  pixel $\hat{v}' \in [0,255]^3$}
\end{algorithmic}
\end{algorithm}

\begin{table}[t]
\caption{Fraction of pixels clamped in artefact amplification of Figure\,\ref{vis_ampli}.}
\label{clampratio}
\centering
%\setlength\tabcolsep{4.5pt}
%\begin{adjustbox}{angle=90}
%\resizebox{1.3\textwidth}{!}{ 
\begin{tabular}{c |c  c  c  | c }
Amplification & \multicolumn{3}{c|}{Channel} & Pixels \\
factor        & red & green & blue & overall\\
\hline
$\alpha = 1.5$ & 0.0012  &  0.0011  &  0.0021 &   0.0022 \\
$\alpha = 2.0$ & 0.0018  &  0.0017 & 0.0038 & 0.0050\\
$\alpha = 3.0$ & 0.0036  & 0.0039  &  0.0083 & 0.0131 \\
$\alpha = 4.0$ & 0.0068  &  0.0068  &  0.0135 & 0.0242\\
$\alpha = 5.0$ & 0.0107  & 0.0097 &   0.0184 & 0.0357
\end{tabular}
%}
%\end{adjustbox}
\end{table}

However, due to the finite range of RGB color components in digital images, the linear scaling is limited and RGB pixel values exceeding the limit must be clamped. Thus, to restrict the RGB components of ${\hat v}'$ to the range $[0,255]$ for 24-bit color images, we reduce $\alpha$ accordingly for those pixels where clamping is needed. Note that this may cause a local nonlinearity and saturation effect of the artefact amplification. See Algorithm\,\ref{alg:boostedpcalg1} for details. 

An example of artefact amplification is shown in Figure\,\ref{vis_ampli}. Comparing the distorted image with the reference image in the first row, the distortion is hardly visible. In contrast, some distortions are noticeable after artefact amplification and become more and more obvious with the increase of amplification factor (top right row). 
%Zooming these images (bottom row) renders the artefacts even more apparent.

\begin{figure*}[t]
\centering{
\includegraphics[width=0.8\textwidth]{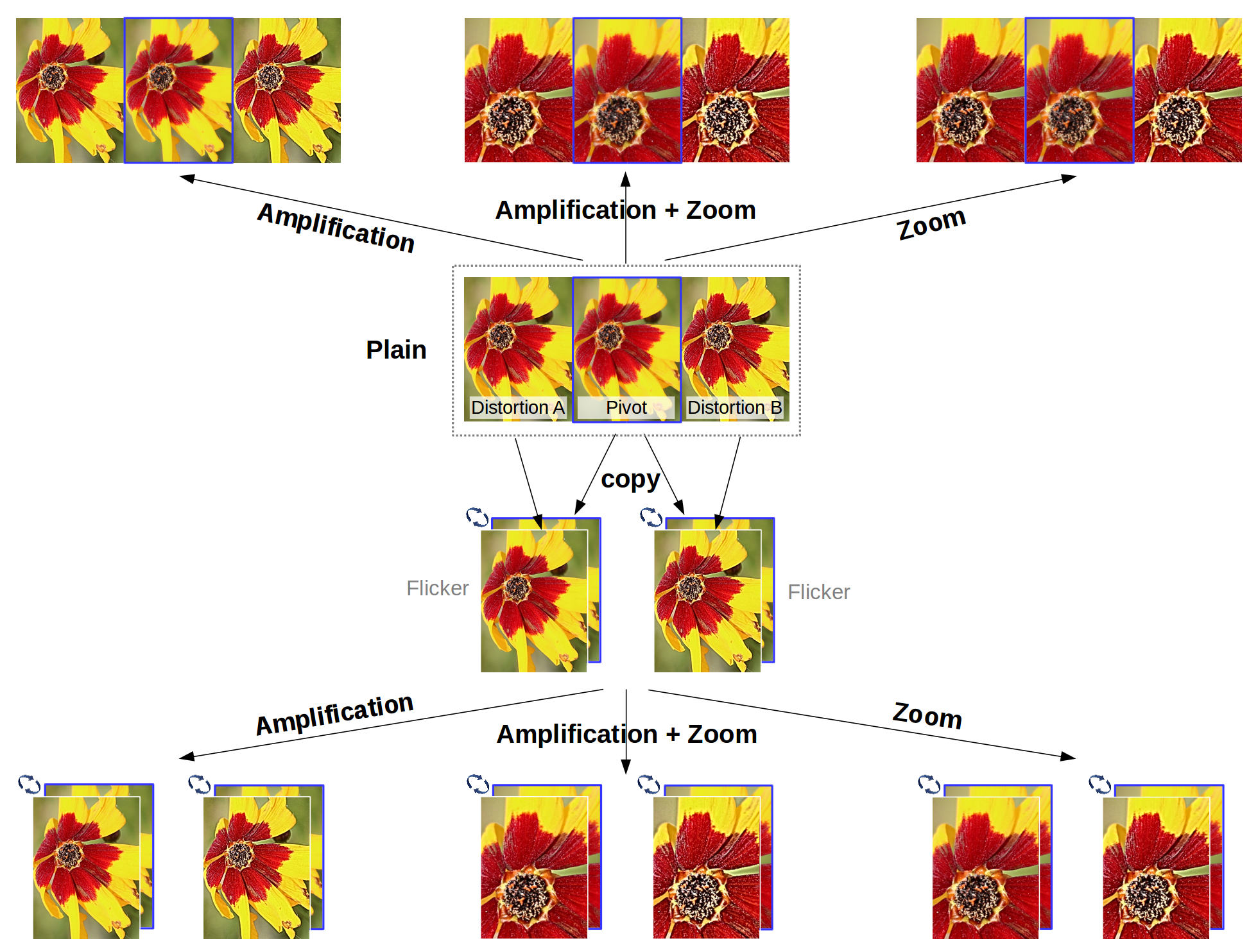}}
\caption{Overview of plain triplet comparison and seven types of boosted triplet comparison. In plain triplet, three stimuli, selected from an image sequence with increasing distortion levels, are displayed side by side. The task is to judge which side is perceptually closer to the pivot in the centre. The three boosting techniques, artefact amplification, zooming, and flicker, help to improve the accuracy, reliability, and speed of the subjective assessment. When boosting with flicker, the left and right images are displayed side by side, each alternating with the pivot eight times per second. In this case, observers judge which side has the stronger flicker effect.}
\label{boostedpcflow}
\end{figure*}

Table\,\ref{clampratio} shows the fraction of the pixel color components and the overall number of pixels that are clamped in the amplification process, for the example shown in Figure\,\ref{vis_ampli}. These fractions are monotonically increasing with the amplification factor $\alpha$. In order to avoid a widespread nonlinearity and saturation effect due to clamping too many pixels, $\alpha$ should be chosen appropriately. Note that, depending on the application, more or less strong amplification can be used. In triplet comparisons, we applied the amplification relative to the pivot stimulus displayed in the centre. Therefore, for baseline triplets of the sort $(i,0,k)$ the differences that are amplified, are typically larger than for most general triplets  $(i,j,k)$ with $i<j<k$ or $i>j>k$. Thus, a more conservative (i.e., smaller) amplification factor $\alpha$ should be used for baseline triplets. In this paper, we set $\alpha=2$ as the artefact amplification factor.  

%-----------------------------------------------------------------------------------
\subsection{Zooming (Z)}\label{sec:zoom}
%-----------------------------------------------------------------------------------
Apart from enlarging color differences by amplifying artefacts, artefacts appear more visible when enlarged spatially. In fact, participants in an IQA experiment may be tempted to enlarge the images displayed in their browser or to move closer to the screen to detect fine differences between images. However, to ensure a uniform and controlled quality assessment, participants are asked to refrain from such adhoc zooming action. Instead, we propose to deliver the displayed images already in a zoomed and cropped fashion. 

Figure\,\ref{vis_ampli} (bottom row) shows an example. Images are cropped to half their linear size and zoomed by a factor of two. The cropped regions were manually selected, and bicubic interpolation was adopted for scaling up. Distortions in the zoomed images are more clearly visible, especially with the increasing artefact amplification factor $\alpha$. 

Similarly to artefact amplification, larger zoom factors are better for visual detection of artefacts, but at some point, undesirable side effects like pixelation set a limit to zooming. Due to the required cropping, only a part of the image content is maintained in the zoomed images, possibly masking image areas with more severe local distortions. In this paper, we chose a fixed zoom factor of two.

%-----------------------------------------------------------------------------------
\subsection{Flicker (F)}\label{sec:flicker}
%-----------------------------------------------------------------------------------
The visibility of distortions can also be enhanced by making use of the flicker effect as explained in Subsection\,\ref{SubsectionFlicker}. In this scenario, a distorted image and its reference are displayed successively at a frequency of 8\,Hz. As already mentioned, for a triplet comparison $(i,j,k)$, using the flicker technique, we show two flickering images side by side, on the left, image $I_i$ alternates with the pivot $I_j$, and on the right $I_k$ alternates with the pivot. Observers are asked to select the one that has a stronger flickering effect. 

%-----------------------------------------------------------------------------------
\subsection{Combinations of Boosting Types}\label{sec:combine_boost}
%-----------------------------------------------------------------------------------
With the three boosting options on hand, we can apply them individually or in combination, for example, zooming together with artefact amplification. This gives rise to seven cases that we abbreviate with the letters A, Z, and F assigned to the boosting methods artefact amplification, zooming, and flicker, respectively. The combinations are A, Z, F, AZ, AF, ZF, and AZF. Without boosting, we obtain \textit{plain} results, which serve as a reference when assessing the performance of the boosting methods. Figure\,\ref{boostedpcflow} shows an overview of the different kinds of boosting options as applied for triplet comparison.

%-----------------------------------------------------------------------------------
%-----------------------------------------------------------------------------------
%-----------------------------------------------------------------------------------
\section{Thurstonian Reconstruction from Triplet Comparisons} \label{sec:tripletrecons}
%-----------------------------------------------------------------------------------
%[Details see ``sub_tex/tripletstudy.tex" ]

Let us consider a set of $M+1$ stimuli. In our experiments, these are a source image $I_0$ together with derived distorted images $I_1,\ldots,I_{M}$. The magnitudes of the perceived stimulus qualities are taken to be unknown latent variables, modelled by normally distributed real-valued random variables $X_0,\ldots,X_{M}$ with variance equal to 1/2. It is the purpose of reconstruction to estimate their means $\bf \mu = (\mu_0,\ldots,\mu_{M}) \in \mathbb{R}^{M+1}$ from a collection of responses to subjective triplet comparisons of stimuli. This setup corresponds to the assumptions Thurstone established as Case V in his analysis for pair comparisons \cite{thurstone1927law}. 

In Subsection\,\ref{sec:thruston_a}, we present formulas for the computation of the probabilities of the responses for the triplet comparisons, followed by MLE of the means in Subsection\,\ref{sec:thruston_b}. In Subsection\,\ref{sec:thruston_c}, we compare the reconstruction performances of MLDS, STE, and our method by means of a simulation with available ground truth data. We also compare the probabilistic model for the decision random variable in MLDS with the uncertainty of the means in the Thurstonian model. This gives rise to a choice of the unspecified variance of the MLDS decision variable such that the reconstructions of the means of the stimuli values on the perceptual scale are given in approximate JND units. 

%-----------------------------------------------------------------------------------
 \subsection{Formulas for the Probabilities of the Responses}
 \label{sec:thruston_a}
%-----------------------------------------------------------------------------------
 We define that for a triplet $t=(i,j,k)$ with $i,j,k \in \left\{0,\ldots,M\right\}$, a subjective comparison yields a response $R_{ijk}= 1$, if the observer judges the left stimulus, numbered $i$, closer to the pivot stimulus $j$ than the right one, $k$. Otherwise, the response is $R_{ijk}= 0$.

From the Thurstonian probabilistic model, it follows that observers act according to the sign of the decision variable 
\begin{equation} \label{equ:z}
    Z_{ijk} = \left|X_k - X_j\right| - \left|X_i - X_j\right|,
\end{equation}
such that
\begin{equation} \label{equ:R}
    R_{ijk}= 
    \begin{cases} 
        1 & \mbox{if } Z_{ijk} > 0 \\ 
        0 & \mbox{if } Z_{ijk} \le 0 
    \end{cases}.
\end{equation}

%Based upon this result and the responses to a set of triplet comparison, we will then formulate the MLE of the means.
%Let us assume the Thurstonian model for stimuli $i,j,k \in \left\{0,\ldots,M\right\}$ having normal distributions on the sensory scale with means $\bfmu = (\mu_0,\ldots,\mu_{M})$ and equal variance 1/2.

Next, we first give an expression, based on a result of \cite{ennis1988variants}, for the probability that the decision variable is positive. It will be a function of the unknown means $\bfmu = (\mu_0,\ldots,\mu_{M})$, so we write it as the conditional probability $\Pr(Z_{ijk} > 0\,|\,\bfmu)$. 
Given a triplet comparison $t=(i,j,k)$, the probabilities for the response $R_{ijk}= 1$ (left stimulus $i$ is closer to the pivot stimulus $j$ than stimulus $k$) and the opposite, $R_{ijk}= 0$, is 
\begin{eqnarray}\label{equ:ProbThurstone}
\nonumber \Pr(Z_{ijk} > 0\,|\,\bfmu) &=& 1 - \Phi(\mu_k-\mu_i) - \Phi\left(\textstyle\frac{\mu_k+\mu_i-2\mu_j}{\sqrt{3}}\right) \\
 && +\, 2\,\Phi(\mu_k-\mu_i)\,\Phi\left(\textstyle\frac{\mu_k+\mu_i-2\mu_j}{\sqrt{3}}\right)\\
\nonumber \Pr(Z_{ijk} \le 0\,|\,\bfmu) &=& 1-\Pr(Z_{ijk} > 0\,|\,\bfmu).
\end{eqnarray}
Algorithm\,\ref{alg:tripletresponse} summarizes the computation.

\begin{algorithm}[t!]
\caption{Probability of a response $R_{ijk} \in \{0,1\}$ to a triplet comparison $(i,j,k)$}
\label{alg:tripletresponse}
\begin{algorithmic}[1]
\State $\bfmu = (\mu_0,\ldots,\mu_{M})$ \Comment{stimuli means in model}
\State $u_0 \gets \mu_k-\mu_i$
\State $v_0 \gets (\mu_k+\mu_i-2\mu_j) / \sqrt{3}$
\State $p \gets 1 - \Phi(u_0) - \Phi(v_0) + 2\Phi(u_0)\Phi(v_0)$
\If{$R_{ijk} = 1$} \Comment{stimulus $i$ closer to $j$ than $k$}
    \State Return $p$
\Else \Comment{stimulus $k$ closer to $j$ than $i$}
    \State Return $1-p$
\EndIf
\end{algorithmic}
\end{algorithm}

Other probabilistic models differ from the above by specifying a different probability for the triplet responses. In MLDS \cite{maloney2003maximum}, we have 
$$
    Z_{ijk} = \left|\mu_k-\mu_j\right| - \left|\mu_i-\mu_j\right| + N_{\sigma}.
$$
In this case,
\begin{equation} \label{equ:ProbMLDS}
    \Pr(Z_{ijk} > 0\,|\,\bfmu) = \Phi\left(\frac{\left|\mu_k-\mu_j\right| - \left|\mu_i-\mu_j\right|}{\sigma}\right).
\end{equation}
The default for the parameter is $\sigma = 1.$
In stochastic triplet embedding (STE, \cite{van2012stochastic}), the probability for a positive response is given directly as
\begin{equation} \label{equ:ProbSTE}
    \Pr(Z_{ijk} > 0\,|\,\bfmu) = \frac{e^{-\alpha(\mu_i-\mu_j)^2}} {e^{-\alpha(\mu_i-\mu_j)^2}+e^{-\alpha(\mu_k-\mu_j)^2}}.
\end{equation}
The parameter $\alpha>0$ is not contained in the original method. Thus, its default value is $\alpha=1$. Here, we have introduced it for the purpose of model calibration.     

In the special case of baseline triplets of the form $(i,0,k)$, we have that the response $R_{i0k}=1$, i.e., that the left stimulus, numbered~$i$, is closer to the pivot 0 than the right stimulus~$k$, may also be interpreted as the judgement that the impairment in stimulus~$k$ is greater than the impairment in stimulus~$i$. In effect, this amounts to a response to a regular pair comparison, and the probabilistic model for the decision random variable simply becomes
\begin{equation} \label{equ:ProbPC}
    \Pr(Z_{i0k} > 0\,|\,\bfmu) = \Phi\left(\mu_k-\mu_i\right).
\end{equation}
The difference to the normal interpretation of a triplet comparison is that here the impairment of the pivot is fixed to be equal to 0. In the general triplet comparison, however, all stimuli are modelled as random variables. 

%-----------------------------------------------------------------------------------
\subsection{Maximum Likelihood Estimation of the Means}
 \label{sec:thruston_b}
%-----------------------------------------------------------------------------------
For the actual reconstruction by the maximum likelihood method we take as input a finite multiset $T$ of annotated triplets, $(i,j,k,R_{ijk})$, where $R_{ijk} \in \{0,1\}$ is the response to the triplet comparison $(i,j,k)$ as in the above part. In subjective quality assessments with triplet comparisons, each triplet may be presented multiple times, collecting a response each time. Thus, $T$ may contain multiple copies of both, $(i,j,k,0)$ and $(i,j,k,1)$. To keep the notation simple, we trust that it is clear from the context what  $R_{ijk}$ refers to in each case. Assuming that the responses are independent, we have that the negative log-likelihood of this data under the model assumptions is given by 
\begin{eqnarray}
%    L(\bfmu) &=& -\sum_{(i,j,k,R_{ijk}) \in T} 
%    \Pr(Z_{ijk} > 0\,|\,\bfmu)^{R_{ijk}} (1 -  \Pr(Z_{ijk}> 0\,|\,\bfmu)^{1-R_{ijk}} \\
    \label{loglikelihood} L(\bfmu) &=& -\sum_{(i,j,k,R_{ijk}) \in T} p^{R_{ijk}} (1 -  p)^{1-R_{ijk}},\\
    \nonumber p &=& \Pr(Z_{ijk} > 0\,|\,\bfmu).
\end{eqnarray}
The MLE estimate of the latent variable then is given by
\begin{equation} \nonumber
    \hat{\bfmu} = \argmin_{\bfmu=(\mu_0,\ldots,\mu_{M})}  L(\bfmu).
\end{equation}
In our experiments, we allowed a third option for triplet question responses, namely an answer \textit{not sure} (see Subsection\,\ref{sec_interface}). To account for such undecided responses, we simply assign the value $R_{ijk} = 1/2$ to the corresponding triplets $(i,j,k)$ and use Equation\,(\ref{loglikelihood}) as given. 

There are many algorithms for such nonlinear optimization problems, and generally, there is no guarantee that the global maximum will be attained. In our computations, we used the ``fmincon" function in Matlab, which is a nonlinear programming solver that finds the minimum of a constrained nonlinear multivariable function.

Solutions to this optimization are unique up to an additive constant. This constant can be chosen arbitrarily, and we have used this option to align all reconstructions such that the reconstructed scale value for the undistorted reference stimulus $I_0$ is $\mu_0 = 0$. These reconstructions of impairment scales are not yet in JND units because their probabilistic model assumed Gaussian distributions with the variance of 0.5. Since the JND unit corresponds to a value of $\Phi^{-1}(0.75) \approx 0.6745$, we divide the results by that to obtain impairment scales in JND units.

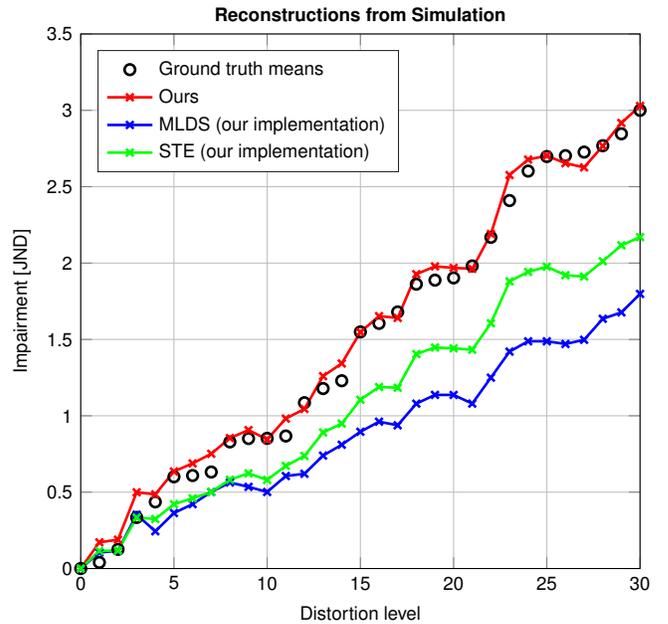
\begin{figure}[t]
\centering{
\begin{tikzpicture}

\begin{axis}[%
width=0.41\textwidth,
height=2.8in,
at={(0.772in,0.516in)},
scale only axis,
xmin=1,
xmax=31,
xtick={1,6,11,16,21,26,31},
xticklabels={{0},{5},{10},{15},{20},{25},{30}},
ymin=0,
ymax=3.5,
ytick={0,0.5,1,1.5,2,2.5,3,3.5},
yticklabels={{0},{0.5},{1},{1.5},{2},{2.5},{3},{3.5}},
axis background/.style={fill=white},
label style={font=\sffamily},
xticklabel style={font=\sffamily\scriptsize},
yticklabel style={font=\sffamily\scriptsize},
xlabel style={font=\sffamily\scriptsize},
ylabel style={font=\sffamily\scriptsize},
title style={yshift=-1ex,font=\bfseries\sffamily\scriptsize},
title={Reconstructions from Simulation},
ylabel = {Impairment [JND]},
xlabel = {Distortion level},
xmajorgrids,
ymajorgrids,
legend style={at={(0.03,0.97)}, anchor=north west, legend cell align=left, align=left, draw=white!15!black}, font = \sffamily\scriptsize
]
\addplot [color=black, line width=1.0pt, only marks, mark=o, mark options={solid, black}]
  table[row sep=crcr]{%
1	0\\
2	0.0398603241533799\\
3	0.125510607458163\\
4	0.334431413888523\\
5	0.436388193437871\\
6	0.600441428383186\\
7	0.608874687047042\\
8	0.631835551456055\\
9	0.829127684806957\\
10	0.851166161549611\\
11	0.852483858215839\\
12	0.86767031229411\\
13	1.08561734088813\\
14	1.17844907102385\\
15	1.22947687442351\\
16	1.54928185531691\\
17	1.60502042429833\\
18	1.67979971010673\\
19	1.8618065621294\\
20	1.88809461062064\\
21	1.90285281328238\\
22	1.98125576492291\\
23	2.16853340361049\\
24	2.40927658453024\\
25	2.60225326129925\\
26	2.69752273026749\\
27	2.70328765318224\\
28	2.72661088417446\\
29	2.76848069574384\\
30	2.84638951113454\\
31	3\\
};
\addlegendentry{Ground truth means}

\addplot [color=red, line width=1.0pt, mark=x, mark options={solid, red}]
  table[row sep=crcr]{%
1	0\\
2	0.171585633156883\\
3	0.188780177446083\\
4	0.499193016387284\\
5	0.485740578668422\\
6	0.635729344767131\\
7	0.688432722910064\\
8	0.751171510792217\\
9	0.854271243940795\\
10	0.90618182571592\\
11	0.845747602272178\\
12	0.981911255956583\\
13	1.04471225013016\\
14	1.25985714079496\\
15	1.34265167455886\\
16	1.54819923371452\\
17	1.6523759021929\\
18	1.64097882842387\\
19	1.92721198635857\\
20	1.97814275239387\\
21	1.96858331562182\\
22	1.96258269777585\\
23	2.19068668783567\\
24	2.57641187138581\\
25	2.67802323806989\\
26	2.70260411891276\\
27	2.65375812829499\\
28	2.62644286050355\\
29	2.76656318045598\\
30	2.9174781876452\\
31	3.02846911518746\\
};
\addlegendentry{Ours}

\addplot [color=blue, line width=1.0pt, mark=x, mark options={solid, blue}]
  table[row sep=crcr]{%
1	0\\
2	0.104731298975224\\
3	0.116550798173121\\
4	0.352914529016558\\
5	0.243906076915106\\
6	0.363263220962804\\
7	0.421596516976487\\
8	0.501142445616134\\
9	0.563258127676351\\
10	0.534761128271693\\
11	0.501190384306964\\
12	0.605669591716816\\
13	0.620316789627016\\
14	0.739417869792446\\
15	0.810476266856608\\
16	0.895363888546472\\
17	0.961519112731673\\
18	0.937845686470426\\
19	1.0794504741555\\
20	1.13701308842587\\
21	1.13665028563325\\
22	1.07994396100255\\
23	1.24984250348903\\
24	1.42028966413575\\
25	1.48794804866246\\
26	1.48761973785821\\
27	1.47026432634477\\
28	1.49754814395012\\
29	1.63516608239657\\
30	1.67719311539741\\
31	1.79806400763885\\
};
\addlegendentry{MLDS (our implementation)}

\addplot [color=green, line width=1.0pt, mark=x, mark options={solid, green}]
  table[row sep=crcr]{%
1	0\\
2	0.114916783267309\\
3	0.117546379038754\\
4	0.334317921734277\\
5	0.323814119300265\\
6	0.421297413890238\\
7	0.459744825827441\\
8	0.501046101711018\\
9	0.58119354115561\\
10	0.622995360316862\\
11	0.580285705261802\\
12	0.672081371310655\\
13	0.737148117345313\\
14	0.891221618379101\\
15	0.94887901091114\\
16	1.10532015386391\\
17	1.18875885212272\\
18	1.18367017274776\\
19	1.40410203301061\\
20	1.44711750615121\\
21	1.44145479674214\\
22	1.43263600239821\\
23	1.60597784266443\\
24	1.87997034899635\\
25	1.94223733380505\\
26	1.97626814135479\\
27	1.92039384895481\\
28	1.91183878560713\\
29	2.01179295912314\\
30	2.1165527687008\\
31	2.16986403941048\\
};
\addlegendentry{STE (our implementation)}

\end{axis}
\end{tikzpicture}%
}
\caption{Reconstructions from a sample of \num{20000} simulated random triplet question responses for 31 stimuli, spread over a range of 3\,JND units.  All three reconstruction methods yielded excellent correlations of 0.99, see Table\,\ref{corr_simulatedata}, but only our reconstruction also reproduced the correct range of the means.}
\label{simulateres}
\end{figure}

\begin{table*}[t]
\caption{Simulation for 31 stimuli over a range of 3\,JND.\linebreak
Correlation with ground truth, and range of reconstructed means, averaged over 1000 repetitions.}
\label{corr_simulatedata}
\centering
\begin{tabular}{r | c  c | c c |c c  }
Triplet responses	& \multicolumn{2}{c|}{MLDS ($\sigma =1$)} 	& \multicolumn{2}{c|}{STE ($\alpha =1$)}	& \multicolumn{2}{c}{Ours}	\\		
per sample	&	SROCC	&	Range (JND)	&	SROCC	&	Range (JND)	&	SROCC	&	Range (JND)	\\
\hline
1000 	&	$	0.922	\pm	0.030	$	&	$	2.055	\pm	0.429	$	&	$	0.917	\pm	0.053	$	&	$	2.214	\pm	0.402	$	&	$	0.913	\pm	0.064	$	&	$	3.153	\pm	0.652	$	\\
2500	&	$	0.964	\pm	0.012	$	&	$	1.885	\pm	0.257	$	&	$	0.967	\pm	0.010	$	&	$	2.187	\pm	0.211	$	&	$	0.967	\pm	0.010	$	&	$	3.068	\pm	0.326	$	\\
5000	&	$	0.979	\pm	0.008	$	&	$	1.840	\pm	0.177	$	&	$	0.980	\pm	0.006	$	&	$	2.185	\pm	0.137	$	&	$	0.981	\pm	0.006	$	&	$	3.050	\pm	0.215	$	\\
\num{10000} 	&	$	0.987	\pm	0.005	$	&	$	1.808	\pm	0.126	$	&	$	0.988	\pm	0.004	$	&	$	2.171	\pm	0.097	$	&	$	0.988	\pm	0.004	$	&	$	3.024	\pm	0.151	$	\\
\num{20000} 	&	$	0.992	\pm	0.003	$	&	$	1.797	\pm	0.090	$	&	$	0.993	\pm	0.003	$	&	$	2.168	\pm	0.067	$	&	$	0.993	\pm	0.003	$	&	$	3.015	\pm	0.105	$	\\
\hline							
	&	\multicolumn{2}{c|}{MLDS ($\sigma =1.6594$)}	&		\multicolumn{2}{c|}{STE ($\alpha =0.5316$)}	&		\\
\num{20000} & $0.992	\pm 0.003$ & $2.989 \pm	0.150$ & $0.993 \pm	0.003$ &	$2.974 \pm 0.092$ \\
\end{tabular}
%}
%\end{adjustbox}
\end{table*}

%\begin{figure*}[t]
%\centering{
%\includegraphics[width=0.32\textwidth]{images/ProbThurstone.png}
%\includegraphics[width=0.32\textwidth]{images/ProbSTE.png}
%\includegraphics[width=0.32\textwidth]{images/ProbMLDS.png}\\
%\includegraphics[width=0.32\textwidth]{images/ProbOptSigma.png}
%\includegraphics[width=0.32\textwidth]{images/ProbDefectSTE.png}
%\includegraphics[width=0.32\textwidth]{images/ProbDefectMLDS.png}
%}
%\caption{The top row shows plots of isolines of the probabilites $\Pr(Z_{ijk} > 0\,|\,\bfmu)$ for the Thurstonian  (left), STE (middle), and MLDS (right) models. The parameters $\alpha=0.5316$ for STE and $\sigma=1.6594$ for MLDS were obtained by ensuring that the range of the corresponding reconstructed scales are equal to 3\,JND.
%%minimizing the root-mean-square error of STE, resp.\ MLDS with respect to the ground truth given by the Thurstonian model, on the square $(\mu_i-\mu_j,\mu_k-\mu_j) \in [-1,1]^2$ as shown in the figure. 
%The bottom row (center and right) shows the difference between the resulting probabilities of STE resp.\ MLDS and those of the Thurstonian model. The model fit for STE is much closer to the Thurstonian reference than that of MLDS. The bottom right plot shows the parameter $\sigma$ for MLDS that locally yields equality with the Thurstonian probabilities. \textcolor{red}{[font sizes of axis labels and titles must be improved. LaTeX must be invoked for the math symbols. The two spurious legends must be removed.]}}
%\label{fig:ProbFitting}
%\end{figure*}

%-----------------------------------------------------------------------------------
\subsection{Comparison of Triplet Reconstruction Algorithms}
 \label{sec:thruston_c}
%-----------------------------------------------------------------------------------
Given that there are a number of available methods for the reconstruction of latent scale values from triplet comparisons, the question arises of which of them is the most suitable one to process subjective responses to triplet comparisons for visual quality assessment. For this purpose, we consider in this subsection simulated responses to triplet comparisons based on the Thurstonian model, i.e.\ the impairment scale of a distorted image is given by a normally distributed random variable with a corresponding mean and variance equal to 1/2. 

For the reconstruction by MLE, we compute the likelihoods from one of the equations (\ref{equ:ProbThurstone}), (\ref{equ:ProbMLDS}), and (\ref{equ:ProbSTE}), corresponding to our proposed reconstruction, MLDS, and STE, respectively. We expect that our proposed method gives the most accurate approximations because Equation \,(\ref{equ:ProbThurstone}) is directly derived from the Thurstonian model and the others are not. However, in terms of time complexity, each evaluation of the decision probability $\Pr(Z_{ijk} > 0\,|\,\bfmu)$ requires two evaluations of the normal CDF, while MLDS needs only one, and STE none. 

For baseline triplets of the form $(i,0,k)$, where the pivot is given by the undistorted source image $I_0$ as a reference, we may interpret the response also as a response to a traditional pair comparison $(i,k)$. In this case, we can also apply the usual Thurstonian reconstruction method for pair comparison. 

For our simulation, we firstly considered an artificial sequence of 31 stimuli for ground truth, with impairments on the perceptual scale, ranging from $\mu_0=0$ to $\mu_{30}=2.0235$, which corresponds to $2.0235/\Phi^{-1}(0.75) \approx 3$\,JND. We randomly sampled from the uniform distribution on the interval $[0,\mu_{30}]$ to obtain the remaining 29 stimulus means.  

%The means are:
%0.0000	0.0269	0.0847	0.2256	0.2943	0.4050	0.4107	0.4262	0.5592	0.5741	0.5750	0.5852	0.7322	0.7949	0.8293	1.0450	1.0826	1.1330	1.2558	1.2735	1.2835	1.3364	1.4627	1.6251	1.7552	1.8195	1.8234	1.8391	1.8673	1.9199	2.0235

The triplets $(i,j,k)$ were randomly sampled with the constraint $i \ne j \ne k \ne i$. The baseline triplets $(i,0,k)$ were chosen randomly with $i,k \ne 0$ and $i \ne k$. In five rounds we drew \num{1000} to \num{20000} triplets of each kind and generated one response per triplet according to the Thurstonian probabilistic model. We then applied all triplet reconstruction methods for the responses to triplets of general type $(i,j,k)$. For the baseline triplets $(i,0,k)$ we carried out the reconstruction according to pair comparison and our proposed reconstruction. We repeated the procedure \num{1000} times.

The results of our simulation are presented in Figure\,\ref{simulateres} and Table\,\ref{corr_simulatedata}. As expected, it is confirmed that our reconstruction does faithfully reconstruct the ground truth means from the triplet comparisons that were generated by the same  Thurstonian probability model that underlies our reconstruction method. For baseline triplet comparisons, the reconstruction for pair comparisons from \num{20000} ratings gave very similar results in terms of correlation as the reconstruction for our triplet comparisons, an SROCC of 0.996 (not shown in the table).

Even more notable are the findings that the other algorithms, MLDS and STE, also produced excellent results in terms of the Pearson linear as well as the Spearman rank-order correlation. However, the reconstruction ranges are around 2\,JND, thus, well below the correct 3\,JND. 

To calibrate the STE and MLDS methods to give results in JND units, one could tune their parameters $\alpha$ and $\sigma$ such that the range of reconstructed impairment values is equal to 3. For our simulation, we applied the bisection method to determine these parameters and obtained $\alpha=0.5316$ and $\sigma=1.6594$. The resulting correlations are excellent again (see Table\,\ref{corr_simulatedata}). The RMSE over all 30 reconstructed impairments are 0.1089 for MLDS, 0.0646 for STE, while for our method (without tuning a parameter), we obtained an RMSE of 0.0520.

\begin{figure*}[t]
\centering{\input{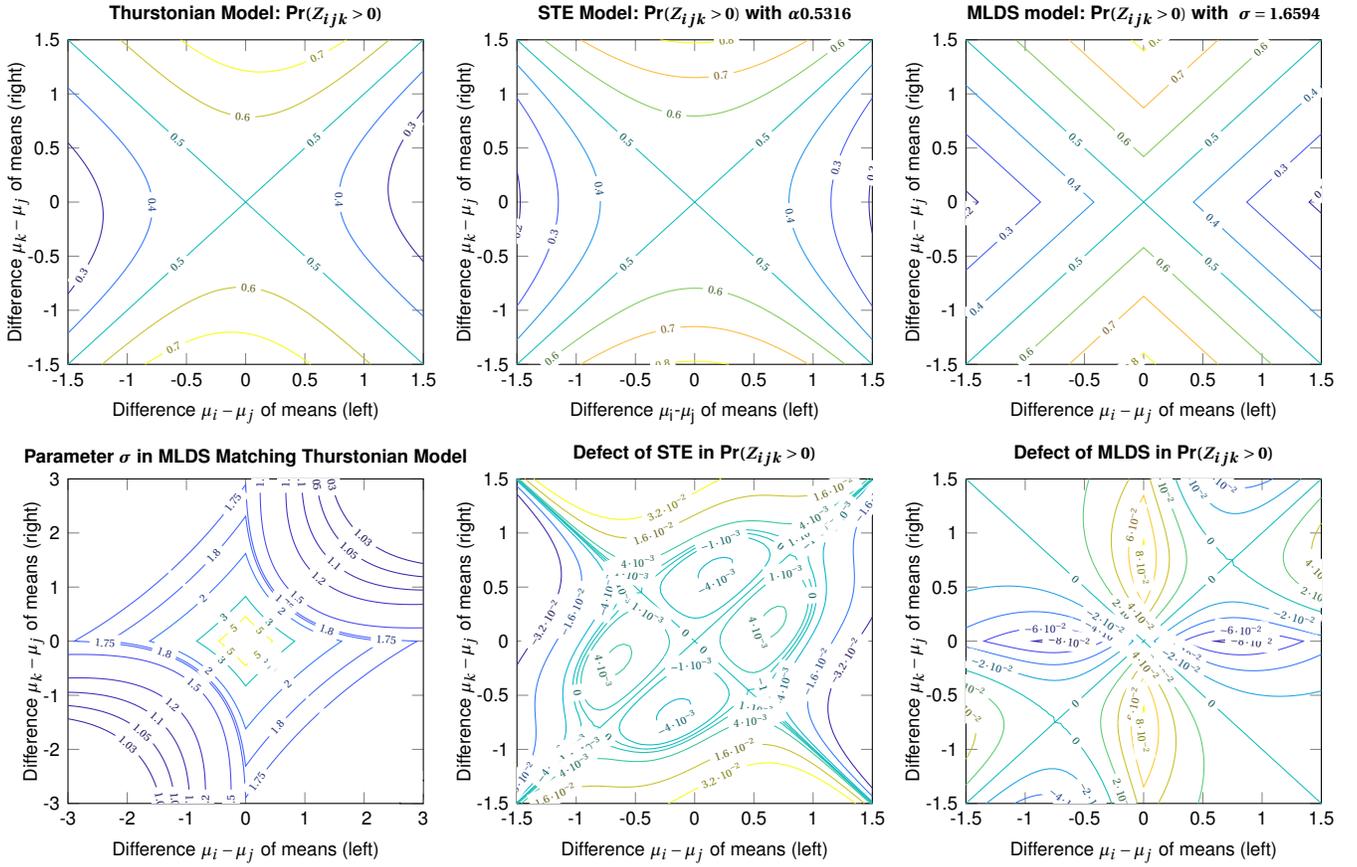}}
\caption{The top row shows plots of isolines of the probabilities $\Pr(Z_{ijk} > 0\,|\,\bfmu)$ for the Thurstonian  (left), STE (middle), and MLDS (right) models. The parameters $\alpha=0.5316$ for STE and $\sigma=1.6594$ for MLDS were obtained by ensuring that the range of the corresponding reconstructed scales is equal to 3\,JND. The bottom row (centre and right) shows the difference between the resulting probabilities of STE resp.\ MLDS and those of the Thurstonian model. The model fit for STE is much closer to the Thurstonian reference than that of MLDS. The bottom right plot shows the parameter $\sigma$ for MLDS that locally yields equality with the Thurstonian probabilities.}
\label{fig:ProbFitting}
\end{figure*}

Of course, the above procedure is not feasible in general because the ground truth range is not known as in our simulation here. One would have to resort to an estimation of a suitable parameter $\alpha$ for MLDS or $\sigma$ for STE. To this end, we propose two approaches.
\begin{enumerate}
    \item Estimate the range of the expected scale values. In our simulation, it was 3\,JND, for example. Then proceed as in our simulation. Randomly choose a sequence of scale values in the selected range, and then tune the parameter $\alpha$, respectively $\sigma$, to achieve a reconstruction by MLDS, resp.\ STE, to match the selected range. 
    \item Minimize the mean square error of the MLDS probabilities for a response $R_{ijk}=1$ by selecting $\hat{\sigma}$,
    $$
        \hat{\sigma} = \argmin_{\sigma>0} \int_{-\Delta}^{\Delta} \int_{-\Delta}^{\Delta} 
         |e(r,s\,|\,\sigma)|^2 dr\, ds
    $$
    where the error $e(r,s\,|\,\sigma)$ is given by
    $$
       \Phi\left(\textstyle\frac{\left|s\right| - \left|r\right|}{\sigma}\right) - 1 + \Phi(s) + \Phi\left(\textstyle\frac{r+s}{\sqrt{3}}\right)
      -\, 2\,\Phi(s)\,\Phi\left(\textstyle\frac{r+s}{\sqrt{3}}\right).
    $$
    Here, $r=\mu_i-\mu_j$ and $s=\mu_k-\mu_j$ denote the left and right differences of impairments in the triplet $(i,j,k)$, ranging over the square domain $[-\Delta,\Delta] \times [-\Delta,\Delta]$. Similarly, one can do the same for STE.
    In Figure\,\ref{fig:ProbFitting}, we visualize the probability functions according to the Thurstonian model along with their approximation in the MLDS and STE methods and the corresponding errors $e(r,s\,|\,\sigma)$ resp.\ $e(r,s\,|\,\alpha)$. Inspecting this figure, it becomes apparent that the globally optimal parameter $\sigma$ will be difficult to obtain as it strongly varies locally.
\end{enumerate}

In summary, all three methods produced excellent results. For baseline triplet comparisons, reconstruction by the traditional Thurstonian approach  (with MLE) was as good as our method for triplet construction. If one needs to have results on the perceptual scale given in JND units, then our proposed reconstruction should be applied. 

\section{Experimental Setup: Materials and Procedures}
\label{sec:Experimental_Setup}
%-----------------------------------------------------------------------------------
The purpose of our experimental studies was to investigate the potential and limitations of the proposed boosting strategies in the application of subjective full-reference image quality assessment. In current FR-IQA datasets, the main approaches have been DCR and PC with the reference image shown additionally, i.e., a case of baseline triplet comparison (Table\,\ref{tb:ivqadatabase}). For both of these, boosting of the underlying image distortions can be applied. Thus, we carried out three main experiments, starting out with baseline triplets, which we then extended to general triplets, finally followed by a smaller study for DCR. In the following, we refer to these as Experiments I, II, and III.

In order to evaluate aspects like accuracy, reliability, and convergence, a large number of comparisons are beneficial. Therefore, our subjective IQA experiments were conducted via crowdsourcing. For the study, a set of original pristine source images, each distorted by various types of distortions, was selected. For each source image and each distortion type, a sequence of increasingly distorted images was generated. By means of a pilot study, we took care to calibrate our dataset such that each such image sequence uniformly spans a perceptual quality range of approximately 3\,JND. In the following subsections, we briefly describe our setup and procedure to achieve these goals. 

%-----------------------------------------------------------------------------------
\subsection{Subjective Crowdsourced IQA Study}
%-----------------------------------------------------------------------------------

In terms of experimental methodology, lab studies are well established and considered reliable because the experimental environment can be controlled, and the whole procedure can be monitored. On the other hand, the number of images that can be assessed is limited due to the time requirements as well as the cost. Alternatively, crowdsourcing studies are more economical, more efficient, more scalable, and can have sufficient reliability if the setup, with a quality control mechanism included, is appropriate \cite{saupe2016crowd} and a suitable outlier removal strategy is employed \cite{ribeiro2011crowdsourcing}. We have installed several measures of control to ensure the validity of the results from our crowdsourcing campaigns, described in the following.

The experiments were carried out on the Amazon Mechanical Turk \cite{amt} platform, in which \textit{requesters} create and submit their \textit{human intelligence tasks} (HITs) for \textit{workers} that carry out the subjective quality assessment. Workers receive a monetary reward by completing a HIT. Requesters specify the number of \textit{assignments} for each HIT to control how many workers can submit work for the HIT.

In our experiments with triplet comparisons, a HIT consisted of 20 questions that gave rise to 20 (ternary) \textit{responses} or \textit{answers} from each crowdworker that completed an assignment for that HIT. For the experiment with degradation category ratings, HITs also had 20 questions each, and workers provided corresponding \textit{ratings} on a 5-point DCR quality scale.

\begin{figure}[t]
\centering{\includegraphics[width=0.48\textwidth]{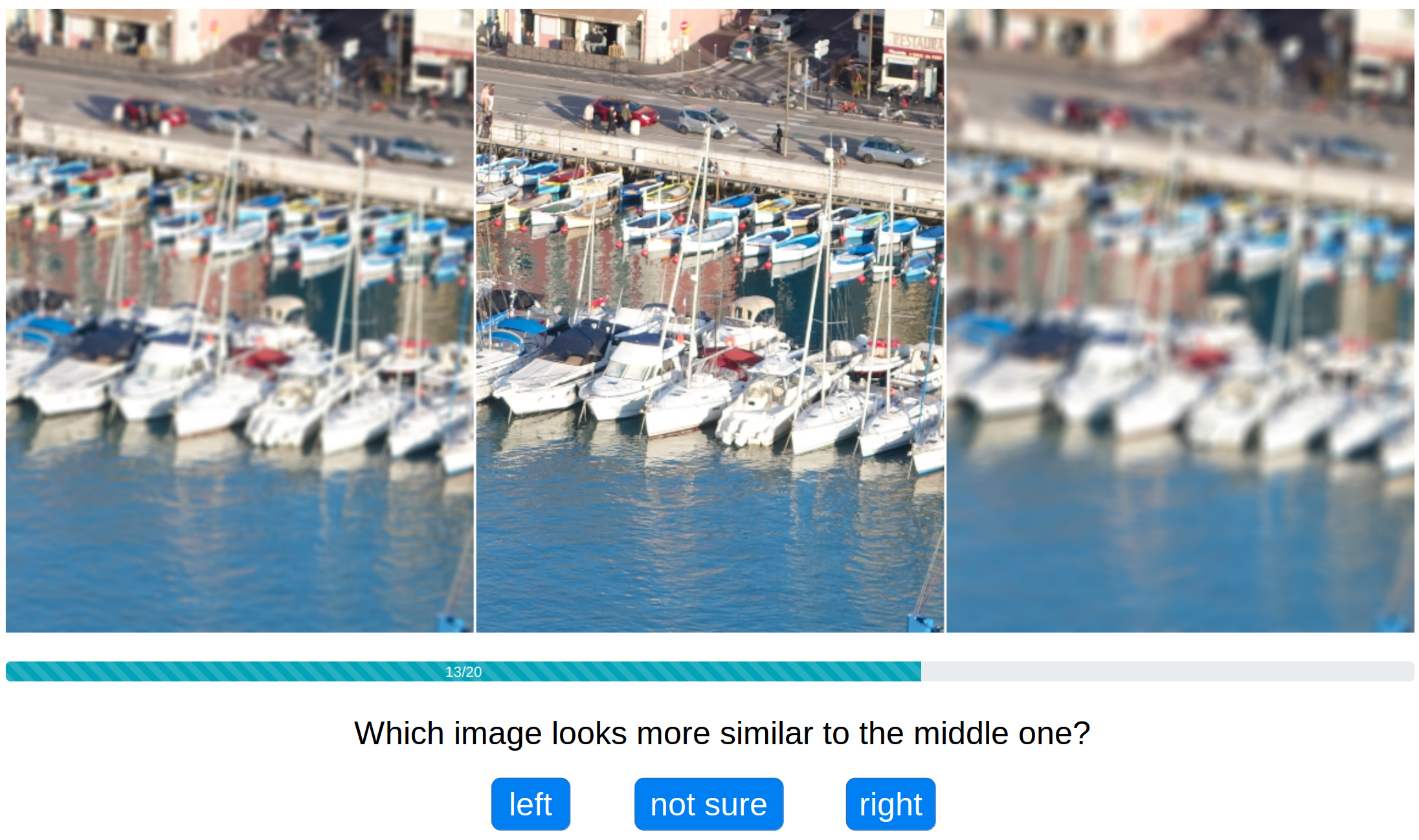}}
%\fbox{\includegraphics[width=0.98\textwidth]{images/interpc_new.eps}}}
\caption{The interface of the triplet comparison experiments without a flickering effect. Crowd workers were asked to select the image that they think looks more similar to the pivot. They could choose ``not sure'' if they could not distinguish the differences. For the experiments with a flickering effect, the interface was the same as the one without a flickering effect shown here, except that three still images were replaced by two flickering images, and the question got changed to ``Which image has a stronger flickering effect?''}
\label{plain_inter}
\end{figure}

%-----------------------------------------------------------------------------------
\subsection{Interface}\label{sec_interface} 
\begin{figure}[t]
%\centering{\includegraphics[width=0.48\textwidth]{images/dcrinterface.png} }
\centering{\includegraphics[width=0.48\textwidth]{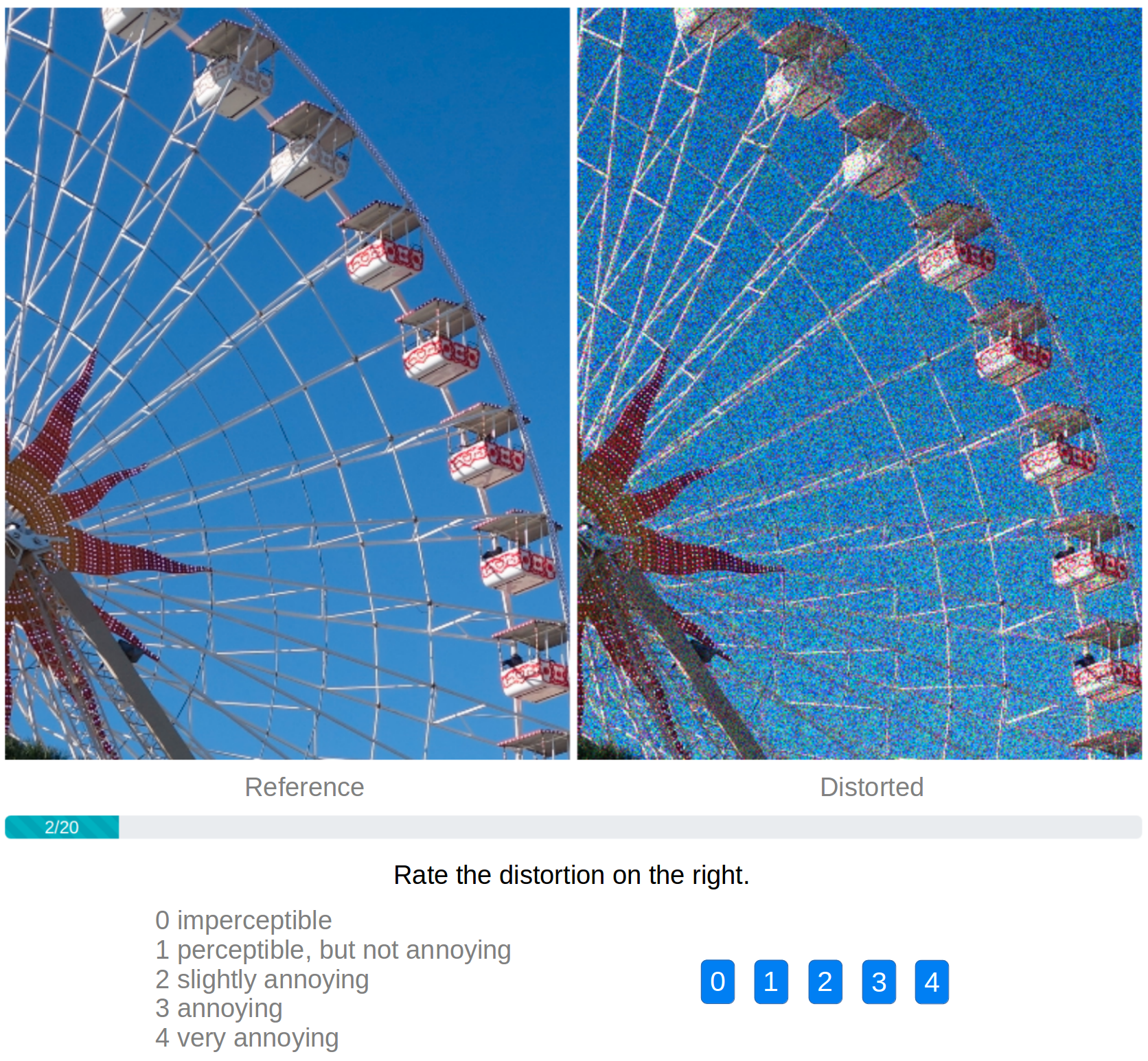} }
\caption{DCR Interface. Two images were placed side by side, with the reference image on the left. Crowd workers were asked to rate the distortion of the right image w.r.t.\ its reference on the left on one of the five categories: 0 (imperceptible), 1 (perceptible but not annoying), 2 (slightly annoying), 3 (annoying), 4 (very annoying).}
\label{dcr_interface}
\end{figure}

% --------------- interface ------------
At the beginning of the experiment, a detailed instruction was shown to the crowd workers, after which they were allowed to start doing assignments. 
\begin{itemize}
        \item In the parts of Experiments I and II that used TC without a flickering effect (Plain, A-, Z-, and AZ-boosted TC), three images were displayed in a row, see Figure\,\ref{plain_inter}. Crowd workers  selected the image that looked more similar to the pivot image in the middle by clicking ``left'' or ``right''. If they could not decide, a third choice, ``not sure'', was available. This option had been introduced in subjective evaluations of the JPEG\,XS image compression \cite{mcnally2017jpeg} and had been found useful to reduce subject stress and fatigue. In an earlier study on the unforced-choice paradigm in applications in audiology, it was also concluded that the efficiency of pair comparison might be compromised when participants are forced to choose between stimuli \cite{punch2001paired}.
        %"Offering a ternary choice to subjects reduces subject stress and fatigue and was deemed beneficial for the reliability of the subjective evaluation results. This in light of the fact that each subject underwent three test sessions of approximately 20 minutes duration during which only a small number of different stimuli had to be evaluated, potentially leading to fatigue and loss of interest." (JPEG XS call for proposals subjective evaluations) mcnally2017jpeg

        \item In the other parts of Experiments I and II that used TC with a flickering effect (F-, AF-, ZF-, and AZF-boosted TC), two flickering images were displayed side by side. Crowd workers selected the image with a perceived stronger flickering effect by clicking ``left'' or  ``right'', or use the ``not sure'' option.

        \item For Experiment III using DCR, two images were displayed side by side, the reference image on the left and the test image on the right, see Figure\,\ref{dcr_interface}. Crowd workers rated the distortion of the test image on the 5-scale category ratings ranging from 0 (imperceptible) to 4 (very annoying).
\end{itemize}
For each of the 20 questions in one assignment, crowd workers had eight seconds to enter their responses. The images were shown only during the first five seconds. In case no answer was given by the crowd worker within the eight seconds, the response was labelled as ``skipped''. Thus, the total time for an assignment was 2m 40s.  

%-----------------------------------------------------------------------------------
\subsection{HIT-Level Quality Control}\label{sec:qualitycontrol}
%-----------------------------------------------------------------------------------
Subjective assessment of image quality through crowdsourcing may pose some challenges due to the lack of control over the experimental environment, lack of knowledge about the background of the workers, and limited reliability of the experimental results. Therefore, we need to detect and filter out low-quality responses. Unreliable responses may be caused by technical problems with the workers' screens and devices, misunderstanding of the subjective task, e.g., limited English proficiency of some international workers, and lack of attention. In addition, some of the workers may try to answer quickly to get the maximum payment in a shorter time, resulting in responses of insufficient quality.

We ensured the quality of workers' answers in the crowdsourcing studies by monitoring the number of questions in each assignment that the workers skipped and including one hidden test question in each assignment. For example, to select suitable test questions for Experiment~I, we proceeded as follows. Based  on our pilot study (see Subsection\,\ref{sec:prestudyquality} below), we first chose baseline triplets with distorted images left and right, for which the perceptual difference in distortion was relatively large, about 1.75\,JND. By a following visual inspection, we then discarded triplet comparisons that did not seem to suggest a straightforward correct response. For Experiments II and III, we proceeded similarly.

% ------------- info for test question
%\textcolor{blue}{We chose the triplets of which the distortion level is of a span of 7 as the candidates (350 in total). We further inspected all the candidate triplets and discarded the ones whose differences can hardly be distinguished, resulting in 199 triplets remained as test questions.} 
%\textcolor{red}{We chose the triplets of which the distortion level is of a span of $k$ ( $k =7$ in Pre-study and Experiment I, $k = 19$ and $9$ in plain and boosted TC in Experiment II, respectively) as the candidates for TC experiments and chose the distorted images at levels 10 -- 12  as the candidates for DCR study. We then discard the ones that are not that easy to be distinguished via visual inspection.}

If a worker skipped (did not answer) more than three questions in an assignment or the hidden test question was answered incorrectly, the assignment was rejected, and all of the responses were discarded. We did not pay the workers for their rejected assignments. 

Responses of rejected assignments were re-collected by making the HITs available again for new workers. Such a rejection and re-collection procedure was carried out for three rounds. We did not reject any assignments in the last, smallest re-collection round, regardless of the performance on test questions and the number of provided responses. By this procedure, we filtered out the low-quality responses at the assignment level and ensured that the number of desired assignments was achieved. 

Accepted assignments might still contain outlier responses. In the next subsection, we describe the procedure for removing such outliers. The statistics of the rejected assignments and outliers for all experiments are provided in Table\,\ref{info1}.

\subsection{Robust HIT-Level Outlier Removal}
\label{sec:outlierremoval}
%-----------------------------------------------------------------------------------
%In general, outliers can be removed based on the ability that a subject makes correct identifications \cite{itu1997methods}. Outliers could also be removed based on the inconsistencies compared with the mean result or based on the ability that a subject makes correct identifications \cite{hossfeld2013best}.
%Just a remark: These two above citations do not support the text, at least I could not find anything in these papers for that purpose

During data collection in each experiment,  unreliable assignments for HITs were discarded and penalized as described above in Section \ref{sec:qualitycontrol}. However, there may still have been uncovered unreliable data left that should be identified as outliers and removed before reconstruction of quality scales. It seems inappropriate to classify individual responses to triplet questions as outliers since the answers are not on an interval scale but ternary (``left'', ``right'', and ``not sure''). Therefore, we considered outlier removal at the HIT assignment level, requiring a multivariate method. After the quality control during each experiment, each assignment carried 16--19 answers for the 19 triplet comparisons per assignment, not counting the single test question per assignment and allowing for skipping up to three questions. 

To this end, we aim at a robust multivariate outlier detection method that flags a prescribed percentage of assignments that markedly differ from the consensus given by the remaining majority of assignments. A robust approach deviates from conventional ones in that the statistics used to identify outliers do not suffer from the influence of the outliers themselves.

The most common and recommended multivariate outlier detection method in this spirit is a fast version of the minimum covariance determinant (Fast-MCD) approach \cite{leys2018detecting}. However, the MCD method operates with Mahalanobis distances that are not suitable in our case since the multivariate data are not only vectors of ternary decisions rather than from a real Euclidean vector space, but also of variable dimension (16--19).

A similar approach was given by the $k$-means{--} algorithm \cite{chawla2013k}. This modification of the classical $k$-means clustering algorithm takes a desired number of outliers into account that is farthest from the cluster centres. Convergence of local optima was proven. Cluster centres are given by the means of the cluster data points. For our application, the method would have to be run for a single cluster ($k=1$). However, this does not work since data from HIT assignments are not simply vectors of some vector space, and these data cannot be averaged. 

To remove HIT assignments as outliers, we propose an adaptation of the above two robust detection methods. 

Firstly, we define the consensus of a subset of assignments for a HIT due to the reconstructed impairment scale values for all stimuli involved in the corresponding experimental study. This consensus replaces the cluster means as used in Fast-MCD and $k$-means{--}. 

Secondly, we need an algorithm to compute the distance of each assignment to the consensus given by the impairment scales reconstructed from the corresponding majority of assignments. A small distance should indicate that the responses collected in a HIT assignment agree well with the reconstructed impairments. Large distances suggest strong disagreement with the consensus and that the corresponding HIT assignments may be regarded as outliers. 

%-----------------------------------------------------------------------------------
\subsubsection{Distance to Consensus for an Assignment of Triplet Comparisons}\label{subtripdist}
For the case of triplet comparisons, we define this distance for an assignment as the complementary weighted true positive rate of the corresponding $N$ responses with respect to the given consensus as follows. Considering the $n$-th response to a triplet question of type $(i,j,k)$ for a particular reference image $I_0$, a given distortion type, and corresponding distorted images $I_i,I_j,I_k$ that make up the triplet question, we compare it with the corresponding impairment scale values $\hat{\mu}_i,\hat{\mu}_j,\hat{\mu}_k$ from the current consensus. If the answer 
``left'' (resp.\ ``right'') for this $n$-th triplet question is in accordance with the consensus, we assign a score of value $v_n = 1$ to it. The answer ``not sure'' earns a score of $v_n = 0.5$. The following table completes the definition of the score $v_n$ for the response to the $n$-th triplet question $(i,j,k)$.
\begin{center}
\small
\begin{tabular}{c  c  c}
Response & $|\hat{\mu}_k - \hat{\mu}_j| \ge |\hat{\mu}_i - \hat{\mu}_j|$	
         & $|\hat{\mu}_k - \hat{\mu}_j| < |\hat{\mu}_i - \hat{\mu}_j|$\\
\hline
left & 1 & 0 \\
right  & 0 & 1 \\
not sure & 0.5 & 0.5 
\end{tabular}
\end{center}

The difficulty of triplet questions varies according to the difference of the left and right differences of impairment scales, $D_l = |\hat{\mu}_i - \hat{\mu}_j|$ and $D_r = |\hat{\mu}_k - \hat{\mu}_j|$. If $D_r \approx D_l$, the decision which is perceptually the smaller one is hard. In this case, an answer that disagrees with the consensus should not be penalized severely. On the other hand, if $|D_r - D_l|$ is large, a wrong decision should be penalized more strongly. Therefore we introduce the weight 
$$
    w_n = |D_r - D_l| = |\hat{\mu}_k - \hat{\mu}_j| - |\hat{\mu}_i - \hat{\mu}_j|
$$
for the $n$-th response and define the distance of the assignment w.r.t.\ the consensus as
\begin{equation}\label{distTrip}
    d = 1 - \frac{\sum_{n=1}^{N} w_n v_n}  {\sum_{n=1}^{N} w_n}.
\end{equation}
We have that $0 \le d \le 1$, and the maximal distance of $d=1$ implies that all responses in that assignment were against the consensus of the majority. 

\begin{figure*}[t]
\centering{
\includegraphics[width=1\textwidth]{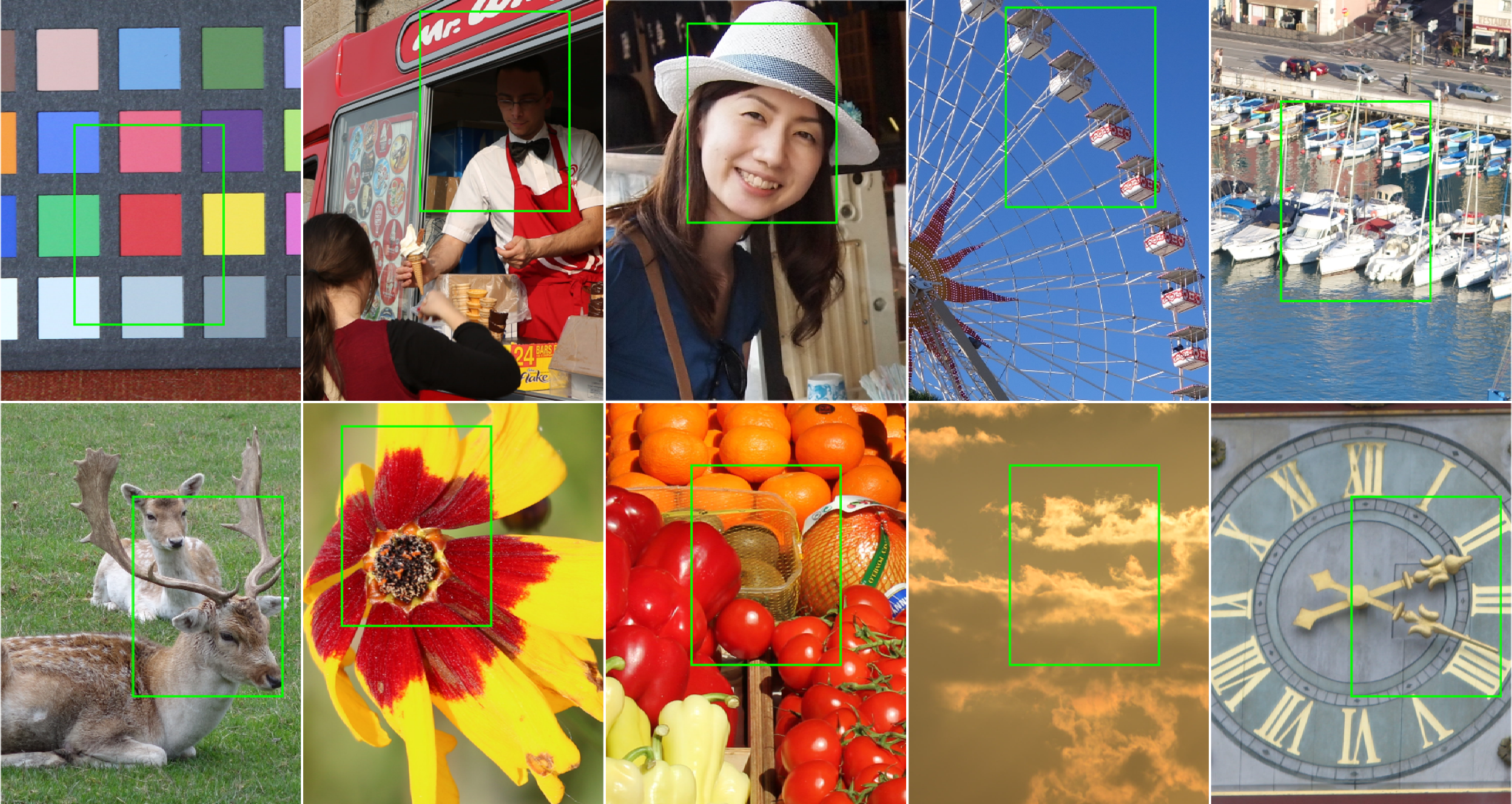}
}
\vspace{-5pt}
\caption{The (cropped) source images for our experiment. From upper left to lower right: source images with a resolution of $512 \times 384$ cropped from the images in MCL-JCI dataset with the following IDs: SRC01, SRC03, SRC06, SRC07, SRC09, SRC17, SRC28, SRC31, SRC45, SRC50, respectively. The green inset rectangles ($256 \times 192$ pixels) indicate the regions used for the boosting methods with zooming.}
\label{source images}
\end{figure*}

%-----------------------------------------------------------------------------------
\subsubsection{Distance to Consensus for an Assignment of DCR Questions}
In case of an assignment of DCR questions, the consensus produced by a subset of HIT assignments is given by the corresponding DMOS values for all test images involved in the experiment. In a HIT assignment with ratings $r_n$ for pairs $\left(I_0^n,I_{k(n)}^n\right)$ and corresponding DMOS values $\hat{\mu}\left(I_{k(n)}^n\right), n=1,\ldots,N,$ from the consensus, we define the distance of the given ratings to the consensus as the mean of the absolute differences to the consensus,
\begin{equation}\label{distDCR}
   d = \frac{1}{N}\sum_{n=1}^N \left|r_n - \hat{\mu}\left(I_{k(n)}^n\right)\right|.
\end{equation}

With these definitions made, we now give the iterative, robust outlier removal algorithm.
\begin{enumerate}
    \item Input: $M$ HIT assignments with responses, a target $L<M$ of assignments to be kept.
    \item Start with the subsample of all $M$ HIT assignments.
    \item Compute the consensus of the subsample: reconstruction of all impairment scales.
    \item Compute the distances of all $M$ assignments from the consensus by Equation\,(\ref{distTrip}), resp.\,(\ref{distDCR}).
    \item Choose the $L$ smallest distances and create a new subset.
    \item Repeat steps 3 to 5 until convergence (new subset is the same as the old one) or a timeout.
\end{enumerate}
In our experiments, we removed a fraction of 5\% of HIT assignments as outliers and observed convergence in just 4--7 iterations.

%   Guidelines to write figure captions professionally:  
%   https://www.internationalscienceediting.com/how-to-write-a-figure-caption/ 

%-----------------------------------------------------------------------------------
\subsection{Source Images}
%-----------------------------------------------------------------------------------
Ten source images were selected from the MCL-JCI dataset \cite{jin2016statistical}, whose original resolution is $1080 \times 1920$. In our subjective study, the original resolution is too large to display on the screens of crowd workers. To ensure that a triplet can be displayed without image re-scaling, we manually cropped each image to $512 \times 384$ pixels. We chose to crop portrait-mode subimages because triplets of such images better utilize screen space.  We further cropped the images to $256 \times 196$ pixels for experiments with boosting by zooming and subsequently upscaled them back to $512 \times 384$ pixels for display. Figure\,\ref{source images} shows the ten (cropped) source images and their parts used for zooming.

\begin{figure*}[t]
\centering{
\begin{tikzpicture}

\begin{axis}[%
width=1.72in,
height=2.2in,
at={(0in,0in)},
scale only axis,
axis on top,
xmin=0.5,
xmax=384.5,
tick align=outside,
y dir=reverse,
ymin=0.5,
ymax=512.5,
axis line style={draw=none},
ticks=none,
title style={font=\bfseries\sffamily\scriptsize, yshift=-1ex},
title={JPEG 2000}
]
\addplot [forget plot] graphics [xmin=0.5, xmax=384.5, ymin=0.5, ymax=512.5] {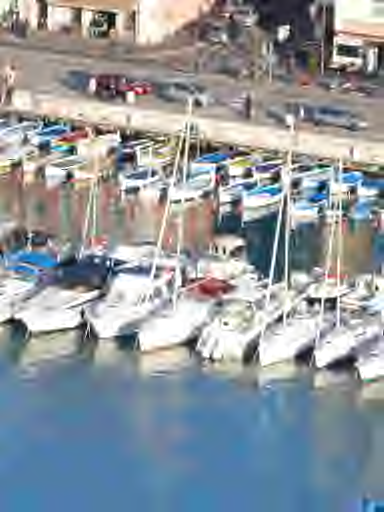};
\end{axis}

\begin{axis}[%
width=1.72in,
height=2.2in,
at={(0in,2.45in)},
scale only axis,
axis on top,
xmin=0.5,
xmax=384.5,
tick align=outside,
y dir=reverse,
ymin=0.5,
ymax=512.5,
axis line style={draw=none},
ticks=none,
title style={font=\bfseries\sffamily\scriptsize, yshift=-1ex},
title={Reference Image}
]
\addplot [forget plot] graphics [xmin=0.5, xmax=384.5, ymin=0.5, ymax=512.5] {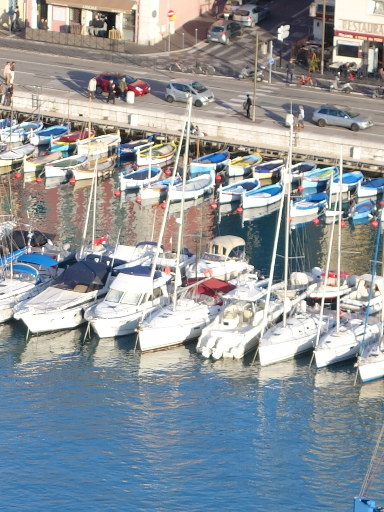};
\end{axis}

\begin{axis}[%
width=1.72in,
height=2.2in,
at={(1.74in,0in)},
scale only axis,
axis on top,
xmin=0.5,
xmax=384.5,
tick align=outside,
y dir=reverse,
ymin=0.5,
ymax=512.5,
axis line style={draw=none},
ticks=none,
title style={font=\bfseries\sffamily\scriptsize, yshift=-1ex},
title={Lens Blur}
]
\addplot [forget plot] graphics [xmin=0.5, xmax=384.5, ymin=0.5, ymax=512.5] {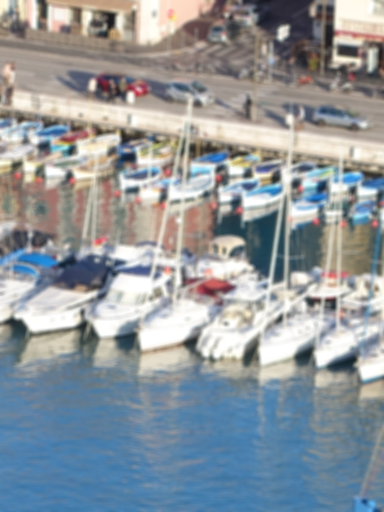};
\end{axis}

\begin{axis}[%
width=1.72in,
height=2.2in,
at={(1.74in,2.45in)},
scale only axis,
axis on top,
xmin=0.5,
xmax=384.5,
tick align=outside,
y dir=reverse,
ymin=0.5,
ymax=512.5,
axis line style={draw=none},
ticks=none,
title style={font=\bfseries\sffamily\scriptsize, yshift=-1ex},
title={Color Diffusion}
]
\addplot [forget plot] graphics [xmin=0.5, xmax=384.5, ymin=0.5, ymax=512.5] {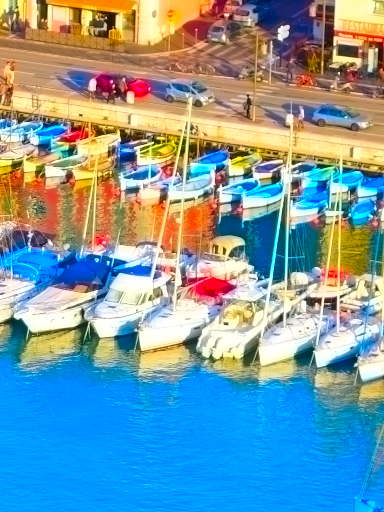};
\end{axis}

\begin{axis}[%
width=1.72in,
height=2.2in,
at={(3.48in,0in)},
scale only axis,
axis on top,
xmin=0.5,
xmax=384.5,
tick align=outside,
y dir=reverse,
ymin=0.5,
ymax=512.5,
axis line style={draw=none},
ticks=none,
title style={font=\bfseries\sffamily\scriptsize, yshift=-1ex},
title={Motion Blur}
]
\addplot [forget plot] graphics [xmin=0.5, xmax=384.5, ymin=0.5, ymax=512.5] {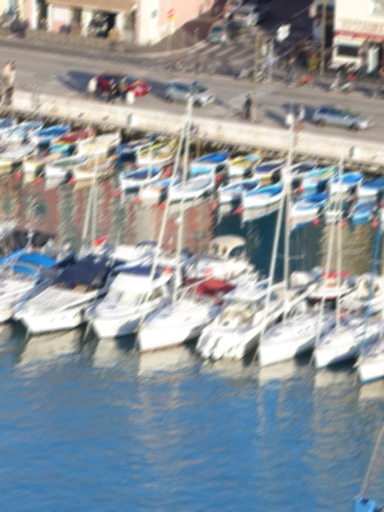};
\end{axis}

\begin{axis}[%
width=1.72in,
height=2.2in,
at={(3.48in,2.45in)},
scale only axis,
axis on top,
xmin=0.5,
xmax=384.5,
tick align=outside,
y dir=reverse,
ymin=0.5,
ymax=512.5,
axis line style={draw=none},
ticks=none,
title style={font=\bfseries\sffamily\scriptsize, yshift=-1ex},
title={High Sharpen}
]
\addplot [forget plot] graphics [xmin=0.5, xmax=384.5, ymin=0.5, ymax=512.5] {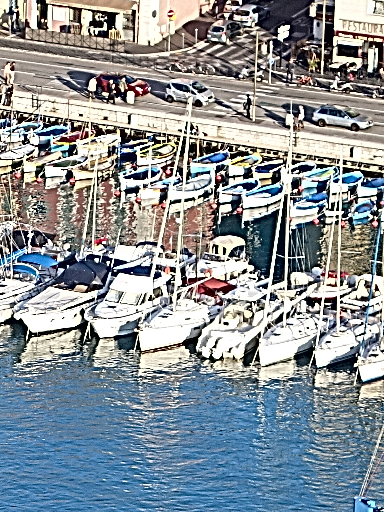};
\end{axis}

\begin{axis}[%
width=1.72in,
height=2.2in,
at={(5.22in,0in)},
scale only axis,
axis on top,
xmin=0.5,
xmax=384.5,
tick align=outside,
y dir=reverse,
ymin=0.5,
ymax=512.5,
axis line style={draw=none},
ticks=none,
title style={font=\bfseries\sffamily\scriptsize, yshift=-1ex},
title={Multiplicative Noise}
]
\addplot [forget plot] graphics [xmin=0.5, xmax=384.5, ymin=0.5, ymax=512.5] {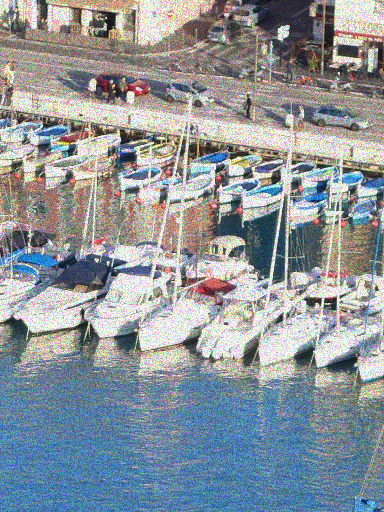};
\end{axis}

\begin{axis}[%
width=1.72in,
height=2.2in,
at={(5.22in,2.45in)},
scale only axis,
axis on top,
xmin=0.5,
xmax=384.5,
tick align=outside,
y dir=reverse,
ymin=0.5,
ymax=512.5,
axis line style={draw=none},
ticks=none,
title style={font=\bfseries\sffamily\scriptsize, yshift=-1ex},
title={Jitter}
]
\addplot [forget plot] graphics [xmin=0.5, xmax=384.5, ymin=0.5, ymax=512.5] {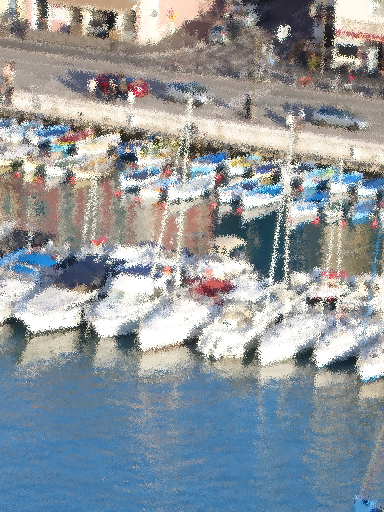};
\end{axis}
\end{tikzpicture}%
}
\caption{One of the source images (upper left) is distorted by each of the seven considered types of distortion. The degree of distortion is the largest used in this study. It corresponds to the third JND w.r.t.\ the source image.}
\label{alldistype}
\end{figure*}

%-----------------------------------------------------------------------------------
\subsection{Distortion Types}
%-----------------------------------------------------------------------------------
For our validation experiments, the source images were degraded by seven distortions, selected from \cite{lin2019kadid,IQAExperts300,hosu2018expertise}. All distortions were implemented in Matlab, with source code made available by the authors. 

\begin{itemize}
   \item \textbf{Color Diffusion} converts an image from RGB to CIELAB color space, where a 2D Gaussian smoothing kernel is used to blur the $a$ and $b$ channels. Its distortion magnitude is determined by the standard deviation of the kernel.
   \item \textbf{High Sharpen} applies the unsharp masking method to sharpen an image. An image is sharpened by subtracting a blurred (unsharp) version of the image from itself. 
   Its distortion magnitude is determined by the parameter of the strength of the sharpening effect.
   \item \textbf{Jitter} warps an image according to two matrices describing random local shifts of each pixel in horizontal and vertical directions. The amount of distortion is determined by the magnitude of the shift.  
    \item \textbf{JPEG\,2000} is an image compression standard with distortion magnitude determined by the compression ratio. 
   \item \textbf{Lens Blur} performs spatial 2D filtering on each color channel of an image with a circular kernel, whose distortion magnitude is determined by the radius of the kernel. 
   \item \textbf{Motion Blur} performs spatial 2D filtering on each color channel of an image with a kernel oriented at 45 degrees. The distortion magnitude is determined by the size in the direction of motion.
   \item \textbf{Multiplicative Noise} adds speckle noise to an image $I$, obtaining $I + n \odot I$, where $n(x,y)$ are i.i.d.\  random variables, uniformly distributed with zero mean and an adjustable variance. The magnitude of the distortion is determined by the variance. 
\end{itemize}
Figure\,\ref{alldistype} shows one of our source images together with distorted images for all of the seven distortion types.

%-----------------------------------------------------------------------------------
\subsection{Distortion Levels: Design}
\label{sec:leveldesign}
%-----------------------------------------------------------------------------------
For each combination of a source image and a distortion type, a sequence of increasingly distorted images is to be defined. In previous FR-IQA datasets, only 4--6 levels of distortion were considered (Table\,\ref{tb:ivqadatabase}). The boosting techniques are expected to help differentiate between distorted images that look the same at first glance. Thus, for our first main experiment, we strove to generate image sequences such that the perceptual difference of any two successive images is fixed at only 0.25\,JND. This is a very small difference: According to the Thurstonian probabilistic model, in a 2AFC experiment, the fraction of observations that correctly identify the stimulus with better quality is  $\Phi(0.25\,\Phi^{-1}(0.75)) \approx 0.5670$. The corresponding detection rate is only $2 \cdot 0.5670 -1 \approx 0.1339$. Our second main experiment shrunk the spacing even more, to 0.1\,JND, corresponding to an expected detection rate of merely 0.0538. However, in this case, we considered only one particular type of distortion to keep the overall cost within our budget for the crowdsourcing. 

Having defined the psychovisual spacing of distortion in each image sequence, the question remained over what range of distortions the sequence should span. In a recent study on just noticeable differences in video sequences, it was found that the perceptual quality at the third JND is between fair and poor on the 5-point ACR scale \cite{wang2017videoset}. The authors concluded that distortions stronger than 3\,JND are not acceptable by today’s viewers. Thus, we set the range of impairment for each image sequence to 3\,JND. For content providers of high-quality media, we believe the first third of this range, from lossless compression up to 1\,JND is most relevant.

Therefore, in our first main experiment, an image sequence consisted of the pristine, original image together with 12 distorted versions at impairments of $0.25\, k$\,JND, $k=1,\ldots,12$. In the second experiment, we had even 30 distorted images uniformly ranging over 3\,JND. In many of our figures, we show results over these 12, respectively 30, distortion levels. Since all reference source images and all distortion types test images at the same distortion level correspond to nearly the same perceived magnitude of distortion, we can average results over these image sequences for each distortion level. 

%----------------------------------------------------------------------------------
\subsection{Test Image Generation}\label{sec:testimage}
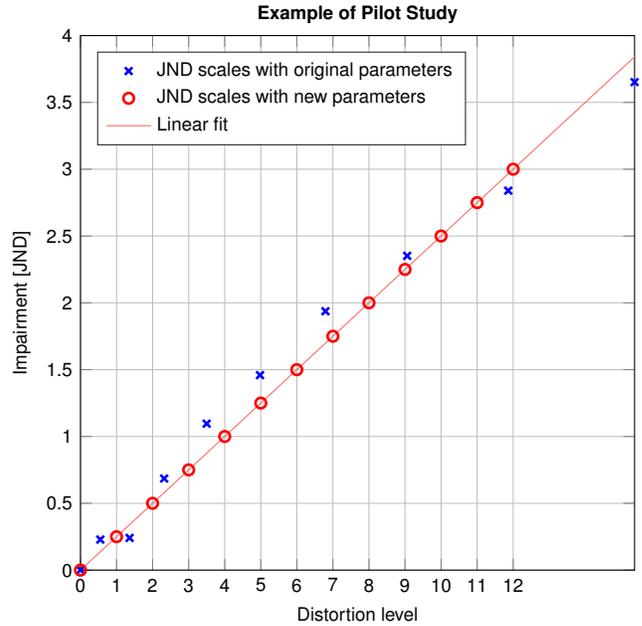
\begin{figure}[t]
\centering{
\begin{tikzpicture}

\begin{axis}[%
width=2.9in,
height=2.8in,
at={(0in,0.516in)},
scale only axis,
xmin=0,
xmax=4.9479,
label style={font=\sffamily},
xticklabel style={font=\sffamily\scriptsize},
yticklabel style={font=\sffamily\scriptsize},
xlabel style={font=\sffamily\scriptsize},ylabel style={font=\sffamily\scriptsize},
title style={yshift=-1ex,font=\bfseries\sffamily\scriptsize},
ytick = {0, 0.5, 1, 1.5, 2, 2.5, 3, 3.5, 4},
yticklabels = {{0}, {0.5}, {1}, {1.5}, {2}, {2.5}, {3}, {3.5}, {4}},
xtick={0,0.321997275016312,0.643994550032625,0.965991825048937,1.28798910006525,1.60998637508156,1.93198365009787,2.25398092511419,2.5759782001305,2.89797547514681,3.21997275016312,3.54197002517944,3.86396730019575},
xticklabels={{0},{1},{2},{3},{4},{5},{6},{7},{8},{9},{10},{11},{12}},
xlabel={Distortion level},
ymin=0,
ymax=4,
ylabel={Impairment [JND]},
axis background/.style={fill=white},
title={Example of Pilot Study},
xmajorgrids,
ymajorgrids,
legend style={at={(0.03,0.97)}, anchor=north west, legend cell align=left, align=left, draw=white!15!black, font = \sffamily\scriptsize}
]
\addplot [color=blue, line width=1pt, only marks, mark=x, mark options={solid, blue}]
  table[row sep=crcr]{%
0	0\\
0.1771	0.228860676330079\\
0.43864	0.241070408850966\\
0.74726	0.6854895069647\\
1.1267	1.09600426925855\\
1.6031	1.45910306569522\\
2.1874	1.93738190619698\\
2.917	2.35184819466875\\
3.82	2.83957755722041\\
4.9479	3.65182751247467\\
};
\addlegendentry{JND scales with original parameters}

\addplot [color=red, line width=1pt, only marks, mark=o, mark options={solid, red}]
  table[row sep=crcr]{%
0	0\\
0.321997275016312	0.25\\
0.643994550032625	0.5\\
0.965991825048937	0.75\\
1.28798910006525	1\\
1.60998637508156	1.25\\
1.93198365009787	1.5\\
2.25398092511419	1.75\\
2.5759782001305	2\\
2.89797547514681	2.25\\
3.21997275016312	2.5\\
3.54197002517944	2.75\\
3.86396730019575	3\\
};
\addlegendentry{JND scales with new parameters}

\addplot [color=white!40!red]
  table[row sep=crcr]{%
0	0\\
0.1771	0.137501163628779\\
0.43864	0.340561888278529\\
0.74726	0.580175717296675\\
1.1267	0.874774483684613\\
1.6031	1.24465339025011\\
2.1874	1.6983062976939\\
2.917	2.264770718832\\
3.82	2.96586360848072\\
4.9479	3.84156977706852\\
};
\addlegendentry{Linear fit}

\end{axis}
\end{tikzpicture}%
}
\caption{Example of the linear fitting procedure in the pilot study for the image sequence with source image SCR03 and distortion type ``High Sharpen''. On the horizontal axis, the physical distortion parameter for the selected distortion type increases linearly. 13 impairment values (blue crosses) were reconstructed, the last 4 of them being greater than 3\,JND. The last 3 of these are disregarded and not shown. From the remaining 10 points, the line fit was generated. Using this result, 13 equally spaced physical distortion parameters, as shown, were obtained, which span the range of 3\,JND on the perceptual impairment scale. These parameters define the twelve distortion levels for the image sequence used in Experiments I and III. For this example, a difference of one distortion level corresponds to a difference of 0.322 for the physical distortion parameter of the type ``High Sharpen''.}
%The MSE and confidence interval of the linear fit are 0.0268 and $[0.7274, 0.8254]$, respectively.}
\label{linearfitvis}
\end{figure}

%-----------------------------------------------------------------------------------
In order to determine sequences of distorted images with the desired equal spacing on the perceptual scale, we carried out a pilot study. The overall procedure for each of the 70 image sequences (10 source images, 7 distortion types) was as follows:
\begin{enumerate}
\item Generate a sequence of 12 distorted images by increasing the corresponding scalar distortion parameter $\lambda$. The parameters are chosen by the method of bisection to yield an image sequence that is equally spaced according to the structural similarity index SSIM \cite{wang2004image}.
\item Run a crowdsourcing study using pair comparison with additional display of the corresponding source image, which is equivalent to the baseline triplet comparison strategy. 
\item Reconstruct the impairment scales from the comparison data in JND units. Calibrate the scales such that the impairment for the pristine source image is equal to 0. The result is a sequence of impairments, parametrized over the corresponding distortion parameter $\lambda$ for the given distortion type.
\item Truncate this sequence such that only the last remaining impairment value is outside of the range from 0 to 3\,JND. Fit a straight line to these data points, constrained to pass through the point for the source image. Without loss of generality, we may assume that this is  the origin, $(0,0)$. Let the slope of the line fit be $s > 0$. Then the expected impairment at parameter $\lambda$ according to the line fit is $s\lambda$.
\item Define the sequence of distortion parameters as $$\lambda_k= \frac{k}{4s},\quad k=0,\ldots,12.$$
\item Generate the sequence of distorted images according to these parameter settings. As a result, the impairment scale of the $k$-th image will be approximately $0.25\,k$\,JND, and the distortions span a total range of about 3\,JND.
\end{enumerate}
Figure\,\ref{linearfitvis} shows an example of impairment reconstructions for one image sequence together with the  line fit and the resulting choices of 12 physical distortion parameters, which we can expect to correspond to distorted images uniformly spaced 0.25\,JND apart from each other. 

For the second experiment the intended spacing in impairment is 0.1\,JND, and the procedure is the same as above, except for Step 5, where we set $\lambda_k= \frac{k}{10s}, k=0,\ldots,30$. This was done for 10 image sequences from the 10 source images, but only one distortion type, motion blur.

In the rest of this subsection, we briefly describe the details of the selection of the baseline triplet comparisons in the pilot study, the numbers of collected responses, and the quality control. 

\subsubsection{Sampling Strategy}\label{sec:prestudysampling}
%-----------------------------------------------------------------------------------
Our sampling strategy proceeded in three rounds of data collection. The first round was for initialization. For each of the 70 image sequences, we randomly sampled pairs of the 13 test images (including the source image) by choosing the edges of a random sparse graph with vertex degree of six and with nodes corresponding to images. Thus, each image was randomly compared to six other images of the same sequence. For all sequences together, this resulted in $39 \times 70 = \num{2730}$ triplets of images. We used baseline triplet comparisons, so the pivot image in the centre of the triplets was fixed as the corresponding source image.

In the second and third rounds, we applied an active sampling strategy. A minimal spanning tree connecting 13 nodes (i.e., test images) was produced for each sequence, . The 12 edges then yielded the test images for the triplet questions and were chosen based on maximizing the expected information gain, following \cite{li2018hybrid}. This resulted in $12 \times 70 = \num{840}$ triplets of images in each round.

Each triplet question was presented to 30 crowd workers. Hence, overall $(2730 + 2 \times 840) \times 30 = \num{132300}$ responses were collected.

%Comparing all pairs of test images for all 70 sequences would require $70{13 \choose 2}=5460$ comparisons. 

\subsubsection{Quality Control}\label{sec:prestudyquality}
%-----------------------------------------------------------------------------------
The quality of the experiment was controlled as described in Section\,\ref{sec:qualitycontrol} and by a simplified outlier removal process.  There were 1035 assignments that were rejected and recollected because the test question was answered incorrectly, or more than three questions were skipped. 

The outlier detection in the pilot study was chosen differently from that in the main experiment. This is because the pilot study's purpose was merely to help generate test image sequences with prescribed impairment scales. This required accurate reconstructions of these scales for the test images used in the pilot study. To identify HIT assignments with unreliable responses, we, therefore, could rely on the ground truth, given by the ordering of the test stimuli on the physical distortion scale. 

Consider the $n$-th baseline triplet comparison  $(i,0,k)$ of a HIT assignment, where $i$ and $k$ denote the indices of the test images in a given sequence.
Similar to Subsection\,\ref{subtripdist}, we gave a score $u_n \in \{0, 0.5, 1\}$ to its response as specified in the following table.
\begin{center}
\small
\begin{tabular}{c  c  c}
Response & $i < k$	 & $k < i$\\
\hline
left & 1 & 0 \\
right  & 0 & 1 \\
not sure & 0.5 & 0.5 
\end{tabular}
\end{center}
Note, that by construction, $i \ne k$. The normalized sum of the scores in an assignment, $\frac{1}{N}\sum_{n=1}^{N} u_n$ can be called the true positive rate. 
The distance of an assignment w.r.t.\ the ground truth ordering is defined as
\begin{equation}\label{distPilot}
    d = 1 - \frac{1}{N}\sum_{n=1}^{N} u_n. 
\end{equation}
After computing these distances to ground truth ordering for all assignments, we sorted them according to distance and removed the last 10\% of them as outliers. 

%----------------------------------------------------------------------------------
\subsection{Comparison of Resolutions of FR-IQA Datasets}\label{sec:compdatasets}
%-----------------------------------------------------------------------------------
%    TID2013 where 4-5 test images where spaced at 33, 30, 27, 24, 21 dB PSNR. 
%    But our graph shows PSNR from 24 to 36 
%    We expect that the density of our B set is 2.5 times as large as for the A set, but the curves show a factor larger than 4
% The best image qualities in KADID are only a mere 28 dB? Can that be true?

\begin{figure}[t]
\centering{
    \input{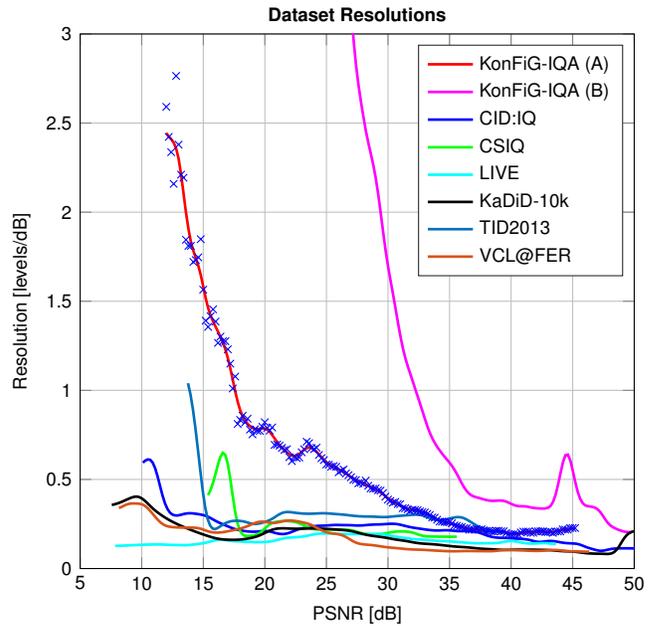}
}
\vspace{-10pt}
\caption{The resolution of FR-IQA datasets, defined by the number of distortion levels per dB on the PSNR scale, generally varies between 0.2 and 0.5 levels/dB for conventional datasets. Our new fine-grained datasets, KonFiG-IQA, Parts A and B, have much greater resolution over a large portion of the entire PSNR range. The data points (black crosses) are samples of the resolution function, computed for KonFiG-IQA, Part A. The curves in the figure were obtained by a gaussian averaging filter of width 2\,dB. 
}
\label{fig:resolutionKonFiG}
\end{figure}

Our dataset is the first one designed based on perceptual criteria that had been assessed in a pilot study. For each combination of a source image and a distortion type, the goal was to generate a sequence of distorted images with equal increments of perceived impairment. In Part A of our dataset, this increment is 0.25\,JND, and in Part B, it is 0.1\,JND. In other datasets, the distortion parameters either were selected manually (e.g., LIVE, VCL@FER, CID:IQ, MDID, and KADID-10k) or according to a plan w.r.t.\ increments of PSNR or bitrate (e.g., TID2008 and TID2013). 

For an image sequence in a fine-grained FR-IQA dataset, we expect a large number of distortion levels spread over the respective ranges of distortion, in our case 3\,JND. To quantify and compare different FR-IQA datasets in this regard, we introduce the notion of \textit{dataset resolution}. As a reference scale, we adopt the PSNR in dB. The resolution for an FR-IQA dataset then is a function of the PSNR, which indicates how close PSNRs of consecutive images from image sequences are in the neighbourhood of the given PSNR. We chose to define resolution locally this way because of the nonlinear relationship between PSNR and perceived quality which causes the resolution function to vary significantly, in particular for our perceptually guided FR-IQA datasets.

TID2013 adopted a spacing of 3\,dB PSNR between consecutive images in a sequence, so its dataset resolution is 1/3 level per dB PSNR, in this case uniformly over the entire range of distortion. And a resolution of 0.4\,levels/dB at 28\,dB PSNR means that the average length of intervals of PSNR values of consecutive images in a sequence including 28\,dB, is 2.5\,dB, the inverse of the resolution of 0.4 levels/dB.

\begin{table*}[t]
\caption{Overview of All Subjective Studies}
%\vspace{-3pt}
\label{info2}
%\centering
%\setlength\tabcolsep{4.5pt}
\resizebox{1\textwidth}{!}{ 
\begin{tabular}{c |c  c  c c c c c c c }
%  & 
%  Sources & 
%  \makecell{Distortion\\types} & 
%  \makecell{Distortion\\levels} & 
%  \makecell{Boosting\\types}  & 
%  \makecell{Responses/triplet\\Ratings/DCR} & 
%  \makecell{Reponses/ratings\\per sequence} & 
%  \makecell{Total number of\\reponses/ratings} & 
%  \makecell{Reponses/ratings\\after outlier removal}\\
& \multirow{2}{*}{Sources} & Distortion & Distortion & Boosting & Responses/triplet & Reponses/ratings & Total number of & Reponses/ratings \\
&  & types & levels & types & Ratings/DCR & per sequence & reponses/ratings & after outlier removal \\
\hline \\
\multirow{2}{*}{Pilot study} & \multirow{2}{*}{10} & \multirow{2}{*}{7} & \multirow{2}{*}{13} & \multirow{2}{*}{Plain} & \multirow{2}{*}{30} & 1890 & \num{132300} & \multirow{2}{*}{\num{119092}} \\
& & & & & &  $((39 + 12 + 12 )\times 30$) & 
$((39 + 12 + 12 )\times 30 \times 70$) &  \\
\\
\multirow{2}{*}{Experiment I} & \multirow{2}{*}{10} & \multirow{2}{*}{7} & \multirow{2}{*}{13} & Plain, A, Z, AZ & \multirow{2}{*}{20} & 1360  & \num{761600} & \multirow{2}{*}{\num{706914}}\\
& & & & F, AF, ZF, AZF & & ($68 \times 20$) & ($68 \times 70 \times 8 \times 20$) & \\ 
\\
Experiment II& \multirow{2}{*}{10} & \multirow{2}{*}{1} & \multirow{2}{*}{31} & \multirow{2}{*}{Plain}  & \multirow{2}{*}{9} & \num{29070} & \num{290700} & \multirow{2}{*}{\num{271510}} \\
Plain TC& &  & &  & & ($3230 \times 9$) & ($3230 \times 10 \times 9$) & \\
\\
Experiment II & \multirow{2}{*}{10} & \multirow{2}{*}{1} & \multirow{2}{*}{31} & \multirow{2}{*}{AZF} & \multirow{2}{*}{9} & 
9585 & \num{95850} & \multirow{2}{*}{\num{89034}}\\
Boosted TC&  & & & & & ($1065 \times 9$) & 
($1065 \times 10 \times 9$) & \\
\\
Experiment III &  \multirow{2}{*}{10} &  \multirow{2}{*}{7} &  \multirow{2}{*}{13} & Plain &  \multirow{2}{*}{50} & 
650 & \num{91000} &  \multirow{2}{*}{\num{76646}}\\
DCR & & & & AZ &  & ($13 \times 50$)  &  ($13 \times 70 \times 2 \times 50$) & \\
\end{tabular}
}
\end{table*}

\begin{table}[t!]
\caption{Number of HIT Assignments in All Experiments}
\begin{center}
\begin{threeparttable}
\label{info1}
\begin{tabular}{r r r r r r}
% \makecell{Assign- \\ ments\tnote{$\ast$}} & \makecell{Pilot\\study} & \makecell{ I \\ Baseline TC} & \makecell{ II \\General TC} & \makecell{ III \\DCR} & \makecell{Alto- \\ gether} \\
\multicolumn{1}{c}{Assign-} & \multicolumn{1}{c}{Pilot} & \multicolumn{1}{c}{I}  & \multicolumn{1}{c}{II}  & \multicolumn{1}{c}{III}  & Alto- \\
\multicolumn{1}{c}{ments\tnote{$\ast$}} & \multicolumn{1}{c}{study} & \multicolumn{1}{c}{Baseline TC} & \multicolumn{1}{c}{General TC} & DCR & \multicolumn{1}{c}{ gether} \\
\hline \\ [-1mm]
Total &8000 & \num{48469}$\quad$ & \num{24234}$\quad$ & 5387 & \num{86090} \\ [2mm]
Rejected\tnote{$\star$} & 1035 & \num{8385}$\quad$ & 3889$\quad$ & 597 & \num{13906} \\
Discarded\tnote{$\dagger$}& -- & 826$\quad$ & 366$\quad$ & 539 & \num{1708} \\
Outliers & 697 & 2052$\quad$ & 1443$\quad$ & 217 & \num{4419} \\  [2mm]
Kept\tnote{$\ddagger$} & 6268 & \num{37206}$\quad$ & \num{18536}$\quad$ & 4034 & \num{66057}
\end{tabular}
\smallskip
\begin{tablenotes}
\scriptsize
\item[$\ast$] Each assignment has 20 questions.
\item[$\star$] Assignments were rejected during the experiment because of failing the test questions or skipping more than three questions. Assignments were re-collected after rejection.
\item[$\dagger$] Assignments that were discarded before outlier removal because of line clicking (all 20 responses were the same).
\item[$\ddagger$] Assignments are remaining for analysis.
\end{tablenotes}
\end{threeparttable}
\end{center}
\end{table}

\begin{figure}[t]
\centering{
\begin{tikzpicture}

\begin{axis}[%
width=0.42\textwidth,
height=2.8in,
at={(0.772in,0.516in)},
scale only axis,
xmin=1,
xmax=13,
label style={font=\sffamily},
xticklabel style={font=\sffamily\scriptsize},
yticklabel style={font=\sffamily\scriptsize},
xlabel style={font=\sffamily\scriptsize},ylabel style={font=\sffamily\scriptsize},
title style={font=\bfseries\sffamily\scriptsize, yshift = -1ex},
xtick={1,2,3,4,5,6,7,8,9,10,11,12,13},
xticklabels={{0},{1},{2},{3},{4},{5},{6},{7},{8},{9},{10},{11},{12}},
ytick = {0,1,2,3,4,5,6,7,8, 9},
yticklabels = {{0}, {1}, {2}, {3}, {4}, {5}, {6}, {7}, {8}, {9}},
xlabel={Distortion level},
ymin=0,
ymax=9,
ylabel={Impairment [JND]},
axis background/.style={fill=white},
title = {Main Result of Experiment I},
xmajorgrids,
ymajorgrids,
legend image post style={scale=0.7},
legend style={at={(0.03,0.97)}, anchor=north west, legend cell align=left, align=left, draw=white!15!black, font = \sffamily\scriptsize}
]
\addplot [color=red, line width=1.0pt, mark=square, mark options={solid, red}]
  table[row sep=crcr]{%
1	0\\
2	2.59706831661087\\
3	4.26553170105846\\
4	5.2242876126013\\
5	5.96073436743364\\
6	6.55751727935511\\
7	7.01959678127063\\
8	7.46029696635724\\
9	7.79929496399081\\
10	8.06486070750066\\
11	8.36316658019236\\
12	8.55848026988876\\
13	8.72402923519651\\
};
\addlegendentry{AZF}

\addplot [color=blue, line width=1.0pt, mark=asterisk, mark options={solid, blue}]
  table[row sep=crcr]{%
1	0\\
2	0.721194688263639\\
3	1.72787108979394\\
4	2.62210257120091\\
5	3.35509013130071\\
6	3.94695097383813\\
7	4.43519049279175\\
8	4.93188998243433\\
9	5.37273501971997\\
10	5.76967456389788\\
11	6.11516451251138\\
12	6.47039140057971\\
13	6.719583428525\\
};
\addlegendentry{ZF}

\addplot [color=green, line width=1.0pt, mark=o, mark options={solid, green}]
  table[row sep=crcr]{%
1	0\\
2	0.935052347088913\\
3	1.87777537755407\\
4	2.61237359383661\\
5	3.30479391354477\\
6	3.82733483939497\\
7	4.32581796741648\\
8	4.7391748128178\\
9	5.17893912988685\\
10	5.52814534376671\\
11	5.87179615024154\\
12	6.18377934640356\\
13	6.41634384097018\\
};
\addlegendentry{AF}

\addplot [color=red, dashed, line width=1.0pt, mark=square, mark options={solid, red}]
  table[row sep=crcr]{%
1	0\\
2	0.736805927670893\\
3	1.49946373391861\\
4	2.0535542080172\\
5	2.47069532118887\\
6	2.7694737619148\\
7	3.05857579031971\\
8	3.35930675595318\\
9	3.50815306326633\\
10	3.67196890576876\\
11	3.83569550033255\\
12	3.98687882951444\\
13	4.09115228161582\\
};
\addlegendentry{AZ}

\addplot [color=black, line width=1.0pt, mark=x, mark options={solid, black}]
  table[row sep=crcr]{%
1	0\\
2	0.400759325619599\\
3	0.998162604787465\\
4	1.55091362226597\\
5	2.09429824317605\\
6	2.56661884428813\\
7	2.97415113329135\\
8	3.35794089488612\\
9	3.74057186641199\\
10	4.07079899222581\\
11	4.46237006815846\\
12	4.77820959417372\\
13	5.03657206236911\\
};
\addlegendentry{F}

\addplot [color=green, dashed, line width=1.0pt, mark=o, mark options={solid, green}]
  table[row sep=crcr]{%
1	0\\
2	0.483498206741097\\
3	1.0038487344009\\
4	1.53130510009873\\
5	1.85748080166338\\
6	2.25456948538335\\
7	2.56937825192623\\
8	2.87516524573699\\
9	3.13164579864366\\
10	3.35638993442658\\
11	3.56991032859619\\
12	3.7720503887635\\
13	3.89957517389654\\
};
\addlegendentry{A}

\addplot [color=blue, dashed, line width=1.0pt, mark=asterisk, mark options={solid, blue}]
  table[row sep=crcr]{%
1	0\\
2	0.379801162977994\\
3	0.738030129284806\\
4	1.21102919599988\\
5	1.61679333070535\\
6	1.96736238610242\\
7	2.28388594360742\\
8	2.59722416499207\\
9	2.85366487098676\\
10	3.134353426522\\
11	3.40207898732996\\
12	3.6424763799186\\
13	3.78502082225534\\
};
\addlegendentry{Z}

\addplot [color=black, dashed, line width=1.0pt, mark=x, mark options={solid, black}]
  table[row sep=crcr]{%
1	0\\
2	0.249239961108294\\
3	0.529838968565233\\
4	0.8087021855847\\
5	1.09470970667619\\
6	1.35950619329395\\
7	1.64425578495864\\
8	1.96021522138968\\
9	2.18085171875008\\
10	2.41887626147279\\
11	2.70709959913946\\
12	2.91718698151299\\
13	3.11501041817669\\
};
\addlegendentry{Plain}

\end{axis}
\end{tikzpicture}%
}
\caption{Average reconstructed impairment scales over 70 sequences for 8 types of TC with baseline triplets. Scales from AZF-boosted TC have the largest range, almost 3 times as large as the range of Plain TC.}
\label{jndscale}
\end{figure}
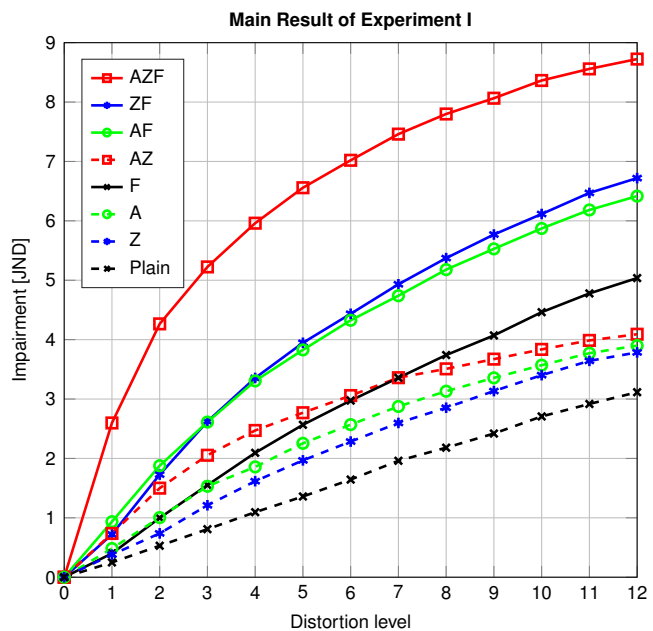

\begin{figure*}[t]
\centering{\input{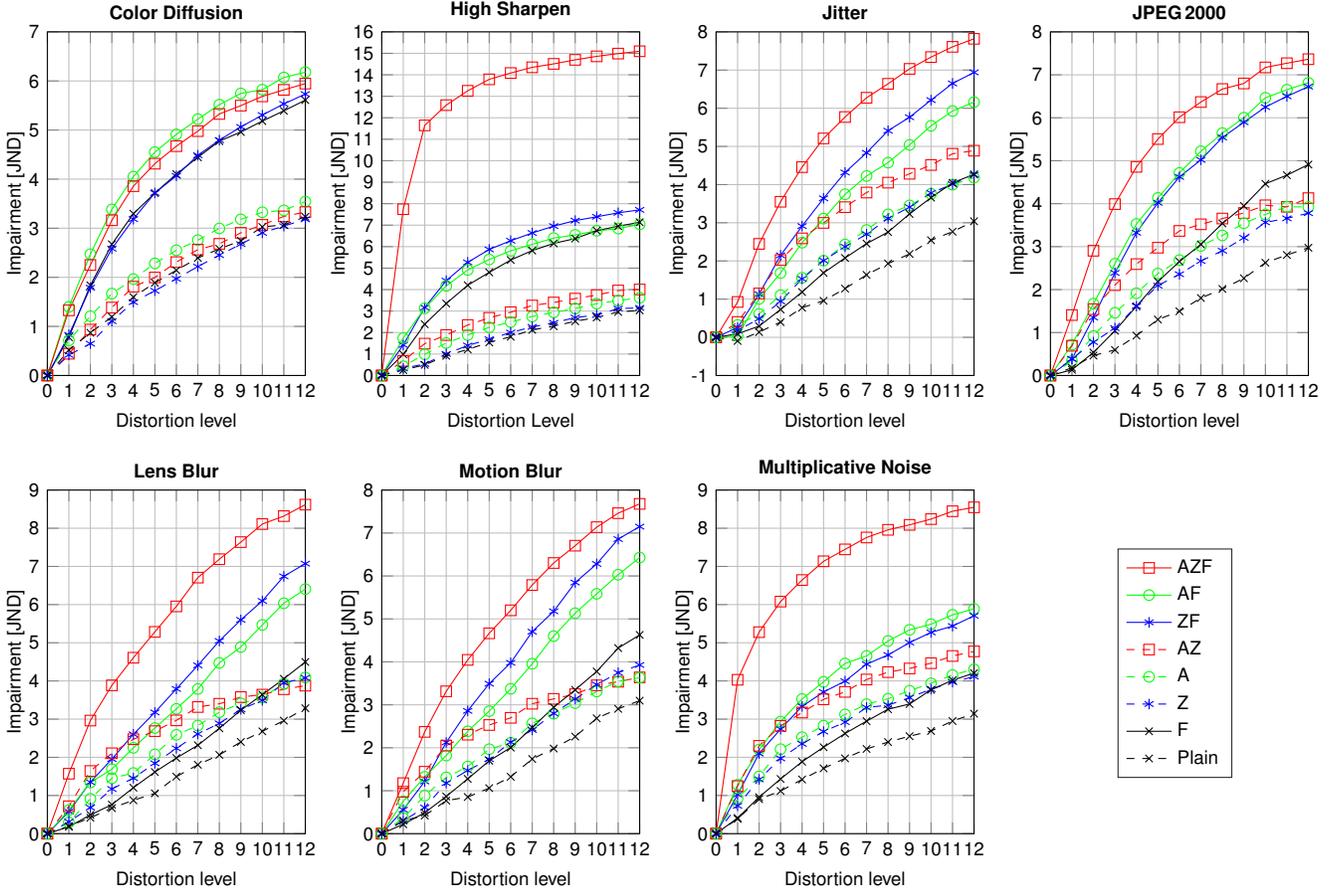}}
\caption{Average reconstructed JND scales as in Figure\,\ref{jndscale} for each type of distortion. Each point on any of these graphs corresponds to the mean impairment scale in JND units of the 10 source images, distorted with the respective type and at the given level.} 
\label{jndscale_per_distortion}
\end{figure*}

We computed the dataset resolution functions for our datasets, KonFiG-IQA Parts A and B, and six other datasets, namely LIVE, CSIQ IQA, VCL@FER, TID2013, CID:IQ, and KADID-10k. For each of these, the procedure was as follows. First, we scanned consecutive images in each image sequence of the FR-IQA dataset and collected the corresponding PSNR intervals. Secondly, we sampled the PSNR scale uniformly with a step size of 0.2\,dB, and for each sample value, we averaged the lengths of all those intervals that contain the PSNR sample. The inverse of this average length is the value of the resolution function at the given PSNR sample value. Due to the discrete nature, the resolution function is only piecewise constant and appears noisy as there are some intervals of very small size. For visualisation, therefore, we show a smooth approximation obtained by a gaussian averaging filter of the width of 2\,dB PSNR. 

Figure\,\ref{fig:resolutionKonFiG} shows the resulting resolution functions for the selected datasets and also the samples collected for KonFiG-IQA (Part A).  The figure confirms that our new fine-grained datasets have much greater resolution over a large portion of the total PSNR range. 

We also note that for TID2013, the design goal to have 3\,dB PSNR between consecutive images in each sequence (0.33 levels/dB) was not strictly followed. The dataset resolution mainly varies between 0.2 and 0.4 levels/dB and exceeds 1 level/dB at the low end of the PSNR scale.

%-----------------------------------------------------------------------------------
%-----------------------------------------------------------------------------------
%-----------------------------------------------------------------------------------
\section{Experiment I: Boosting for Baseline Triplet Comparison}
%-----------------------------------------------------------------------------------
The purpose of our first experiment is to apply the proposed boosting methods to our FR-IQA dataset in a crowdsourcing campaign for the analysis of the performance w.r.t.\ that achieved without boosting as traditionally done. Here we use baseline triplet comparisons, i.e., we present two different distorted test images, $I_i$ and $I_k$, together with the undistorted reference image $I_0$ in the middle of the triplet, which is denoted by $(i,0,k)$. This interface corresponds to that used in TID2008, TID2013, and MDID. 

We recall from the previous section that we have 10 source images, each distorted by 7 types of distortion, giving 70 image sequences, each consisting of a reference (source) image $I_0$ and 12 increasingly distorted versions of the reference image, $I_1,\ldots,I_{12}$. 
%By design, we have that these distortions range over a span of 3\,JND on the perceptual scale, so the perceptual difference between two consecutive images is 0.25\,JND.
By design, the distortions span 3\,JND units on the perceptual scale, so the perceptual difference between two consecutive images is 0.25\,JND.
For each of the 70 sequences, we applied 8 types of baseline triplet comparisons, namely the plain TC without boosting and A-, Z-, AZ-, F-, AF-, ZF-, AZF-boosted baseline TC.

In this setup, for each sequence, there are ${13 \choose 2} = 78$ possible triplet comparisons. Since there is less information gain in responses for triplets for which the correct answer is obvious, we only considered the 68 triplets $(i, 0, k)$ with $|k-i| \le 8$ and $k \ne i$. Thus, triplets $(i, 0, k)$ with a perceptual distance between $I_k$ and $I_i$ greater than 2\,JND were omitted. In this way we generated two groups of $68\times70\times4 = \num{19040}$ triplets each. The first group contained those triplets that were used for comparisons without boosting by flicker. For the other group of triplets, to be used with F-boosting, a different interface had to be applied (see Section\,\ref{sec_interface}), so they were collected in separate HITs in the crowdsourcing. The triplets were randomly oriented (either as $(i, 0, k)$ or $(k, 0, i)$) and shuffled in each group. Finally, they were split up and distributed into crowdsourcing HITs. Each HIT consisted of 19 triplet comparisons and 1 additional triplet comparison from our pool of test questions. 

% ---------------------GAIN
%\begin{figure*}[t]
%\centering{\includegraphics[width=1\textwidth]{imtexfile/conceptgain.eps}}
%\caption{Left: Example of sensitivity gain of baseline triplet (boosted AZ). Right: Gain of main study.
%\textcolor{red}{Left: Make red line for gain more fat. Use "AZ-Boosted Triplets". Legend: Put the gain FIRST. Title = Concept of Gain Factor for Boosting. Right: Gain goes from 0 to 4.5. Use a grid. Make figures left and right the same size. Title = Gain Factor for all Types of Boosting. Legend: Just use abbreviations A, F, Z, AZ, etc.}
%}
%\label{gain_example}
%\end{figure*}

\begin{figure*}[t]
\centering{\input{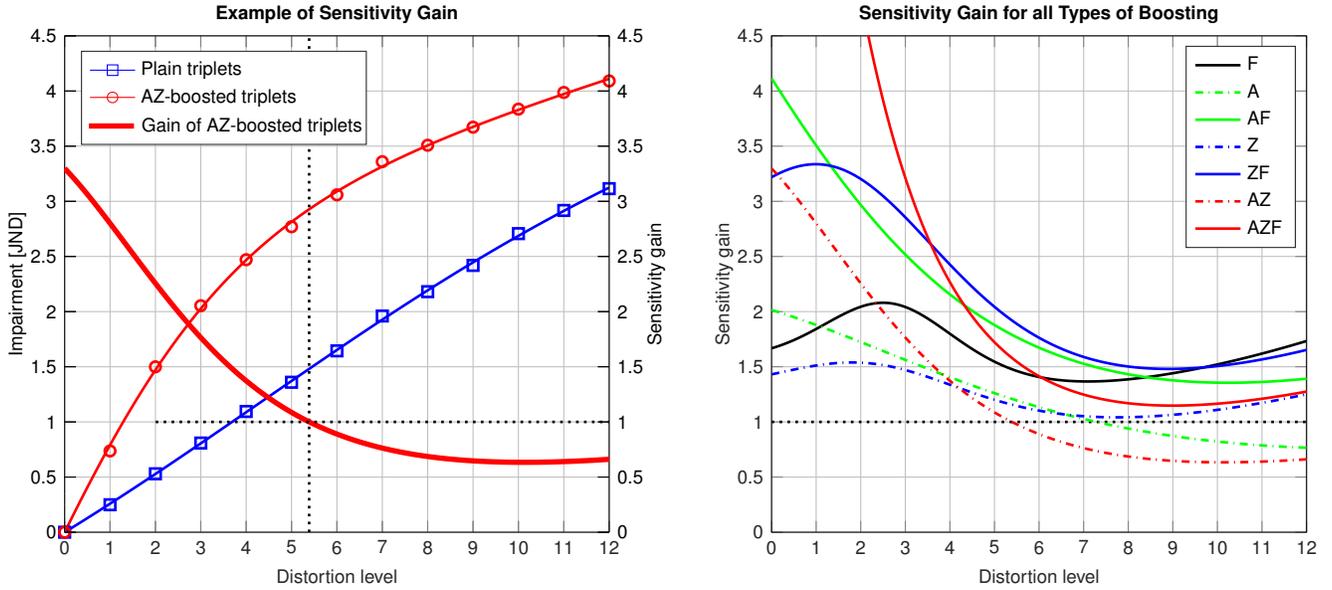}}
\caption{Left: The sensitivity gain of AZ-boosted baseline triplet comparison is illustrated. The gain is the ratio of the derivatives of the fitted impairment scale functions. Thus, the gain is the factor by which an increase of perceived distortion is multiplied when AZ-boosting is applied. Here, the sensitivity gain is greater than 1 for distortion levels up to 5 but less than 1 at larger levels. This indicates that AZ-boosting is more sensitive than plain baseline TC for small distortions up to around 1.25\,JND and less sensitive for larger levels. Right: Sensitivity gain for all seven types of boosted baseline triplet comparisons. It shows that boosting in baseline triplet comparisons is most effective for smaller distortions. 
%\textcolor{red}{Left: Make red line for gain more fat. Use "AZ-Boosted Triplets". Legend: Put the gain FIRST. Title = Concept of Gain Factor for Boosting. Right: Gain goes from 0 to 4.5. Use a grid. Make figures left and right the same size. Title = Gain Factor for all Types of Boosting. Legend: Just use abbreviations A, F, Z, AZ, etc.}
}
\label{gain_example}
\end{figure*}

We spawned 20 assignments per HIT, i.e., we collected 20 responses for each triplet comparison. We controlled the quality of the experiment and removed outliers as described in Sections\,\ref{sec:qualitycontrol} and\,\ref{sec:outlierremoval}. In the end, \num{37206} assignments of HITs with \num{706914} TC responses remained for the reconstruction of impairment scales for 70 image sequences and analysis. For details see Table\,\ref{info2} and\,\ref{info1}. Using the method proposed in Section\,\ref{sec:tripletrecons}, the perceived impairments of the images were reconstructed from the set of TC responses. We evaluated and compared the performances of the eight types of TC by several criteria as follows.

%-----------------------------------------------------------------------------------
\subsection{Results: Reconstructed Impairment Scales}
%-----------------------------------------------------------------------------------
Figure\,\ref{jndscale} summarizes the reconstructed impairment scales for stimuli at the 12 distortion levels. Each curve represents the average taken over all 70 image sequences of seven distortion types. The main findings from this global view over the range of 3\,JND are as follows:

\begin{itemize}
\item
    The result from plain triplet comparisons is given by the black dashed curve. It shows a linearly increasing impairment, reaching 3.1\,JND for distortion level 12. %3.1150 is the exact number
    Thus, Experiment I confirms our expectations from the pilot study quite well, a perceptually linearly increasing impairment over 3\,JND.
\item
    The three colored  dashed curves are for the results without boosting by flicker. They are well above the baseline given by plain TC and extend the range of perceived distortion from 3 to about 4\,JND, with the red curve for combined AZ-boosting yielding the largest increase. Thus, for boosting with artefact amplification, zooming, or their combination, we have gained 1\,JND over the range of the first 3\,JND.
\item
    The four solid curves are for the results with boosting by flicker. These provide an additional large increase in performance. Just the boosting by flicker alone extends the range of perceived distortion to 5\,JND. The combinations with either zooming or artefact amplification provide about 6.5\,JND in place of only 3\,JND, given by traditional plain comparisons. However, the top performing boosting is the combination of all three methods, AZF-boosting, giving close to 9\,JND and providing an overall increase by a factor of almost 3 over plain comparison.
\end{itemize}

These findings hold for the averages taken over all distortion types and source images. The impairment curves for the different distortion types, averaged over the 10 source images for the sequences, show more detailed views of the results and can be found in Figure\,\ref{jndscale_per_distortion}.

%-----------------------------------------------------------------------------------
\subsection{Results: Sensitivity Gain}
%-----------------------------------------------------------------------------------
The strongest effect of boosting strategies can be observed for smaller distortion levels, up to 1\,JND. 
To quantify the performance of boosting also locally at different distortion levels, we introduce the concept of sensitivity and sensitivity gain. In general, a sensitivity analysis determines how changes of an independent variable affect a particular dependent variable. If a differentiable function gives their functional dependence, we can use its derivative as a measure of sensitivity. 

In our case, we think of the distortion level as the independent variable and the corresponding reconstructed impairment scales from the eight types of TC as dependent variables. Although the distortion levels are discrete, they correspond to equally spaced, physical, and real-valued distortion parameters. It may be assumed that the resulting impairments of image quality can be modelled by a continuous and differentiable function. Given our data, we applied the 5-parameter logistic fitting \cite{sheikh2006statistical} for the curves in Figure\,\ref{jndscale} using 
\begin{equation}\label{eq:5para}
    \beta_1 \left(\frac{1}{2} - \frac{1}{1 + \exp{\left(\beta_2 (x-\beta_3)\right)}}\right)  + \beta_4 x + \beta_5,
\end{equation}
where $x$ denotes the distortion level, and $\beta_1$ to $\beta_5$ are the parameters for the fit. In the numerical optimization procedure for the curve fitting, local optima will be obtained, depending on the choice of initial parameters. Therefore, we ran the optimization multiple times with different initial conditions and visually checked the fitting quality.

The derivatives of the fitted functions then model the observed \textit{sensitivities} of the plain and boosting techniques to assess the impairment. The sensitivity is also a function of the physical distortion magnitude, respectively the distortion level, allowing a local analysis.

We define the \textit{sensitivity gain}, provided by a particular boosting method as the quotient of the sensitivity for a boosting method and the sensitivity of the method using plain triplet comparisons. A gain larger than 1 indicates an increase of sensitivity by that factor due to boosting. Figure\,\ref{gain_example} (left) illustrates the procedure for the case of AZ-boosting. Firstly, it clearly shows that the curve fitting yielded visually convincing smooth functional approximations of the empirical data. Secondly, we see that AZ-boosting yielded a sensitivity gain greater than 1 for distortion levels up to 5 (corresponding to about 1.25\,JND), but the method using plain baseline TC was more sensitive for larger levels than the one with AZ-boosted TC.

\begin{table}[t!]
\caption{True Positive Rate and Average Response Times for Triplet Comparisons}
\label{tb:tprtime}
\centering
\begin{tabular}{r|cc|cc}
Method  & TPR & Rank & Response Time (s) & Rank\\
  \hline
Plain TC &  0.7703 & 8 &  2.314 $\pm$ 0.412 & 8\\
A-boosted TC &  0.8001 & 6 &  2.229 $\pm$ 0.410 & 7\\
Z-boosted TC &  0.8002 & 5 &  2.166 $\pm$ 0.408 & 5\\
F-boosted TC &  0.8470 & 4 &  2.225 $\pm$ 0.482 & 6\\
AZ-boosted TC &  0.7791 & 7 &  2.146 $\pm$ 0.409 & 4\\
AF-boosted TC &  0.8707 & 2 &  2.060 $\pm$ 0.438 & 3\\
ZF-boosted TC &  0.8803 & 1 &  2.021 $\pm$ 0.440 & 2\\
AZF-boosted TC &  0.8627 & 3 & 1.998 $\pm$ 0.434 & 1
\end{tabular}
\end{table}

%\begin{table}[t!]
%\caption{Average Response Times for Triplet Comparisons}
%\label{tb:secs}
%\centering
%\begin{tabular}{rc}
%Method  & Avg.\ Response Time (s) \\
%  \hline
%Plain TC &  2.314 $\pm$ 0.412\\
%A-boosted TC &  2.229 $\pm$ 0.410\\
%F-boosted TC &  2.225 $\pm$ 0.482\\
%Z-boosted TC &  2.166 $\pm$ 0.408\\
%AZ-boosted TC &  2.146 $\pm$ 0.409\\
%AF-boosted TC &  2.060 $\pm$ 0.438\\
%ZF-boosted TC &  2.021 $\pm$ 0.440\\
%AZF-boosted TC & 1.998 $\pm$ 0.434 
%\end{tabular}
%\end{table}

The right part of Figure\,\ref{gain_example} shows the sensitivity gain as a function of the distortion level for all seven types of boosting. The curves demonstrate that the boosting methods for baseline triplet comparison are most effective for smaller distortions up to about 1\,JND, which corresponds to distortion level 4. Especially for the boosting with flicker, sensitivity gains larger than 2 were achieved. 

For larger distortion levels near 2\,JND, a kind of saturation effect can be noticed; the gain dropped below 2 but still is larger than 1, except for A- and AZ-boosting. This indicates that perceptually, boosting small distortions to a reference image makes their difference more apparent than boosting large distortions. This effect is particularly strong for artefact amplification among the basic A-, Z- and F-boosting techniques. The sensitivity gain almost linearly drops from 2 at distortion level 0 to 0.8 at distortion level 12. This may, in part, be explained by the nonlinearity of the boosting due to pixel value clamping, see Table\,\ref{clampratio}. Naturally, for small distortions in the test images, there is less clamping to be expected than in test images with large distortions, and such a deficiency does not hold for the other two basic boosting types. 

%-----------------------------------------------------------------------------------
\subsection{Results: True Positive Rate} % within 1\,JND
%-----------------------------------------------------------------------------------
\begin{figure}[t]
\centering{

\begin{tikzpicture}

\begin{axis}[%
width=0.41\textwidth,
height=2.8in,
at={(0.772in,0.516in)},
scale only axis,
xmin=1,
xmax=8,
xticklabel={$\mathsf{\pgfmathprintnumber{\tick}}$},
    yticklabel={$\mathsf{\pgfmathprintnumber{\tick}}$},
label style={font=\sffamily},
every axis label/.append style={font=\sffamily\scriptsize},
xticklabel style={font=\sansmath\sffamily\scriptsize},
yticklabel style={font=\sffamily\scriptsize},
xlabel style={font=\sffamily\scriptsize},ylabel style={font=\sffamily\scriptsize},
title style={font=\bfseries\sffamily\scriptsize, yshift = -1ex},
title = {JND Threshold Detection},
xlabel={Distortion level},
ymin=0.5,
ymax=1,
xtick = {1,2,3,4,5,6,7,8},
xticklabels = {{1},{2},{3},{4},{5},{6},{7},{8}},
ytick = {0.5, 0.55, 0.6, 0.65, 0.7, 0.75 , 0.8, 0.85, 0.9, 0.95, 1},
yticklabels = {{0.5}, {0.55}, {0.6}, {0.65}, {0.7}, {0.75} , {0.8}, {0.85}, {0.9}, {0.95}, {1}},
ylabel={TPR},
axis background/.style={fill=white},
xmajorgrids,
ymajorgrids,
legend style={at={(0.97,0.03)}, anchor=south east, legend cell align=left, align=left, draw=white!15!black, font=\sffamily\scriptsize}
]
\addplot [color=red, line width=1.0pt, mark=square, mark options={solid, red}]
  table[row sep=crcr]{%
1	0.88766303012433\\
2	0.959318486166396\\
3	0.985225103199174\\
4	0.988810506658804\\
5	0.991860902255639\\
6	0.995543024227235\\
7	0.995538478549314\\
8	0.991518624993857\\
};
\addlegendentry{AZF}

\addplot [color=blue, line width=1.0pt, mark=asterisk, mark options={solid, blue}]
  table[row sep=crcr]{%
1	0.728879362811251\\
2	0.859824192835029\\
3	0.930661948990122\\
4	0.960272157108457\\
5	0.977008469042138\\
6	0.983162686127082\\
7	0.986137770897833\\
8	0.989255061673792\\
};
\addlegendentry{ZF}

\addplot [color=green, line width=1.0pt, mark=o, mark options={solid, green}]
  table[row sep=crcr]{%
1	0.777530784918556\\
2	0.874761180788105\\
3	0.927196545284781\\
4	0.963024881005034\\
5	0.96953646370829\\
6	0.969927760577915\\
7	0.983859157698167\\
8	0.986165198535555\\
};
\addlegendentry{AF}

\addplot [color=red, dashed, line width=1.0pt, mark=square, mark options={solid, red}]
  table[row sep=crcr]{%
1	0.67703907440169\\
2	0.870322041097141\\
3	0.921443191311612\\
4	0.90328145731977\\
5	0.930894822841417\\
6	0.962176642586859\\
7	0.968507574082264\\
8	0.954454119367045\\
};
\addlegendentry{AZ}

\addplot [color=green, dashed, line width=1.0pt, mark=o, mark options={solid, green}]
  table[row sep=crcr]{%
1	0.625507518796993\\
2	0.820482701852671\\
3	0.882697153422773\\
4	0.853347461791734\\
5	0.884796466832347\\
6	0.940452448523269\\
7	0.949709383753501\\
8	0.932257635510344\\
};
\addlegendentry{A}

\addplot [color=blue, dashed, line width=1.0pt, mark=asterisk, mark options={solid, blue}]
  table[row sep=crcr]{%
1	0.542084048531693\\
2	0.74297726546759\\
3	0.841549492604059\\
4	0.832748937294216\\
5	0.849115672584823\\
6	0.918173664202523\\
7	0.94984630694383\\
8	0.913912600127771\\
};
\addlegendentry{Z}

\addplot [color=black, line width=1.0pt, mark=x, mark options={solid, black}]
  table[row sep=crcr]{%
1	0.641644828984225\\
2	0.720396272544105\\
3	0.796316460794844\\
4	0.882582471585125\\
5	0.902632346798369\\
6	0.919503397850368\\
7	0.935324033122021\\
8	0.96203788094185\\
};
\addlegendentry{F}

\addplot [color=black, dashed, line width=1.0pt, mark=x, mark options={solid, black}]
  table[row sep=crcr]{%
1	0.520834193326453\\
2	0.699856350434911\\
3	0.751109702652991\\
4	0.758091337412158\\
5	0.770410063148066\\
6	0.852334881320949\\
7	0.880337205057462\\
8	0.876332436483365\\
};
\addlegendentry{Plain}

\end{axis}
\end{tikzpicture}%
}
\caption{For assessment of the JND threshold, distorted images are compared with the source image. The figure shows the corresponding TPRs from our corresponding triplet comparisons, averaged over all 70 image sequences. Corresponding detection rates linearly increase from 0\% for a TPR of 0.5 to 100\% for a TPR of 1. Boosting increases the TPR and reduces the JND threshold, which can be read off the graphs at the TPR of 0.75.
}
\label{boostedpc:dtr}
\end{figure}
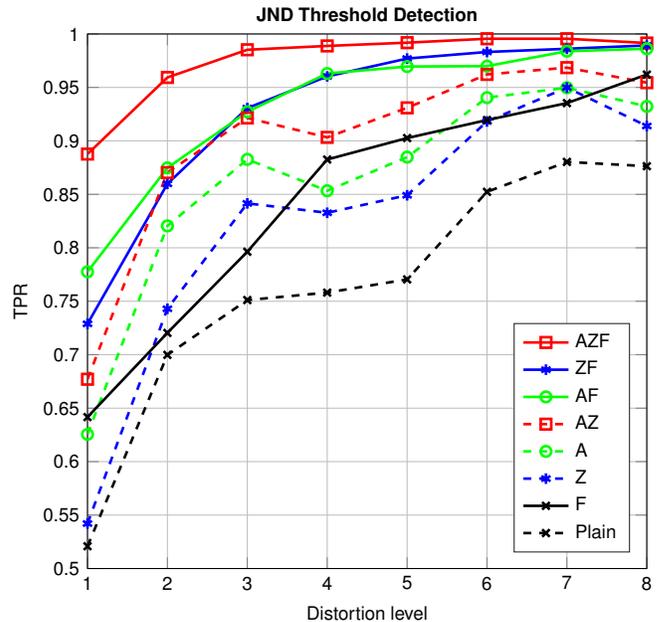

Recall from Section\,\ref{sec:prestudyquality} that the true positive rate (TPR) for a set of baseline triplet responses is the ratio of correct answers w.r.t.\ the ground truth given by the ordering of the stimuli of an image sequence. In this regard, a response of type ``not sure'' scores 1/2 point. Therefore, the TPR can also be considered a criterion for validating the performance improvement by the boosting methods. A TPR of 0.5 can be achieved by guessing alone. The TPR for boosted TC is expected to be larger than for plain TC, as boosting should enable observers to make more correct decisions regarding image quality differences.

Table\,\ref{tb:tprtime} shows the first overall average results for the TPRs in Experiment I (with increasing order) and confirms our expectation. The TPRs show a similar ordering as observed in the average reconstructed impairment scales at larger distortion levels, as shown in Figure\,\ref{jndscale}:
\begin{center}
    Plain, Z, A, AZ, F, AF, ZF, AZF.
\end{center}

For a more detailed perspective, we consider comparisons as typically done to assess the JND threshold in image sequences with increasing distortion. This amounts to comparisons of distorted images with the corresponding source reference images. In terms of triplet comparisons, we therefore look at triplets $(0,0,k)$ (or $(i,0,0)$). We expect that the TPR increases with the distortion level $k$ (resp.\ $i$) and that boosted TCs give rise to larger TPRs.

Figure\,\ref{boostedpc:dtr} shows the TPRs for this analysis, averaged over the 70 sequences in our dataset. Corresponding detection rates linearly increase from 0\% for a TPR of 0.5 to 100\% for a TPR of 1, and the JND threshold on one of the curves is reached at a TPR of 75\%. For the case of plain TC, we see that the JND threshold was reached at distortion level 3. The dataset was designed to have the threshold at distortion level 4, and there the TPR is 0.76, still close to 0.75 as expected.

All seven types of boosting increase the TPR compared to plain TC. To give an example, consider test images having just one distortion level difference (0.25\,JND). Plain comparison yielded a TPR of only 0.52, which is not much better than guessing.\footnote{In Section\,\ref{sec:leveldesign} we showed that the expected TPR of the corresponding 2AFC pair comparison for a perceptual difference of 0.25\,JND is 0.567, which is a bit larger than the empirical TPR of 0.52 observed here. This may be due to the basic assumption in Thurstonian models that the distributions of the latent perceptual image quality scales are perfectly Gaussian. It is unknown to what extent this assumption holds true.}
On the other hand, with combined AZF-boosting, the TPR is 0.88, a very strong improvement.  

On the far end of the scale at levels 6 to 8, corresponding to distortions 1.5 to 2\,JND, the gains in TPR achieved by boosting are smaller. This is due to the saturation effect described earlier, i.e.,  consistent with our findings for the sensitivity gain by boosting. These gains are much more pronounced for small distortions levels. 

%-----------------------------------------------------------------------------------
\subsection{Results: Response Time}
%-----------------------------------------------------------------------------------
Boosting not only helps observers to find the correct answers to comparisons but also reduces their response time. Table\,\ref{tb:tprtime} shows the response times in seconds for all methods with and without boosting, averaged over all baseline triplet comparisons and with corresponding standard deviations. A response for a plain TC required 2.3 seconds on average. All types of boosted TC were faster, with AZF-boosting requiring slightly less than 2 seconds on average. The two-sample $t$-test revealed that these time savings are statistically significant, with very small p-values less than $10^{-11}$.
%\textcolor{blue}{(two-sample $t$-test, of which the null hypothesis is that the data in two groups comes from independent random samples from normal distributions with equal means and equal but unknown variances)}
%---------------Spent time
%\begin{figure*}[t!]
%\centering{
%\includegraphics[width=0.98\textwidth]{images/vistimeall2.eps}}
%\caption{Avg secs-overall.}
%\label{secs-over}
%\end{figure*}
%---------------------------------

%--------------Example - why boosted PC became flattened in the end -2
% -------remove temporally ----- 
%\begin{figure*}[t!]
%\centering{\includegraphics[width=0.95\textwidth]{images/SRC45_motionblur_22.469626-18.891355.png} \\
%\vspace{2pt}
%$~$\centering{\includegraphics[width=0.95\textwidth]{images/Boost-SRC45_motionblur_22.469626-18.891355.png}}
%}
%\caption{[saturation effect - ]Motion blur, $\text{im}_12$, reference image, and $\text{im}_10$. Upper: Plain PC. Lower: Boosted PC. For the pairs on the upper row, it can be distinguished that the left image is more blurry than the right one, while for the pairs in the lower row, the blurry artefact is too much that the differences between the images can hardly be differentiated.  }
%\label{secs-over}
%\end{figure*}

%-----------------------------------------------------------------------------------
%-----------------------------------------------------------------------------------
\section{Experiment II: Boosting for General Triplet Comparison}
%-------------------------------------------------------------------------------
%-------------------------------------------------------------------------------
%-------------------------------------TRIPLET-----------------------------------
%[“Based on the main experiments we identify the single most promising technique. For example, this could be boosting in combination with flickering. We select a single distortion type, for example jitter and a very small range of distortion parameters for 21 images, for example, covering only a range of 1 JND. This is a difficult task, because previously we had 13 images spread over 3 JNDs (5 over 1 JND), so the difference between consecutive images is five times as small (0.05 versus 0.25 JNDs). Then we select a long sequence of random triplets. We present these triplet questions to the crowd only in two ways, with plain comparisons and in the boosted version using the same AMT GUI as before. Then we reconstruct the scale values for both cases. In the analysis, we can easily compare TPRs, the scale value reconstruction, SROCC curves etc for a varying number of ratings used. The boosted version should be highly superior to the plain comparison technique. “]

In the previous section, we noticed that there is a saturation effect for larger distortion levels, and the reason could be that boosting small distortions of two test images makes their difference stand out better than boosting large distortions with the same difference in distortion levels. 

To ameliorate this drop in effectiveness of boosting for larger distortion levels, we consider general triplet comparisons $(i,j,k)$, where the pivot image $I_j$ is not fixed to be the undistorted reference image $I_0$ of a sequence. Instead, we may allow arbitrary triplets with different pairwise stimuli and select the stimulus with the median distortion magnitude as the pivot. 

The artefact amplification then linearly increases the differences with respect to the pivot as before and not the distortions with respect to the undistorted reference images. Typically, these image differences to be enlarged will be smaller than in Experiment I with baseline triplets. Likewise, in the flicker technique, the flicker is between distortion levels that typically are closer together than with baseline triplets. In Experiment I, we have seen larger sensitivity gains for such smaller quality differences. Thus, we expect to arrive at a sensitivity gain that is enlarged over the whole range of distortion levels, thereby reducing or even eliminating the saturation effect observed for baseline triplet comparisons.     

%This arguments not yet so strong, because 0,10,12 does not compare between 10 and 12... 
%Rewrote the text above to make it clearer and dropped this paragraph below.

%We expect an increase in sensitivity gain. For example, consider a baseline triplet $(10,0,12)$ with two images $I_{10}$ and $I_{12}$ which observers found difficult to compare. An observer will have to compare image differences of 10 and 12 levels, which corresponds to a ratio of 1.2. The same triplet as a general triplet reads $(0,10,12)$, and the observer compares image differences of 2 and 10 levels, corresponding to a larger ratio of 5. Following the basic rules of perception set up by Weber and Fechner, it is not the difference between compared stimuli but rather the ratio of intensities of stimuli that is the determinant for detection of a change between the two. Therefore, we expect that the likelihood for a correct response for the triplet comparison $(0,10,12)$ is larger than for the baseline triplet comparison $(10,0,12)$.

The second motivation to conduct Experiment II is to assess and compare the performance of boosted versus plain TC in terms of the convergence as the number of TCs increases. On the one hand, the reconstructed impairment scales converge, and for the analysis, we consider their confidence intervals from samples of TCs as estimates for the precision of the computed impairment scales. On the other hand, for a given set of distorted images derived from a fixed reference image, the ordering of the reconstructed impairment values should converge, and the Spearman rank-order correlation (SROCC) is the appropriate measure for this analysis. 

To study such convergence aspects, a very large pool of TC responses and a challenging set of image sequences are desirable. Therefore, we increased the number of distortion levels in each image sequence from 12 to 30, so that the spacing between consecutive test images is only 0.1\,JND. The number of TCs per image sequence was increased from 1360 to 9585 and \num{29070} for boosted and plain TC, respectively. To limit the cost for such an enlarged study, we chose to restrict Experiment II to only one type of distortion, motion blur, and to the most promising type of boosting, AZF-boosting. All 10 source images were used, resulting in 10 sequences of increasing motion blur distortion.

\begin{figure}[t]
\centering{\input{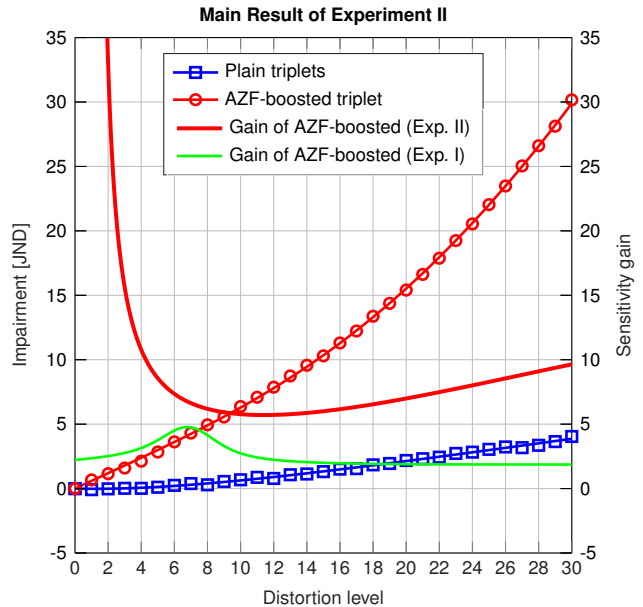} }
\vspace{-10pt}
\caption{Experiment II (general triplet comparisons, motion blur distortion): The reconstruction of impairment scales from plain and AZF-boosted TC are shown by the curves with square and circular markers, respectively. On average, the red curve increases with a slope 7 times as large as that for the blue curve. Thus, the overall gain in sensitivity by AZF-boosting is given by a factor of about 7. The sensitivity gain function for AZF-boosting is also shown (solid red curve). It is globally larger than 5, while the corresponding sensitivity gain function from Experiment I (baseline triplet comparisons, motion blur distortion), shown by the solid green curve, is everywhere below 5.}
\label{trip_score}
\end{figure}

For 31 images $I_0, I_1, ..., I_{30}$ in each of the 10 image sequences, we considered triplets $(i,j,k)$ with $i<j<k$. We further limited the span of these triplets. The span of a triplet $(i,j,k)$ is $S = \max(i,j,k)-\min(i,j,k)$ which in this case is $S=k-i$. For boosted TC, we set the maximal span to 10, and for plain TC to 20, corresponding to 1 and 2\,JND, respectively. We anticipated that with plain TC, small differences were harder to detect, and so larger spans for plain TC than for boosted TC would be advisable for a fair comparison. The number of triplets for each image sequence thus was $\sum_{n=2}^{S}(31-n)(n-1)$, which is 3230 for plain and 1065 for boosted TC. 

For each of the triplet comparisons, we collected 9 responses from the crowd workers. The presentation of the triplets was randomized in sequence and in orientation, showing either $(i,j,k)$ or $(k,j,i)$. The quality control and outlier removal were carried out as in Experiment I. For details regarding the numbers of HITs that were rejected or classified as outliers, see Table\,\ref{info1}.

%\textcolor{green}{+ In total, 15,300 HITs for Plain PC, 5,045 HITs for A+Z+F PC }
%- Rejection (3 rounds fill-up): 2240 HITs rejected in Plain PC (15216 remained), 1,649 HITs rejected (4669 remained)in A+Z+F PC

%2240 HITs were rejected in the Plain TC and 1649 HITs were rejected in the Boosted-AZF TC.

%Because of the test question,  3889 HITs were rejected and not paid. Using the same strategy as in the previous study (see Section \ref{sec:outliermain}), we first removed the HITs where the answers were always the same, resulting in 366 HITs removed. After that, we adopted the iterative outlier removal strategy with 95\% ratings kept, where the metric is using W-TPR-C as in the previous study.  
% 253-plain all same   
% 113 boost all same
   
%----NOTE: [We didn't show the results of span 20 for plain TC, only showed span 10 for both, and actually increasing the span didn't increase the performance of plain TC that much] ----%
%We set $k=20$ for the Plain TC, and $k=10$ for the Boosted-AZF TC. 
%This resulted in 3,230  triplets in Plain TC and 1,065 triplets in Amplified + Zoomed + Flicker TC.
%https://paper.dropbox.com/doc/Boosted-PC-Main--BDsvmDESvjIoPA4Sp5_7741fAg-QAZYlnpVpilTAVxpMTvhF
%- Sample from the full triplets — limit the span of the triplets to k (triplets are (0, ? ,k), namely (k choose 3) = (k+1)k(k-1)/6), spans start at 0 and go to 30-k, i.e. 31-k different spans. Total = (31-k)k(k-1)(k-2)/6   

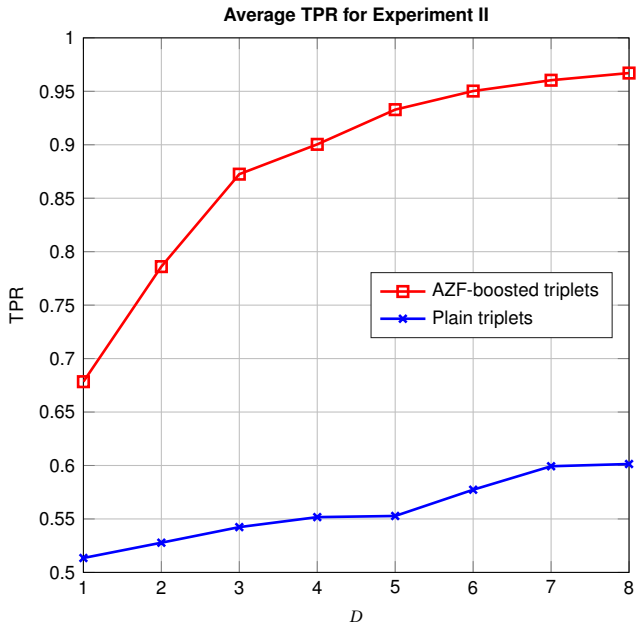
\begin{figure}[t]
%\centering{\includegraphics[width=0.45\textwidth]{images/triplet_tpr_dist_overall_nosamedistance.eps} }
\centering{% This file was created by matlab2tikz.
\begin{tikzpicture}

\begin{axis}[%
width=0.4\textwidth,
height=2.8in,
at={(0.772in,0.516in)},
scale only axis,
xmin=1,
xmax=8,
label style={font=\sffamily},
xticklabel style={font=\sffamily\scriptsize},
yticklabel style={font=\sffamily\scriptsize},
xlabel style={font=\sffamily\scriptsize},ylabel style={font=\sffamily\scriptsize},
title style={font=\bfseries\sffamily\scriptsize, yshift = -1ex},
title = {Average TPR for Experiment II},
xlabel={$D$},
ymin=0.5,
ymax=1,
xtick = {1,2,3,4,5,6,7,8},
xticklabels = {{1}, {2}, {3}, {4}, {5}, {6}, {7}, {8}},
ytick = {0.5, 0.55, 0.6, 0.65, 0.7, 0.75, 0.8, 0.85, 0.9, 0.95, 1},
yticklabels = {{0.5}, {0.55}, {0.6}, {0.65}, {0.7}, {0.75}, {0.8}, {0.85}, {0.9}, {0.95}, {1}},
ylabel={TPR},
axis background/.style={fill=white},
xmajorgrids,
ymajorgrids,
legend style={at={(0.97,0.5)}, anchor=east, legend cell align=left, align=left, draw=white!15!black, font = \sffamily\scriptsize}
]
\addplot [color=red, line width=1pt, mark=square, mark options={solid, red}]
  table[row sep=crcr]{%
1	0.678383552279671\\
2	0.78603712355201\\
3	0.872506891360123\\
4	0.90039585556132\\
5	0.932857249480306\\
6	0.950244781804785\\
7	0.96036885071725\\
8	0.967077404896037\\
};
\addlegendentry{AZF-boosted triplets}

\addplot [color=blue, line width=1pt, mark=x, mark options={solid, blue}]
  table[row sep=crcr]{%
1	0.51347874291683\\
2	0.527772744680724\\
3	0.542367293036949\\
4	0.551708589112281\\
5	0.552785399949745\\
6	0.57733006616365\\
7	0.599294730907972\\
8	0.601390664257831\\
};
\addlegendentry{Plain triplets}

\end{axis}
\end{tikzpicture}%
}
\caption{Average TPR for all triplets $(i,j,k)$ and distance $D = \left| |i - j| - |k - j| \right|$ for all 10 sources. The advantage of boosting by artefact amplification, zooming, and flickering is obvious: The easiest plain triplet comparisons are for $D=8$, and yet they are harder than the hardest AZF-boosted triplet comparisons at $D=1$.}
\label{trip_tpr}
\end{figure}

%-------------------------------------------------------------------------------
\subsection{Results: Reconstruction and Sensitivity Gain}
\label{sec:gainblur}
%-------------------------------------------------------------------------------

We reconstructed the impairment scales for the 10 sequences from the collected responses to the plain and the AZF-boosted triplet comparisons. The results, averaged over the 10 sequences, are shown in Figure\,\ref{trip_score}. For plain TC we obtained a very slowly increasing impairment, reaching 4.0 %\textcolor{blue}{[exactly: 4.0441]}
JND at distortion level 30. For AZF-boosted TC, however, we obtained a range of 30\,JND. Thus, over the range of 3\,JND of motion blur distortion, we recorded an average sensitivity gain of 7.5 %\textcolor{blue}{30.1655/4.0441=7.4591} 
for AZF-boosting. The same boosting for baseline triplets gave an average gain of only 1.8. %\textcolor{blue}{[7.1484 / 4.0441 = 1.7676]}.

Figure\,\ref{trip_score} also displays the achieved local sensitivity gain of AZF-boosted TC over plain TC as functions of the distortion levels. The gain function is shown for baseline triplets used in Experiment I (green curve) and for the general triplets used here (red curve). Over the whole range of distortion levels, the gain achieved in Experiment II is larger than 5. Moreover, the gain is decreasing up until level 12 and then increases again, reaching 9.6 
%\textcolor{blue}{[exactly: 9.6402]} 
at level 30. In contrast, for baseline TC of Experiment I, the gain is limited below 5 and slowly decreases down to 1.9.
%\textcolor{blue}{[exactly: 1.8670]} at level 30. 

Altogether, our first conjecture, to improve on the sensitivity gain by using general triplets in place of baseline triplets, was clearly confirmed with an impressive tripling of performance in sensitivity.

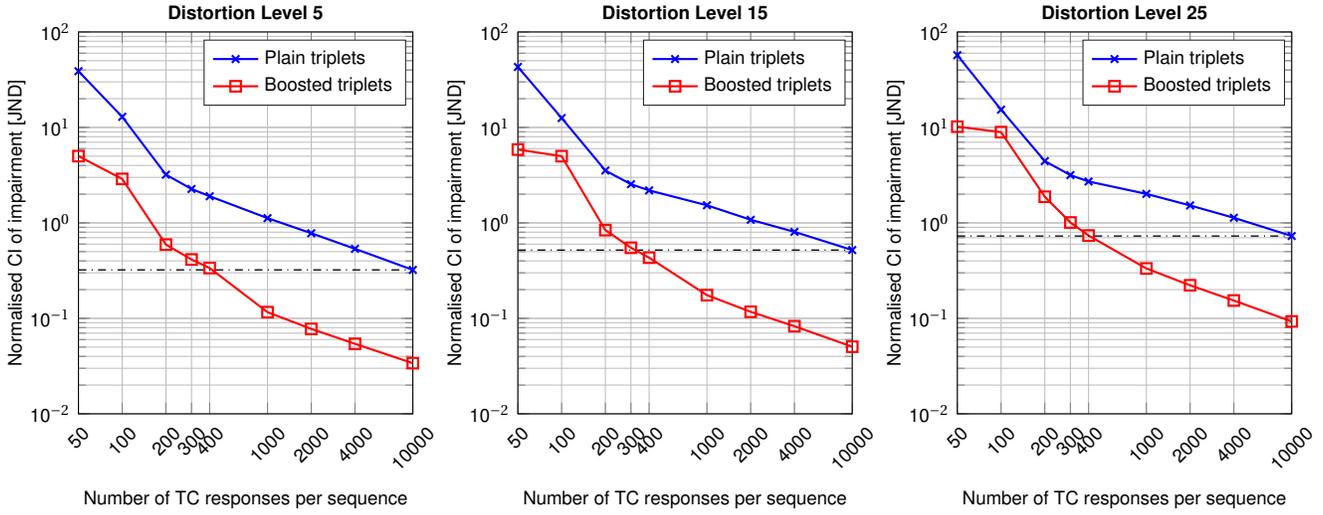
\begin{figure*}[t]
%\centering{\includegraphics[width=1\textwidth]{images/ciscore_500times_newer.eps}}
\centering{% This file was created by matlab2tikz.
\begin{tikzpicture}

\begin{axis}[%
width=1.75in,
height=2in,
at={(0in,0.699in)},
scale only axis,
xmode=log,
xmin=50,
xmax=10000,
label style={font=\sffamily},
xticklabel style={font=\sffamily\scriptsize},
yticklabel style={font=\sffamily\scriptsize},
xlabel style={font=\sffamily\scriptsize},
ylabel style={font=\sffamily\scriptsize, yshift = -1ex},
title style={yshift=-1ex,font=\bfseries\sffamily\scriptsize},
xtick={50,100,200,300,400,1000,2000,4000,10000},
xticklabels={{50},{100},{200},{300},{400},{1000},{2000},{4000},{10000}},
ytick={0.01,0.1,1,10,100},
yticklabels={$\text{10}^{\text\tiny{-2}}$,$\text{10}^{\text\tiny{-1}}$,$\text{10}^{\text\tiny{0}}$,$\text{10}^{\text\tiny{1}}$,$\text{10}^{\text\tiny{2}}$},
xminorticks=true,
xticklabel style={rotate=45},
xlabel={Number of TC responses per sequence},
ymode=log,
ymin=0.01,
ymax=100,
yminorticks=true,
ylabel={Normalised CI of impairment [JND]},
axis background/.style={fill=white},
title={Distortion Level 5},
xmajorgrids,
xminorgrids,
ymajorgrids,
yminorgrids,
legend style={legend cell align=left, align=left, draw=white!15!black, font =\sffamily\scriptsize}
]
\addplot [color=blue, line width=0.8pt, mark=x, mark options={solid, blue}]
  table[row sep=crcr]{%
50	38.8355318590566\\
100	12.9145946185694\\
200	3.20032550927974\\
300	2.26879244864159\\
400	1.9030278525539\\
1000	1.12150909855805\\
2000	0.779638362186285\\
4000	0.535154128950137\\
10000	0.321961595745545\\
};
\addlegendentry{Plain triplets}

\addplot [color=red, line width=0.8pt, mark=square, mark options={solid, red}]
  table[row sep=crcr]{%
50	5.00770687923475\\
100	2.88754199481072\\
200	0.592579845642784\\
300	0.41368317207964\\
400	0.336527625872291\\
1000	0.11611256211974\\
2000	0.077583194298745\\
4000	0.0541434454642638\\
10000	0.0339250630645099\\
};
\addlegendentry{Boosted triplets}

\addplot [color=black, dashdotted, line width=0.5pt, forget plot]
  table[row sep=crcr]{%
50	0.321961595745545\\
100	0.321961595745545\\
200	0.321961595745545\\
300	0.321961595745545\\
400	0.321961595745545\\
1000	0.321961595745545\\
2000	0.321961595745545\\
4000	0.321961595745545\\
10000	0.321961595745545\\
};
\end{axis}

\begin{axis}[%
width=1.75in,
height=2in,
at={(2.3in,0.699in)},
scale only axis,
xmode=log,
xmin=50,
xmax=10000,
label style={font=\sffamily},
xticklabel style={font=\sffamily\scriptsize},
yticklabel style={font=\sffamily\scriptsize},
xlabel style={font=\sffamily\scriptsize},
ylabel style={font=\sffamily\scriptsize, yshift = -1ex},
title style={yshift=-1ex,font=\bfseries\sffamily\scriptsize},
xtick={50,100,200,300,400,1000,2000,4000,10000},
xticklabels={{50},{100},{200},{300},{400},{1000},{2000},{4000},{10000}},
ytick={0.01,0.1,1,10,100},
yticklabels={$\text{10}^{\text\tiny{-2}}$,$\text{10}^{\text\tiny{-1}}$,$\text{10}^{\text\tiny{0}}$,$\text{10}^{\text\tiny{1}}$,$\text{10}^{\text\tiny{2}}$},
xminorticks=true,
xticklabel style={rotate=45},
xlabel={Number of TC responses per sequence},
ymode=log,
ymin=0.01,
ymax=100,
yminorticks=true,
ylabel={Normalised CI of impairment [JND]},
axis background/.style={fill=white},
title={Distortion Level 15},
xmajorgrids,
xminorgrids,
ymajorgrids,
yminorgrids,
legend style={legend cell align=left, align=left, draw=white!15!black, font=\sffamily\scriptsize}
]
\addplot [color=blue, line width=0.8pt, mark=x, mark options={solid, blue}]
  table[row sep=crcr]{%
50	42.9976905692996\\
100	12.5495520229011\\
200	3.53584386907744\\
300	2.54374722718153\\
400	2.19033566797909\\
1000	1.52919198372412\\
2000	1.07829453713992\\
4000	0.806032187486585\\
10000	0.519361832823028\\
};
\addlegendentry{Plain triplets}

\addplot [color=red, line width=0.8pt, mark=square, mark options={solid, red}]
  table[row sep=crcr]{%
50	5.86744271717886\\
100	4.99940373906189\\
200	0.84031337845114\\
300	0.549010536339738\\
400	0.432182966045707\\
1000	0.175123816854119\\
2000	0.11719636617198\\
4000	0.082745056906216\\
10000	0.0504718290049216\\
};
\addlegendentry{Boosted triplets}

\addplot [color=black, dashdotted, line width=0.5pt, forget plot]
  table[row sep=crcr]{%
50	0.519361832823028\\
100	0.519361832823028\\
200	0.519361832823028\\
300	0.519361832823028\\
400	0.519361832823028\\
1000	0.519361832823028\\
2000	0.519361832823028\\
4000	0.519361832823028\\
10000	0.519361832823028\\
};
\end{axis}

\begin{axis}[%
width=1.75in,
height=2in,
at={(4.6in,0.699in)},
scale only axis,
xmode=log,
xmin=50,
xmax=10000,
label style={font=\sffamily},
xticklabel style={font=\sffamily\scriptsize},
yticklabel style={font=\sffamily\scriptsize},
xlabel style={font=\sffamily\scriptsize},
ylabel style={font=\sffamily\scriptsize, yshift = -1ex},
title style={yshift=-1ex,font=\bfseries\sffamily\scriptsize},
xtick={50,100,200,300,400,1000,2000,4000,10000},
xticklabels={{50},{100},{200},{300},{400},{1000},{2000},{4000},{10000}},
ytick={0.01,0.1,1,10,100},
yticklabels={$\text{10}^{\text\tiny{-2}}$,$\text{10}^{\text\tiny{-1}}$,$\text{10}^{\text\tiny{0}}$,$\text{10}^{\text\tiny{1}}$,$\text{10}^{\text\tiny{2}}$},
xminorticks=true,
xticklabel style={rotate=45},
xlabel={Number of TC responses per sequence},
ymode=log,
ymin=0.01,
ymax=100,
yminorticks=true,
ylabel={Normalised CI of impairment [JND]},
axis background/.style={fill=white},
title={Distortion Level 25},
xmajorgrids,
xminorgrids,
ymajorgrids,
yminorgrids,
legend style={legend cell align=left, align=left, draw=white!15!black, font=\sffamily\scriptsize}
]
\addplot [color=blue, line width=0.8pt, mark=x, mark options={solid, blue}]
  table[row sep=crcr]{%
50	57.1623710505734\\
100	15.3527376856739\\
200	4.44299170667002\\
300	3.16752726721091\\
400	2.71517077624422\\
1000	2.01379027472007\\
2000	1.52357976958787\\
4000	1.1316425643835\\
10000	0.728652325501689\\
};
\addlegendentry{Plain triplets}

\addplot [color=red, line width=0.8pt, mark=square, mark options={solid, red}]
  table[row sep=crcr]{%
50	10.185511495807\\
100	8.93253923654134\\
200	1.87940170504633\\
300	1.00987987738442\\
400	0.737177405326869\\
1000	0.333105690244685\\
2000	0.22203147028291\\
4000	0.15375624788652\\
10000	0.0928617533668582\\
};
\addlegendentry{Boosted triplets}

\addplot [color=black, dashdotted, line width=0.5pt, forget plot]
  table[row sep=crcr]{%
50	0.728652325501689\\
100	0.728652325501689\\
200	0.728652325501689\\
300	0.728652325501689\\
400	0.728652325501689\\
1000	0.728652325501689\\
2000	0.728652325501689\\
4000	0.728652325501689\\
10000	0.728652325501689\\
};
\end{axis}
\end{tikzpicture}%
}
\caption{Comparative study on the precision of reconstructions from TCs. Impairment scales were reconstructed for each of the 10 image sequences from sets of 500 random samples of a variable number of responses for plain and AZF-boosted TCs. The figure shows the average lengths of the corresponding 10 confidence intervals for the stimuli at distortion levels 5, 15, and 25 as indicators of precision. The best achievable performance for plain TC, obtained from \num{10000} responses per sequence, was surpassed by reconstructions derived from as few as 400 responses to AZF-boosted TCs.}
%[Confidence intervals for SROCC in Figure\ \ref{trip_ci} were computed using percentile method (95\%)]
\label{trip_ci}
\end{figure*}

%-------------------------------------------------------------------------------
\subsection{Results: True Positive Rate}
%-------------------------------------------------------------------------------
The true positive rate is an indicator of the ease of the corresponding set of triplet comparisons. As for Experiment~I, we computed the true positive rate for plain and boosted triplet comparison, in this case, averaged over all triplets ${i,j,k}$ with $i<j<k$ and span $S = k-i \le 10$. The average TPR for the 10 sequences of motion-blurred images is 0.5583 for plain TC and 0.8810 for AZF-boosted TC. Compared to Experiment I, we see that AZF-boosting brings about an even larger increase of overall TPR. We obtained an improvement of 0.3227 using AZF-boosted TC over plain TC with general triplets. With baseline triplets, the corresponding improvement was smaller, 0.1313, for the case of motion blur, and 0.0924 for all types of distortion on average.
%For baseline triplet, motion blur-plain(0.7701), AZF (0.9014) --> improvement is 0.1313; overall -plain (0.7703), AZF(0.8627) --> improvement is 0.0924.

Figure\,\ref{trip_tpr} shows a more detailed view, breaking up the averages into eight parts, based on the absolute differences $D = \left| |i - j| - |k - j| \right| = 1, \ldots, 8$. Triplet comparisons with larger values of $D$ can be expected to be easier to judge correctly, and this is reflected in the monotonic increase of the TPR. It is remarkable that the TPR for AZF-boosting, even for the smallest difference of just 1 level between perceptual distances of the left and right image to the pivot image, is larger than any of the detailed TPRs for the plain triplet comparison.

%-------------------------------------------------------------------------------
\subsection{Results: Convergence in Precision} \label{sec:ci}
%-------------------------------------------------------------------------------
Assuming that the ratios of responses ``left'', ``right'', and ``not sure'' for each TC $(i,j,k)$ converge as the number of responses tends to infinity, we may expect that the reconstructed impairment scales for the corresponding image sequence also converge. In this subsection, we aim to assess the precision of the reconstructions for given budgets of TCs. For this purpose, we consider the 95\% confidence intervals (CI) for all of the reconstructed impairment scales. 

For each of the 10 image sequences with motion blur, we had collected approximately \num{8900} responses for plain and also for AZF-boosted TC with a span of at most 10 distortion levels. For a given budget of responses, one may create a sample of that size from each of these pools of \num{8900} responses for triplets, using random resampling with replacement. From each such sample, a reconstruction can follow. For each image sequence and each budget of responses, we carried out this procedure and computed the 95\% confidence interval for the resulting impairment scale of each image and recorded their lengths. We used budgets from 50 up to \num{10000} responses per sequence and collected 500 resamplings each time. 

For a fair comparison of the lengths of the confidence intervals derived from plain and boosted TCs, we must take the sensitivity gain of boosted TC into account. The impairment scales reconstructed from boosted TC are approximately 7.5 times larger (Section\,\ref{sec:gainblur}), and therefore the corresponding confidence intervals should be scaled by $1/7.5 \approx 0.13$.

% ----- normalization of CI
%\textcolor{blue}{We used the results from sampling $\num{10000}$ ratings per source to fit the JND scales of AZF and plain via linear fit separately. We then normalized the CI of AZF by a factor of $\lambda = S_{\text{plain}} / S_{\text{AZF}}$, where $S_{\text{plain}}$ and $S_{\text{AZF}}$ are the slope of the linear fits respectively. In our experiment,$S_{\text{plain}} = 0.1131$ and $S_{\text{AZF}} = 0.7120 $, resulting in $\lambda = 0.1588$. After such a normalization, we took the average CI over 10 sources as the final CI at different number of judgments for both of the TC test.}
% ------
Figure\,\ref{trip_ci} shows the resulting (scaled) lengths of the CI for the images at distortion levels 5, 15, and 25, averaged over the 10 sources, on a doubly logarithmic grid. For both methods, plain and boosted TC, the sizes of the confidence intervals shrink by about two orders of magnitude when the budget of responses is increased from 50 to \num{10000}. The precision given by reconstructions from \num{10000} responses for plain TC can be achieved by only 300--400 responses with the AZF-boosted TCs. In other words, in this experiment, a single response for a boosted TC gave as much benefit in terms of resulting precision as 25 to 33 responses for plain TCs.

\begin{table}[t!]
\caption{Convergence in SROCC for AZF-Boosted and Plain TC}
\label{tb:SROCC_convergence}
\centering
\begin{tabular}{r | c c | c c}
TC & \multicolumn{2}{c|}{AZF-boosted TC} & \multicolumn{2}{c}{Plain TC}\\
responses & SROCC &  CI length & SROCC &  CI length\\
 \hline
50 & 0.9372	    &	0.2267	&	0.7340	&	0.7600 \\
100 & 0.9682	&	0.2106	&	0.7409	&	0.7553\\
200 & 0.9915	&	0.1060	&	0.7420	&	0.6689\\
500 & 0.9982	&	0.0273	&	0.7897	&	0.5987\\
1000 & 0.9993	&	0.0036	&	0.8044	&	0.5687\\
2000 & 0.9998	&	0.0013	&	0.8633	&	0.4804\\
5000 & 1    	&	0.0006	&	0.9223	&	0.3513\\
\num{10000} & 1	&	0.0002	&	0.9412	&	0.2452\\
\end{tabular}
\end{table}

\begin{table*}[t]
\caption{Orderings from the Reconstructions for Source Image SRC07 and Motion Blur Distortion.\linebreak 
Last column: Number of Inversions.}
\label{tb:orderings}
\centering
\resizebox{1\textwidth}{!}{ 
\begin{tabular}{lr *{30}{r@{\hspace{1mm}}}r rr}
Method  & Responses & \multicolumn{31}{c}{Ordering} & SROCC & Inv.\\ [1mm]
  \hline \\
Plain TC	&	100	&	\phantom{ } 1	&	\phantom{ } 4	&	18	&	\phantom{ }  8	&	2	&	14	& \phantom{ } 6	&	5	&	12	&	17	&	11	&	0	&	26	&	20	&	15	&	19	&	7	&	16	&	13	&	21	&	10	&	23	&	9	&	3	&	24	&	22	&	29	&	25	&	27	&	30	&	28	&	0.6661	&	117 \\
Plain TC	&	1000	&	6	&	5	&	11	&	1	&	8	&	7	&	0	&	18	&	2	&	3	&	15	&	12	&	4	&	27	&	19	&	26	&	9	&	10	&	13	&	14	&	21	&	16	&	23	&	20	&	17	&	30	&	22	&	25	&	24	&	28	&	29	&	0.7754	&	95\\ 
Plain TC	&	\num{10000}	&	0	&	1	&	3	&	2	&	4	&	5	&	8	&	6	&	7	&	9	&	10	&	11	&	12	&	15	&	13	&	14	&	26	&	17	&	16	&	28	&	23	&	19	&	21	&	18	&	27	&	20	&	24	&	29	&	22	&	25	&	30	&	0.9331	&	40\\ [1mm]
AZF-boosted TC	&	100	&	0	&	2	&	1	&	3	&	13	&	8	&	5	&	6	&	4	&	14	&	11	&	7	&	12	&	9	&	10	&	15	&	16	&	18	&	17	&	22	&	21	&	23	&	24	&	20	&	19	&	26	&	29	&	25	&	28	&	27	&	30	&	0.9484	&	42\\ 
AZF-boosted TC	&	1000	&	0	&	1	&	2	&	3	&	4	&	5	&	6	&	7	&	8	&	9	&	10	&	11	&	12	&	13	&	14	&	15	&	16	&	17	&	18	&	19	&	20	&	21	&	22	&	23	&	24	&	25	&	26	&	27	&	29	&	28	&	30	&	0.9996	&	1\\
AZF-boosted TC	&	\num{10000}	&	0	&	1	&	2	&	3	&	4	&	5	&	6	&	7	&	8	&	9	&	10	&	11	&	12	&	13	&	14	&	15	&	16	&	17	&	18	&	19	&	20	&	21	&	22	&	23	&	24	&	25	&	26	&	27	&	28	&	29	&	30	&	1	&	0
\end{tabular}
}
\end{table*}
%\begin{tablenotes}   for \tnote{a}
%\scriptsize%
%\item[a] Number of inversions.
%\end{tablenotes}

%\begin{figure}[t]
%\centering{\includegraphics[width=0.5\textwidth]{images/vis_snr.eps}}
%\caption{
%SNR of SROCC obtained by resampling different number of ratings (500 times repetition), average over image sequences from 10 sources. Compared to Table~\ref{tb:orderings}, the convergence of the SROCC to its maximum at 1.0 is detailed more clearly by using a logarithmic scale for the SROCC, the signal-to-noise ratio of the collection of SROCC values, given in Equation  \ref{eq:SROCC-SNR}. }
%\label{trip_srocc}
%\end{figure}

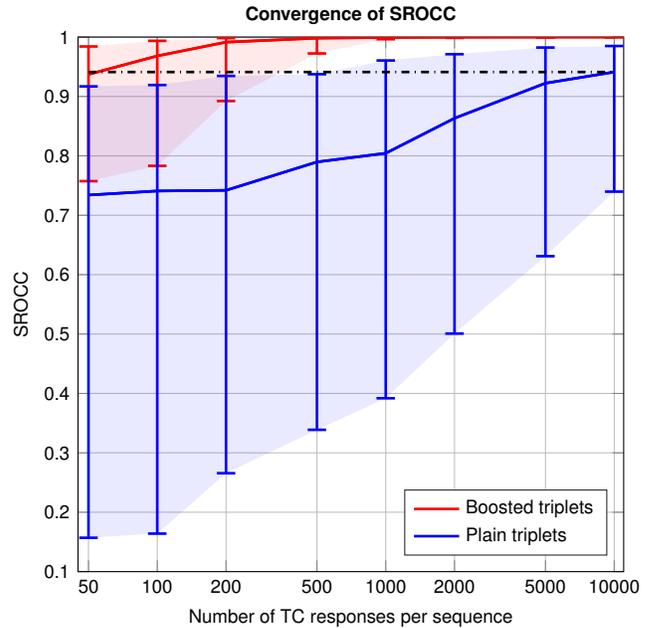
\begin{figure}[t]
%\centering{\includegraphics[width=\columnwidth]{images/srcc_median2.png}}
\centering{% This file was created by matlab2tikz.
\begin{tikzpicture}

\begin{axis}[%
width=0.4\textwidth,
height=2.8in,
at={(0in,0.516in)},
scale only axis,
xmode=log,
xmin=45,
xmax=11000,
label style={font=\sffamily},
xticklabel style={font=\sffamily\scriptsize},
yticklabel style={font=\sffamily\scriptsize},
xlabel style={font=\sffamily\scriptsize},
ylabel style={font=\sffamily\scriptsize},
title style={yshift=-1ex,font=\bfseries\sffamily\scriptsize},
xtick={50,100,200,500,1000,2000,5000,10000},
xticklabels={{50},{100},{200},{500},{1000},{2000},{5000},{10000}},
xminorticks=true,
xlabel={Number of TC responses per sequence},
ymin=0.1,
ymax=1,
ytick={0.1,0.2,0.3,0.4,0.5,0.6,0.7,0.8,0.9,1},
yticklabels={{0.1},{0.2},{0.3},{0.4},{0.5},{0.6},{0.7},{0.8},{0.9},{1}},
ylabel={SROCC},
axis background/.style={fill=white},
title={Convergence of SROCC},
xmajorgrids,
xminorgrids,
ymajorgrids,
legend style={at={(0.97,0.03)}, anchor=south east, legend cell align=left, align=left, draw=white!15!black, font = \sffamily\scriptsize}
]
\addplot [color=red, line width=1.0pt]
  table[row sep=crcr]{%
50	0.937177419354839\\
100	0.968225806451613\\
200	0.991451612903226\\
500	0.998185483870968\\
1000	0.999314516129032\\
2000	0.999798387096774\\
5000	1\\
10000	1\\
};
\addlegendentry{Boosted triplets}

\addplot [color=red, line width=1.0pt,  forget plot] %only marks,
 plot [error bars/.cd, y dir=both, y explicit, error bar style={line width=1.0pt}, error mark options={line width=1.0pt, mark size=3.5pt, rotate=90}]
 table[row sep=crcr, y error plus index=2, y error minus index=3]{%
50	0.937177419354839	0.0470225806451613	0.179717741935484\\
100	0.968225806451613	0.025474193548387	0.18508064516129\\
200	0.991451612903226	0.00694838709677414	0.0990725806451614\\
500	0.998185483870968	0.00161451612903218	0.0257661290322583\\
1000	0.999314516129032	0.000685483870967674	0.00294354838709676\\
2000	0.999798387096774	0.000201612903225845	0.00108870967741936\\
5000	1	0	0.0005645161290323\\
10000	1	0	0.00024193548387097\\
};

\addplot[area legend, dashed, draw=none, fill=red, fill opacity=0.09, forget plot]
table[row sep=crcr] {%
x	y\\
50	0.757459677419355\\
100	0.783145161290323\\
200	0.892379032258064\\
500	0.97241935483871\\
1000	0.996370967741936\\
2000	0.998709677419355\\
5000	0.999435483870968\\
10000	0.999758064516129\\
10000	1\\
5000	1\\
2000	1\\
1000	1\\
500	0.9998\\
200	0.9984\\
100	0.9937\\
50	0.9842\\
}--cycle;
\addplot [color=blue, line width=1.0pt]
  table[row sep=crcr]{%
50	0.734032258064516\\
100	0.740887096774193\\
200	0.741975806451613\\
500	0.789737903225806\\
1000	0.804415322580645\\
2000	0.863346774193548\\
5000	0.922318548387097\\
10000	0.94116935483871\\
};
\addlegendentry{Plain triplets}

\addplot [color=blue, line width=1.0pt, forget plot]
 plot [error bars/.cd, y dir=both, y explicit, error bar style={line width=1.0pt}, error mark options={line width=1.0pt, mark size=3.5pt, rotate=90}]
 table[row sep=crcr, y error plus index=2, y error minus index=3]{%
50	0.734032258064516	0.182967741935484	0.577032258064516\\
100	0.740887096774193	0.178412903225807	0.576887096774193\\
200	0.741975806451613	0.192624193548387	0.476275806451613\\
500	0.789737903225806	0.147762096774194	0.450937903225806\\
1000	0.804415322580645	0.156184677419355	0.412515322580645\\
2000	0.863346774193548	0.107853225806452	0.362546774193548\\
5000	0.922318548387097	0.0600814516129035	0.291218548387097\\
10000	0.94116935483871	0.0438306451612903	0.20136935483871\\
};

\addplot[area legend, dashed, draw=none, fill=blue, fill opacity=0.09, forget plot]
table[row sep=crcr] {%
x	y\\
50	0.157\\
100	0.164\\
200	0.2657\\
500	0.3388\\
1000	0.3919\\
2000	0.5008\\
5000	0.6311\\
10000	0.7398\\
10000	0.985\\
5000	0.9824\\
2000	0.9712\\
1000	0.9606\\
500	0.9375\\
200	0.9346\\
100	0.9193\\
50	0.917\\
}--cycle;
\addplot [color=black, dashdotted, line width=1.0pt, forget plot]
  table[row sep=crcr]{%
50	0.94116935483871\\
100	0.94116935483871\\
200	0.94116935483871\\
500	0.94116935483871\\
1000	0.94116935483871\\
2000	0.94116935483871\\
5000	0.94116935483871\\
10000	0.94116935483871\\
};
\end{axis}
\end{tikzpicture}%
}
\caption{
This figure illustrates the data from Table\,\ref{tb:SROCC_convergence}. Impairment scales were reconstructed from random samples of a variable number of responses for plain and AZF-boosted TCs. We show their (median) rank-order correlation (SROCC) with the ground truth ordering, averaged over image sequences from 10 sources. The upper and lower bound of CI (95\%) was averaged over 10 sequences. The best achievable performance for plain TC required \num{10000} responses but was beaten by reconstructions derived from as few as 100 responses to AZF-boosted TCs.}
\label{trip_srocc}
\end{figure}

%-------------------------------------------------------------------------------
\subsection{Results: Convergence in Ordering}
%-------------------------------------------------------------------------------
To compare the convergence of the quality assessment using boosted TCs with that for plain TCs, we followed the same approach as in the previous Section\,\ref{sec:ci}. Using resampled data for given budgets of TC responses per sequence of 31 images, we computed the impairment scale reconstructions and their corresponding SROCC w.r.t.\ the ground truth determined by the distortion levels of the images. For each size of budget, we produced 500 resamplings and recorded the median SROCC and the corresponding 95\% CI (using the percentile method). Finally, we averaged the median SROCC and lengths of the CIs over the 10 image sequences. 

These results are listed in Table\,\ref{tb:SROCC_convergence} and visualised in Figure\,\ref{trip_srocc}. The SROCC for plain TC is smaller than 0.9 for all sample sizes up to 2000 responses per sequence, while for AZF-boosted TC, the SROCC is above 0.9, even for samples of only 50  responses per sequence. The SROCC of 0.9412 for plain TC using \num{10000} responses is surpassed by the SROCC for boosted TC with as few as 100 responses. 

We demonstrate the advantage of boosted TC over plain TC by means of an example. We pick a source image and its 30 distorted versions, labelled by motion blur distortion levels 0 to 30. Then, for each plain and boosted TC, we choose samples of 100, 1000, and \num{10000} randomly selected responses (with replacement), followed by reconstruction of the 31 impairment scales. Sorting the image labels for each reconstruction according to increasing impairment scales yields permutations of $(0, 1, \ldots, 30)$. See Table\,\ref{tb:orderings} for the results. The table also shows the corresponding SROCC and the number of inversions in the permutations, which express the quality of the orderings (lower is better). The number of inversions is equal to the number of swaps required for sorting a permutation by the Bubble Sort algorithm. 

The best result for plain TC is from a sample of \num{10000} TC responses and has 40 inversions and an SROCC of 0.9331. It has about the same quality as a result obtained from a sample of only 100 AZF-boosted TC responses which has 42 inversions and an SROCC of 0.9484. With 1000 responses for boosted TCs, the ordering of the reconstruction is almost perfect, with only a single inversion left.

To summarise, in this experiment, each of the first 100 responses for AZF-boosted TC was worth more than 100 responses for plain TCs in terms of the resulting SROCC of the reconstruction. To obtain an SROCC of 0.95, our boosting method was 100 times as efficient as plain TC.

\section{Experiment III: Boosting for Degradation Category Rating}
%-------------------------------------------------------------------------------
%\begin{figure}[t]
%\centering{\includegraphics[width=0.45\textwidth]{images/newvis_dcr.png} }
%\caption{The DMOS of the DCR study of Experiment 3 show a sensitivity gain when assessing the perceptual image quality with small distortions up to about 0.75 JND (corresponding to level 3). The DMOS values are averaged over 70 sequences. 95\% confidence intervals were computed for the DMOS of each distorted image. For each of the distortion levels the average length of the coresponding 70 CIs was computed. The figure shows the centered CIs with these average lengths. The solid red curve is the corresponding sensitivity gain function.}
%\label{dcr_score}
%\end{figure}

%\begin{figure}[t]
%\centering{\input{imtexfile/newvis_dcr.tex} }
%\caption{The DMOS of the DCR study of Experiment 3 show a sensitivity gain when assessing the perceptual image quality with small distortions up to about 0.75 JND (corresponding to level 3). The DMOS values are averaged over 70 sequences. 95\% confidence intervals were computed for the DMOS of each distorted image. For each of the distortion levels the average length of the coresponding 70 CIs was computed. The figure shows the centered CIs with these average lengths. The solid red curve is the corresponding sensitivity gain function.}
%\label{dcr_score}
%\end{figure}

\begin{figure}[t]
\centering{\input{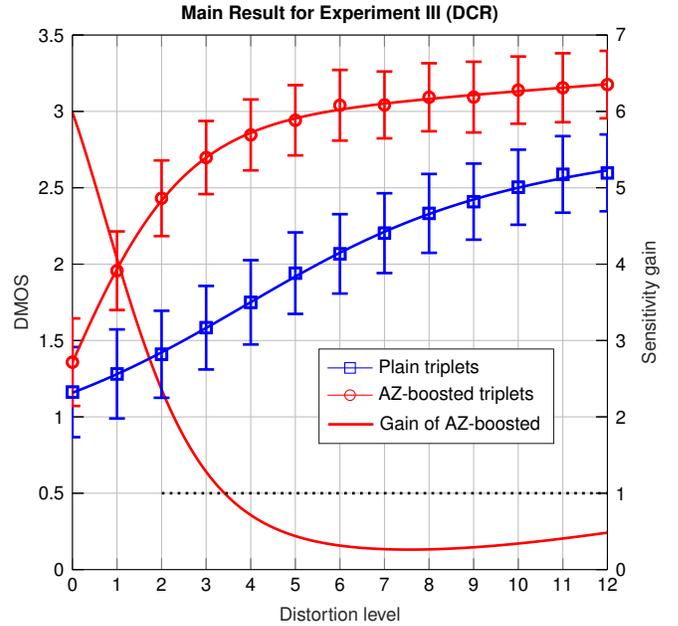} }
\vspace{-5pt}
\caption{The DMOS of the DCR study of Experiment III show a sensitivity gain when assessing the perceptual image quality with small distortions up to about 0.75\,JND (corresponding to level 3). The DMOS values are averaged over 70 sequences. 95\% confidence intervals were computed for the DMOS of each distorted image. For each of the distortion levels, the average length of the corresponding 70 CIs was computed. The figure shows the centred CIs with these average lengths. The solid red curve is the corresponding sensitivity gain function.}
\label{dcr_score}
\end{figure}

In our third and last experiment, we briefly tested the potential of boosting together with degradation category rating (DCR), which is one of the standard methods for subjective FR-IQA. 

Since in the conventional DCR approach, the distorted and reference images are displayed side by side, it is not appropriate to show a flickering image in such a scenario. Hence, only still images can be considered. In Experiment I, AZ-boosted TC provided the largest sensitivity gain for TC among the non-flickering boosting techniques. Therefore, here we repeated Experiment I, investigating the performance of plain and AZ-boosted comparisons applied in the DCR setting, denoted as Plain-DCR and AZ-DCR, respectively.

The experimental setup, the quality control, and the outlier removal for this study were as described earlier in Section\,\ref{sec:Experimental_Setup} and shown in Figure\,\ref{dcr_interface}. As in Experiment I, we used all 70 image sequences from 10 sources and 7 distortion types, each containing a source reference image and 12 increasingly distorted images. The set of all Plain-DCR and AZ-DCR questions was shuffled and distributed into a sufficient number of HITs, each one also containing one test question for quality control. For each image, we collected 50 ratings. The statistics of the collected data is detailed in Tables\,\ref{info1} and\,\ref{info2}.

%----------------------------------------------------------------------------
\subsection{Results:  Reconstruction  and  Sensitivity  Gain}
%-------------------------------------------------------------------------------
Figure\,\ref{dcr_score} shows the DMOS for the two types of DCR, averaged over the 70 sequences. Both methods worked well. The DMOS curve for AZ-DCR is above that for Plain-DCR and spans over a larger interval. Thus, on average, AZ-boosting provided an increased sensitivity as anticipated. However, the gain in sensitivity is restricted to small distortions, up to level 3 (0.75\,JND). For larger distortions, the sensitivity gain is less than 1, showing a saturation effect similar to A- and AZ-boosted baseline triplet comparison, compare Figure\,\ref{gain_example}.

%---- no DMOS 0 at level 0
Note that the DMOS should ideally be equal to $0$ at distortion level 0, since when the reference image is compared to itself, there is no difference between the two, and the distortion should be rated ``imperceptible'' with a score of~0. However, from Figure\,\ref{dcr_score}, the DMOS at level 0 is not equal to 0 but even larger than $1$ for both of the DCR tests. We think the reason for this outcome is that in this experiment, observers were instructed to expect to see distorted images on the right side, and during the work on the assignments, this expectation was fulfilled, almost always. Therefore, the participants of the study may have been hesitant to declare that they could not detect any difference. So many rated the distortion for a displayed test image as ``perceptible, but not annoying'' or even worse, although it actually was identical to the reference image.

%\begin{table}[t!]
%\caption{Average of Median SROCCs over 70 sequences for Plain- and AZ- DCR by sampling different number of ratings per image (bootstrapping 1000 times).}
%\label{tb:SROCC_dmos}
%\centering
%\begin{tabular}{l c c }
%& 13 levels, 50 ratings & 5 levels, 20 ratings \\
%\hline
%Plain  & 0.9991 &  0.9953 \\
%Boost & 0.9986 & 0.9986 \\
%\end{tabular}
%\end{table}

%----------------------------------------------------------------------------
\subsection{Results: Precision}
%-------------------------------------------------------------------------------
% Note that in this case here, bootstrapping is not required to assess the CIs. Thus, we should not  tell such a  story here:

%We evaluated the precision of the acquired DMOS results by bootstrapping. The DMOS of a distorted image is obtained by the average of $N$ ratings, where $N \le 50$ depends on how many outliers etc.\ were removed. By resampling with replacement $N$ ratings from this pool one obtains additional DMOS values that are also valid estimates for the population DMOS of the corresponding image. The standard deviation of these bootstrapped DMOS serves as a measure for precision of the quality assessment method. Smaller standard deviation means higher precision. We estimated the standard deviation from 1000 bootstraps. 

%The standard deviation of the bootstrapped means is closely related to the corresponding confidence interval of the DMOS estimate. Let us assume the ratings are i.i.d.\ random variables with mean $\mu$ and standard deviation $\sigma$. The Central Limit Theorem states that as $N \rightarrow \infty$ the distribution of the means of the ratings tends to the normal distribution $N(\mu,\sigma^2/N)$. Then the 95\% confidence interval of the mean is given by $\pm 1.96\, \sigma\sqrt{N}$ which is asymptotically proportional to the standard deviation of the bootstrapped means.

We evaluated the precision of the acquired DMOS results by computing their 95\% confidence intervals. Figure\,\ref{dcr_score} shows the results for Plain- and AZ-DCR, where we have averaged the full width of the confidence intervals over the 70 image sequences. These average confidence intervals, based on up to 50 collected ratings per test image, range from $\pm 0.220$ to $\pm 0.295$ on the 5-point DCR impairment scale. The confidence intervals for boosted AZ-DCR are slightly smaller than for Plain-DCR.

%-------------------------------------------------------
%------------------------------------------------------------
%------------------------------------------------------------
\section{Rescaling Boosted Impairments by Hybrid Triplet Comparisons}
\label{sec_hybrid}
%------------------------------------------------------------
Our boosting techniques of artefact amplification, zooming, and flickering were designed to perceptually magnify differences between compared stimuli so that human observers are enabled to better distinguish fine-grained distortion levels. The experiments of the previous sections have confirmed this intended effect.

However, boosting amounts to a nonlinear scaling of perceptual distortion, as already apparent from Figure\,\ref{jndscale}. Moreover, this nonlinearity may depend on the distortion type and the content of the source images. For example, using boosting by zooming and flicker, the impairment range of 3\,JND units for plain TC was stretched to 7\,JND for the jitter distortion, but only to 5.5\,JND for color diffusion, see Figure\,\ref{jndscale_per_distortion}.

This nonlinear scaling of perceptual distortion could be disregarded when building FR-IQA datasets. After all, it is a commonly accepted practice to use DCR or reconstructions from pair comparison for subjective quality assessment, although also the DMOS scale is not perceptually linear and is also not proportional to the reconstruction from pair comparisons. Moreover, as shown in Table\,\ref{tb:ivqadatabase}, the creators of FR-IQA and FR-VQA datasets have applied various other methods for assessment of impairment scales, but there is no agreed upon standard that provides any particular scale as a common ground that other scales can be related to. Only recently, procedures were proposed to merge impairment scale values from different methodologies (pair comparison and category rating) and from different datasets, see \cite{ye2012unsupervised,perez2019pairwise,kaipio2020}. 

In this section, we propose a similar approach to transform impairment scale values from boosted triplet comparisons back to scales obtained in the traditional way without boosting  perceptual discrimination power. For this purpose, we construct a (nonlinear) monotonic transformation of boosted scales for each image sequence. Thereby, we preserve the discrimination of fine-grained distortion differences achieved by the boosting approach while simultaneously ensuring that the ranges of transformed absolute impairment scales match those that are obtained when comparing distorted images without boosting. In other words, the transformed impairment values reflect the perceived qualities of the original distorted images rather than the qualities of the images with perceptually boosted distortions. 

Towards this end, we propose a hybrid method by allowing for a fraction of triplet comparisons without boosting. The scale reconstruction from these plain comparisons provides a rough estimate of the desired impairment scales.  We then fit a smooth scalar transformation for each sequence of distorted images that maps the boosted scales to the target scales in the least-squares sense. This transformation should be smooth and monotonic so that the high-quality relative differences of close scale values and the overall ordering of the image sequences obtained from boosted comparisons are maintained. In this way, we recalibrate the scales from boosted comparisons to follow those from standard comparisons without boosting. By construction, this recalibration is adaptive w.r.t.\ the source image contents and the type of distortion.

In Algorithm\,\ref{alg_hybrid}, we outline the hybrid method in the form of a pseudo code\footnote{In lines 8 and 9, the functions $f_{\gamma}$, resp.\ $f_{\hat{\gamma}}$, are applied to each component of their arguments.}. Then we provide the results of it when applied to the comparisons that we had collected in Experiment~I.

\begin{figure}[t]
\centering{
\definecolor{mycolor1}{rgb}{0.00000,1.00000,1.00000}%
\begin{tikzpicture}

\begin{axis}[%
width=2.9in,
height=2.8in,
at={(0.04in,0.06in)},
scale only axis,
xmin=1,
xmax=13,
xtick={1,2,3,4,5,6,7,8,9,10,11,12,13},
xticklabels={{0},{1},{2},{3},{4},{5},{6},{7},{8},{9},{10},{11},{12}},
ytick = {0,1,2,3,4,5,6,7,8,9,10},
yticklabels = {{0},{1},{2},{3},{4},{5},{6},{7},{8},{9},{10}},
xlabel style={font=\color{white!15!black}\sffamily\scriptsize},
ylabel style={font=\color{white!15!black}\sffamily\scriptsize},
xticklabel style={font=\color{white!15!black}\sffamily\scriptsize},
yticklabel style={font=\color{white!15!black}\sffamily\scriptsize},
xlabel={Distortion level},
ymin=0,
ymax=10,
ylabel={Impairment [JND]},
axis background/.style={fill=white},
title style={font=\bfseries\sffamily\scriptsize, yshift = -1ex},
title={Impairment Scale Recalibration},
xmajorgrids,
ymajorgrids,
legend style={at={(0.03,0.97)}, anchor=north west, legend cell align=left, align=left, draw=white!15!black, font = \tiny\sffamily}
]
\addplot [color=blue, mark=o, mark options={solid, blue} , line width = 1pt]
  table[row sep=crcr]{%
1	0\\
2	0.853761766007312\\
3	2.44730130967042\\
4	3.81928031135622\\
5	4.11629412333653\\
6	5.91312279058666\\
7	6.87836683233013\\
8	7.69713023920672\\
9	8.02216335936501\\
10	9.09796106367277\\
11	9.18755207635811\\
12	9.32326241805646\\
13	9.81472289518602\\
};
\addlegendentry{Boosted scales from 200 responses}

\addplot [color=green, mark=o,dashed,  mark options={solid, green}, line width = 1pt]
  table[row sep=crcr]{% make it dashed
1	0\\
2	0.874791194908422\\
3	0.265714225516325\\
4	0.420579287277431\\
5	0.783026628918232\\
6	0.461704742315134\\
7	1.85299884281452\\
8	2.10951541906675\\
9	2.07669952640394\\
10	2.63918040245287\\
11	2.42046279824426\\
12	4.00671897103573\\
13	3.18353618668278\\
};
\addlegendentry{Plain scales from 200 responses}

\addplot [color=black, mark=x, mark options={solid, black}, line width = 1pt]
  table[row sep=crcr]{%
1	0\\
2	0.101334269108357\\
3	0.344500961912821\\
4	0.626826243466304\\
5	0.698943034315519\\
6	1.24032498007852\\
7	1.62228528776544\\
8	2.00896788925091\\
9	2.18069602050185\\
10	2.83404130147859\\
11	2.89490480073015\\
12	2.98914433710768\\
13	3.35196110913797\\
};
\addlegendentry{Recalibration}

\addplot [color=red, mark=asterisk,dashed, mark options={solid, red}, line width = 1pt]
  table[row sep=crcr]{% dashed
1	0\\
2	0.539545740412104\\
3	0.362940740983105\\
4	0.585832624782792\\
5	1.08562580397669\\
6	1.21995452439265\\
7	1.3226859132296\\
8	2.01471282296769\\
9	2.33641908520384\\
10	2.44613605940394\\
11	2.91660019535004\\
12	3.25880874701545\\
13	3.42543372282146\\
};
\addlegendentry{Plain scales from 1360 responses}

%\addplot [color=black, mark=o, mark options={solid, black}]
%  table[row sep=crcr]{%
%1	0\\
%2	0.135706916531429\\
%3	0.398632177685418\\
%4	0.64948040374927\\
%5	0.709635161737628\\
%6	1.16670909625946\\
%7	1.52872985147655\\
%8	1.93752779410559\\
%9	2.12915895954121\\
%10	2.8666118583836\\
%11	2.93302196774237\\
%12	3.03429424762313\\
%13	3.40260726088965\\
%};
%\addlegendentry{optimal map [from boost (1000 response) to plain (full response)]}

\end{axis}
\end{tikzpicture}% 
}
\vspace{-10pt}
\caption{Example of the impairment scale recalibration by the hybrid method with parameters $K=400$ and $\alpha=0.5$. The data stems from the image sequence for source image SRC06 and distortion type jitter, assessed by plain and AZF-boosted triplet comparisons in Experiment I. The hybrid method fits the reconstruction from $(1-\alpha) K = 200$ boosted TCs (blue line with circles) to that from $\alpha K = 200$ plain TCs (green dashed line), with the result shown by the black line with crosses. For comparison, the reconstruction from all 1360 plain TC responses is also shown (red dashed line).
}
\label{fig_hybrid}
\end{figure}
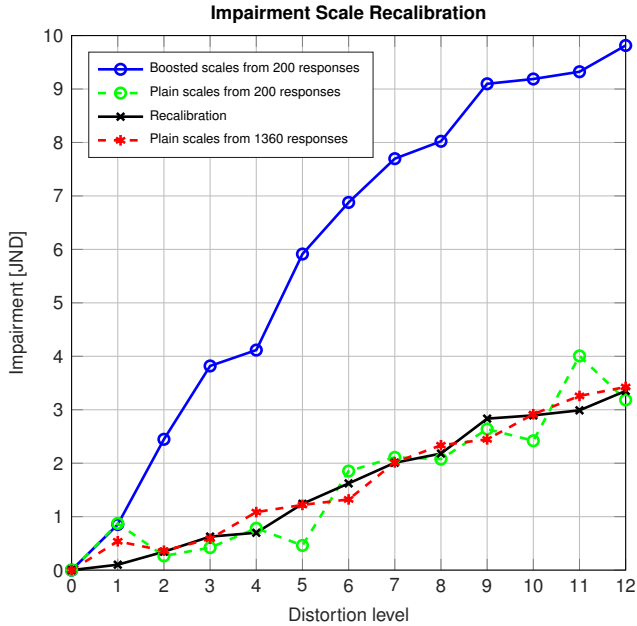

\begin{algorithm}[t!]
\caption{Hybrid method: Re-calibration of scales from boosted triplet comparisons}
\label{alg_hybrid}
\begin{algorithmic}[1]
\State Input: $I_0,\ldots,I_N$ 
    \Comment{sequence with $N$ distorted images}
\State Parameters: $K, \alpha$ 
    \Comment{budget of comparisons, $0<\alpha<1$}
\State Do $(1- \alpha)K$ boosted triplet comparisons 
\State Result: $\mu_0^\text{boost},\ldots,\mu_N^\text{boost}$ 
    \Comment{reconstructed scales}
\State Do $\alpha K$ plain triplet comparisons 
\State Result: $\mu_0^\text{plain},\ldots,\mu_N^\text{plain}$ 
    \Comment{reconstructed scales}
\State $f_{\gamma}:\mathbb{R} \rightarrow \mathbb{R}$ 
    \Comment{select family of monotonic functions}
\State $\hat{\gamma}=\arg\min_{\gamma}||f_{\gamma}(\mu^\text{boost}) - \mu^\text{plain}||^2$
    \Comment{function fitting}
\State Output: $f_{\hat{\gamma}}(\mu^\text{boost})$ 
    \Comment{transformed impairment scales}
\end{algorithmic}
\end{algorithm}

For our implementation of the hybrid method we chose the 5-parameter logistic function of Equation\,(\ref{eq:5para}) as $f_{\gamma}$. We forced it to be monotonically increasing by the constraint $\beta_1, \beta_2, \beta_4 \ge 0$. 

Figure\,\ref{fig_hybrid} shows the results for the example of one image sequence. For each of the 70 image sequences of Experiments I, we sampled $K=400$ random triplet comparisons, half of them as plain triplet comparisons in Step 4 ($\alpha = 0.5$). Since the result of the hybrid method depends on the chosen samples for the TCs, we repeated the hybrid procedure (random sampling and reconstruction, followed by recalibration) 100 times and kept only the median result w.r.t.\ the mean-square difference between the recalibrated reconstructed scales from boosted TC and the scales from plain TC. 

\begin{figure}[t]
\centering{
\input{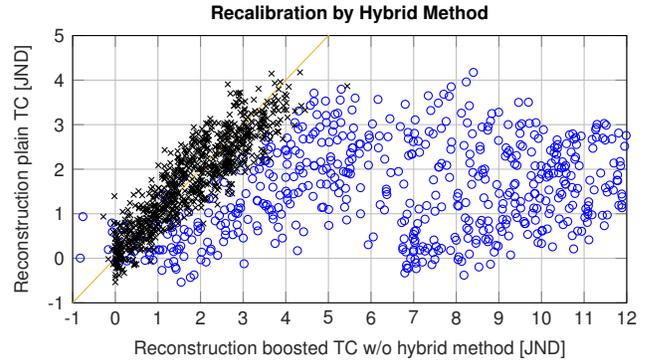} 
}
\vspace{-10pt}
\caption{Recalibration of the scale reconstruction from boosted triplet comparison of Experiment I by the hybrid method. The right point cloud (blue circles) is the scatter plot of the reconstruction from the AZF-boosted triplet comparison versus those from plain triplet comparisons. The recalibration by the hybrid method produced the left part of the scatters plot along the diagonal as intended for the recalibration. Note that for visual clarity, the figure truncates the reconstructions from boosted comparison to a maximum of 12\,JND. There are additional points (blue circles) further to the right, which are not shown. See the main text for more details. 
}
\label{fig_hybrid_scatterplot}
\end{figure}

Figure\,\ref{fig_hybrid_scatterplot} illustrates the resulting rescaled impairment scales for all distorted images from Part A of our dataset KonFiG-IQA.  In Experiment I, we had obtained 1360 responses for each  type of triplet comparison and  each of the 70 sequences with 12 distorted images. The figure shows two scatter plots together. The first one (blue circles) is for the scales, reconstructed from a random sample of only 200 AZF-boosted TCs ($K=400, 1-\alpha=0.5$) versus those reconstructed from all 1360 responses per sequence from plain TCs. Due to the boosting, the resulting impairment scales are much larger than those from plain comparison. 

\begin{table}[t!]
\caption{Recalibration for 200 AZF-boosted TCs per Sequence}
\label{table_recalibration_results}
\centering
\begin{tabular}{r c c c c}
Recalibration & RMSE & MAE & PLCC & SROCC\\
 \hline
before & 11.1439 & 8.9748 & 0.3274 & 0.3300 \\
after  & \multicolumn{1}{r}{0.4836} & 0.3916 & 0.8995 & 0.9051 
\end{tabular}
\end{table}

\begin{table*}[t]
\caption{Performance of the Recalibration in the Hybrid Method per Source Image and Distortion Type: Median RMSE (and STDDEV) w.r.t.\ Scales from 1360 Plain TC per Sequence.}
\label{table_mapping_rmse}
\centering
\begin{tabular}{l  l  l  l  l l l l l}
	&	Motion Blur		&		Lens Blur &			Color Diffusion &		Jitter		&	Multiplicative Noise &							High Sharpen	&	JPEG 2000		&		\textit{Average} \\
	\hline
SRC09	&	0.43	$\pm$	0.91	&	0.42	$\pm$	0.24	&	0.42	$\pm$	0.82	&	0.39	$\pm$	0.29	&	0.40	$\pm$	0.26	&	0.43	$\pm$	0.28	&	0.47	$\pm$	0.24	&	\textit{0.42} \\
SRC17	&	0.34	$\pm$	0.61	&	0.45	$\pm$	0.45	&	0.44	$\pm$	0.31	&	0.44	$\pm$	0.25	&	0.45	$\pm$	0.30	&	0.48	$\pm$	0.24	&	0.45	$\pm$	0.37	&	\textit{0.44}\\
SRC28	&	0.43	$\pm$	0.27	&	0.36	$\pm$	0.18	&	0.39	$\pm$	0.27	&	0.47	$\pm$	0.82	&	0.41	$\pm$	0.30	&	0.46	$\pm$	0.24	&	0.54	$\pm$	0.29	&	\textit{0.44}\\
SRC31	&	0.39	$\pm$	0.20	&	0.49	$\pm$	0.23	&	0.40	$\pm$	0.22	&	0.47	$\pm$	0.18	&	0.44	$\pm$	0.29	&	0.50	$\pm$	0.22	&	0.41	$\pm$	0.32	&	\textit{0.44}\\
SRC07	&	0.45	$\pm$	0.23	&	0.44	$\pm$	0.26	&	0.38	$\pm$	0.28	&	0.48	$\pm$	0.29	&	0.51	$\pm$	0.33	&	0.45	$\pm$	0.31	&	0.47	$\pm$	0.30	&	\textit{0.45}\\
SRC01	&	0.45	$\pm$	0.21	&	0.40	$\pm$	0.26	&	0.47	$\pm$	0.26	&	0.53	$\pm$	0.35	&	0.49	$\pm$	0.21	&	0.46	$\pm$	0.35	&	0.50	$\pm$	0.25	&	\textit{0.47}\\
SRC50	&	0.39	$\pm$	0.25	&	0.43	$\pm$	0.28	&	0.51	$\pm$	0.61	&	0.47	$\pm$	0.25	&	0.50	$\pm$	0.28	&	0.44	$\pm$	0.24	&	0.57	$\pm$	0.31	&	\textit{0.47}\\
SRC03	&	0.46	$\pm$	0.35	&	0.49	$\pm$	0.25	&	0.48	$\pm$	0.35	&	0.42	$\pm$	0.21	&	0.39	$\pm$	0.50	&	0.47	$\pm$	0.22	&	0.58	$\pm$	0.38	&	\textit{0.47}\\
SRC45	&	0.51	$\pm$	0.29	&	0.44	$\pm$	0.28	&	0.50	$\pm$	0.32	&	0.44	$\pm$	0.26	&	0.50	$\pm$	0.32	&	0.53	$\pm$	0.26	&	0.48	$\pm$	0.26	&	\textit{0.49}\\
SRC06	&	0.45	$\pm$	0.23	&	0.50	$\pm$	0.41	&	0.59	$\pm$	0.65	&	0.46	$\pm$	0.20	&	0.52	$\pm$	0.29	&	0.46	$\pm$	0.27	&	0.67	$\pm$	0.41	&	\textit{0.52}\\
\textit{Average}	&\textit{0.43}			&	\textit{0.44}			&	\textit{0.46}			&	\textit{0.46}			&	\textit{0.46}			&	\textit{0.47}			&	\textit{0.51}			&	
\end{tabular}
\end{table*}

These raw scales from $(1-\alpha)K=200$ boosted TCs were adaptively recalibrated by the hybrid method, using an additional random sample of $\alpha K=200$ responses (per sequence) from plain TC. As in the previous figure, we kept only the (mean-square difference) median result from recalibrated reconstructions of 101 random samples of the budget of $K=400$ TC responses. The corresponding scatters plot for the recalibrated results is also included in the graph (black crosses).  The recalibration translates each blue circle horizontally to the corresponding black cross. Thus, the recalibration from boosted TCs conforms to the range of scales obtained for plain TCs, i.e., represent the perceptual qualities corresponding to the original distorted images without any boosting. Table\,\ref{table_recalibration_results} gives the corresponding numerical results for the fitting procedure in the hybrid method. RMSE denotes the root-mean-square difference, MAE stands for the mean absolute difference. 

We also examined the extent to which the performance of the recalibration by the hybrid method depends on the type of distortion and the choice of the source image. For this purpose, we computed the RMSE between the corresponding recalibrated scales and those derived from all 1360 responses per sequence, using plain comparison. The results are shown in Table\,\ref{table_mapping_rmse}. The medians of these mean-square differences indeed do not vary strongly between source images or distortion types. However, with only 200 boosted and plain TC responses per sequence of 13 images, the variation between different random samples is larger, as shown by the standard deviations listed in the table. 

To judge the usefulness of the hybrid method, it is important to keep in mind that the purpose of the recalibration is not to achieve a ``perfect'' result of zero RMSE or an SROCC equal to 1, but only to match the range of the scales reconstructed from plain comparisons: Due to the increased sensitivity of the boosted image quality assessment, we expect the details of the reconstructed scales from boosted TCs to be more accurate than those reconstructed from plain TCs without boosting.

%\begin{figure}[t]
%\centering{
%\input{imtexfile/hist_maperror} 
%\input{imtexfile/cdf_maperror_median_400}
%}
%\caption{CDF of the absolute error between the JND scales of the hybrid method ($y'$) and the %reconstructions using full responses ($p$) of the median case (the right figure in %Figure\ \ref{fig_hybrid_scatterplot}).}
%\label{fig_hist_maperror}
%\end{figure}

%-------------------------------------------------------
%------------------------------------------------------------
%------------------------------------------------------------
\section{Discussion, Limitations, and Future Work}
\label{sec_discussion}
%------------------------------------------------------------

%------------------------------------------------------------
\subsection{Other Options for Boosting in FR-IQA}
\label{sec:boostopt}
%------------------------------------------------------------
To exploit the full potential of boosting in FR-IQA, other options for boosting can be investigated. 

\subsubsection{Type of artefact amplification}
For the artefact amplification, we have worked with the RGB color space. The RGB color space corresponds well to how images are technically displayed on devices and how colors are processed in the human visual system. However, the RGB space is not a perceptually uniform color space and is not well suited for human interaction.  The HSV color space (hue, saturation, value) corresponds better to how people experience color \cite{plataniotis2013color}. It separates the chromatic (hue, saturation) from the achromatic (value) color components. In the context of artefact amplification, this implies that the clamping of the value component in HSV space does not affect the color appearance, while when working in RGB space, clamping of an R, G, or B component does. Another option is to extend the technique to a context-dependent one, which takes into account the local JND (e.g.,~\cite{wu2017enhanced}) when amplifying the image distortion at each pixel. 

\subsubsection{Amplification and zoom factors}
The effects of amplification and zoom factors could be explored by conducting subjective tests with different values.  

\subsubsection{Flicker frequency}
In our flickering study, the displayed image buffer was swapped between the reference and the distorted image 8 times per second. In other words, the frequency of the visual signal was 4\,Hz. This is different from ~\cite{watson1986temporal}, in which the temporal TCSF suggests a contrast threshold of flickering frequency of 8\,Hz. However, it should be noted that images are different from the test stimuli used in the experiments regarding the TCSF. An interesting future work, therefore, is to characterize the TCSF for our IQA application. Furthermore, buffer swap rates, different from 8 times per second, may increase the sensitivity for subjective IQA even more. 
%------------------------------------------------------------
\subsection{Limitation of General Triplet Comparison}
\label{sec_multidimensional_scaling}
%------------------------------------------------------------
In Experiment II, we introduced general triplet comparisons that provided an improved sensitivity gain compared to baseline triplet comparisons, especially for larger distortions (2 to 3\,JND). However, there is an important limitation of this method. Such triplet comparisons aim at capturing relations between perceptual distances of stimuli. The reconstruction of the corresponding quality scales w.r.t.\ a reference then relies on the assumption that the given sequence of stimuli can be modelled as a subset of a one-dimensional Euclidean space such that distances properly add up. Thus, if $I_a, I_b, I_c$ denote three stimuli with increasing impairment scales, then we expect for the perceptual distances that $d(I_a, I_b) + d(I_b, I_c)= d(I_a, I_c)$ holds. This assumption appears natural for each image sequence derived for a single type of distortion, and therefore general triplet comparisons, with or without boosting, may be expected to provide meaningful results.

However, general triplet comparisons are no longer applicable if we mix several distortion types in one image sequence as was done in MDID \cite{sun2017mdid}, for example. In this case, consider two images with equal impairment scale, derived from the same reference image, but for two different types of distortion like JPEG compression and color diffusion. Then, clearly, these two distorted images are perceptually noticeably different from each other, yet their difference in impairment is equal to 0. In \cite{perez2019pairwise} it was therefore suggested to rename the JND measurement units of impairment scales to ``just objectionable differences''. 

The reason for this discrepancy is that a set of distorted images, derived from different types of distortions or with mixed distortions together, cannot be expected to lie on a one-dimensional continuum in image space. In our future work, we will study to what extent multidimensional scaling (MDS) methods can facilitate the application of general triplet comparisons for impairment scale reconstruction also for image sets derived from a reference image by multiple distortions. For example, one can consider maximum likelihood difference scaling (MLDS, \cite{maloney2003maximum}) or stochastic triplet embedding (STE, \cite{van2012stochastic}) for MDS, select a suitable embedding dimension, and finally derive impairment scales from the distances in the multidimensional embedding space. The results can be compared and validated with corresponding DMOS values. 

\subsection{Optimal Allocation of Triplet Comparisons in the Hybrid Method}
\label{sec_optimal_hybrid}
%------------------------------------------------------------
In Section\,\ref{sec_hybrid} we introduced a hybrid method that combined boosted and plain triplet comparisons in order to recalibrate reconstructed impairment scales to the traditional impairment ranges achieved without boosting. A fraction $\alpha$ of a fixed budget of $K$ comparisons was devoted to plain comparisons. For our empirical analysis, we have used $\alpha=0.5$. The more responses we collect from boosted TCs, the better the accuracy and precision of the resulting reconstruction. But increasing $\alpha$ reduces the number of auxiliary plain TCs and worsen the alignment with the scale range valid for impairment without boosting. 

To investigate the tradeoff between accuracy, respectively precision, on the one side and the alignment with the ground truth result achievable without boosting on the other side, we will carry out suitable simulations and experiments. As a first step we will rerun the computations as provided in Figures\,\ref{fig_hybrid},\,\ref{fig_hybrid_scatterplot}, and Table\,\ref{table_mapping_rmse} with variable parameters $K$ and $\alpha$. After defining a suitable target functional, which combines accuracy, respectively precision, with alignment quality, we can estimate an optimal fraction $\alpha$ of plain TCs for any given budget of $K$ TCs.

%------------------------------------------------------------
\subsection{Adaptive Sampling Strategies}
\label{sec_AdaptiveSampling}
%------------------------------------------------------------
The confidence intervals of the reconstructed impairment of quality scales generally will shrink as more and more responses to triplet comparison are collected. In our experiments, we used TCs $(i,j,k)$ with an equal number of target responses per TC. The triplets were moderately restricted, e.g., to spans of 10 or 20 distortion levels in Experiment~II. However, an adaptive sampling strategy based on information theoretic considerations may reduce the number of responses required to achieve the same quality of the result. In such a procedure, the expected information gain is considered that can be derived for the next response for a TC or  the responses for a batch of several TCs. This expected information gain may be maximized, thereby determining the next TCs to be posted to observers.
 
 Such adaptive sampling approaches have already been proven useful for pair comparisons, see, e.g., \cite{ye2012unsupervised} and \cite{li2018hybrid}. A similar general adaptive framework for the assessment of psychometric functions is QUEST+ \cite{watson2017quest}. We propose to tailor such adaptive sampling strategies for the case of triplet comparisons which would further improve the performance for impairment scale assessment by boosted triplet comparisons.

%------------------------------------------------------------
\subsection{Application Scenario}
\label{sec_application}
%------------------------------------------------------------
In general, boosting strategies help to evaluate more conservative JND thresholds and increase the discrimination power of subjective image quality assessment. In the existing FR-IQA datasets, the distorted images were typically generated by applying only a few sparse distortion levels to a reference stimulus. In such cases, it may be expected that subjective FR-IQA by comparisons without boosting already provides reliable results. However, in many applications, assessing slight quality differences is desirable. For example, in video compression, fine-grained quality assessment for small impairment scales up to 1\,JND  would be desirable for content providers that strive to satisfy most of their consumer clients.  Our boosting strategies enable faster and less expensive reconstruction for such small distortion scales. If needed, the reconstructed scores of the stimuli with boosted distortions subsequently can be mapped back to the JND scale for the stimuli without boosted distortions, as shown in Section\,\ref{sec_hybrid}. We are currently applying the flicker technique for subjective assessment of the JND and the satisfied user ratio (SUR) in image and video compression methods.

Another application of boosting strategies is robust watermarking. Recently, several methods using JND assessment have been proposed for robust image watermarking \cite{wan2020jnd,wang2020novel,zhou2020spatial,qin2019new}. Our boosting strategies would increase the visibility of the distortions caused by a watermarking algorithm.  Therefore, the watermarking algorithm can be optimized so that the watermark distortions would not be visible even when the distortion of the stimulus is boosted. Thus, this procedure would result in more robust watermarking.

A further future research direction is the use of boosting strategies for the subjective evaluation of other imaging modalities, such as stereoscopic imaging and screen content.

%In general, boosting helps to evaluate more conservative JND thresholds and increase the inter-rater agreements. The application is not so strong for current datasets where only a few levels of distortions are considered. Here subjective IQA can be done fine without boosting. But in more challenging datasets with many more levels (like 30 what we have), it becomes essential. Or if one uses combined distortions of several types. There we do not have a sequence per source, but an unordered set of distorted images per source. MDID is an example, 80 images/source. We could even try it ourselves, using the existing AMT interface that we used here. 

%------------------------------------------------------------
\subsection{KonFiG-IQA: The Konstanz Fine-Grained IQA Dataset}
\label{sec_datasets}
%------------------------------------------------------------

Our Konstanz Fine-Grained IQA dataset (KonFiG-IQA, Parts A and B) will be publically available online in our dataset repository \texttt{http://database.mmsp-kn.de}. Part A contains the 10 source images and corresponding distorted versions (7~distortion types at 12 levels, spaced at 0.25\,JND), resulting in 840 distorted images in total. We supply the (MATLAB) code to boost the distortions with respect to a reference image by artefact amplification, zooming, and flicker. Part B provides the distorted images for motion blur only, however, with 30 levels of distortion, spaced at 0.1\,JND.

KonFiG-IQA also includes a large number of subjective responses to triplet comparison and DCR ratings. Only those responses and ratings are provided that were remaining after the data cleaning process and outlier removal. For each TC response, we give a record 
\begin{center}
    \texttt{[source\_id, distortion\_type, (i,j,k), response, time\_stamp, time\_used, worker\_id]},
\end{center}
where  \texttt{response} denotes the response \texttt{left}, \texttt{not sure}, \texttt{right}, and \texttt{worker\_id} is an anonymous identifyer of the corresponding observer. In total, there are \num{706914} and \num{360544} responses for triplet comparisons for Parts A and B, respectively.

For Part A we also conducted a DCR study, yielding \num{360554} ratings. For each one we provide a record
\begin{center}
    \texttt{[source\_id, distortion\_type, distortion\_level, rating, time\_stamp, time\_used, worker\_id]}.
\end{center}

The last part of the dataset consists of the impairment scales, reconstructed from the TC responses and DCR ratings. 

The source code for generating the distorted images and for reconstructing impairment scales from triplet responses will be provided on \texttt{GitHub}.

% Full noisy data we collected (before removing the HITS having the same answers, before outlier removal, and contains some triplets that has no answer (skipped by the observer)) Useful data (removed the HITs that has the same answers, before outlier removal, containing some triplets that has no answer (skipped by the observer)) After outlier removal (containing some triplets that has no answer (skipped by the observer) Clean data after outlier removal (removed the triplets that has no answer (skipped by the observer)

%For dataset A, we generated 3640 images from 10 source images, each of which was distorted by 7 types of distortions at 13 levels and processed by 4 strategies. We provided subjective opinions of these images in two forms, one is reconstructed quality scores in the unit of JND obtained by TC from a total of 706914 responses, the other is DMOS by a DCR test with 76646 ratings. 
%For dataset B, we generated 650 images from 10 sources images, each of which was distorted by 1 type of distortion at 31 levels and processed by 2 strategies. For these images, we provided reconstructed quality scores in the unit of JND obtained by TC from a total of 360554 responses. 

%-------------------------------------------------------
%------------------------------------------------------------
%------------------------------------------------------------
\section{Conclusion}
\label{sec_conclusion}
%------------------------------------------------------------
For many applications of full-reference image quality assessment, reliable and precise subjective image quality measurement for small distortions near and also below the just noticeable difference on the perceptual quality scale is important. In this contribution, we showed that the sensitivity of conventional assessment methods w.r.t.\ small changes in perceptual quality can be strongly enlarged by several boosting methods. For this purpose, we proposed artefact amplification, zooming, the flicker test, and their combinations. 

We proposed triplet comparisons for the subjective quality assessment and provided the details required to reconstruct impairment scales for distorted images. This reconstruction is obtained by maximum likelihood estimation, based on the Thurstonian probabilistic model of image quality, thus compatible with the state-of-the-art method applied for classical pair comparison.

To assess the potential of the approach, we have created the first FR-IQA dataset, KonFiG-IQA, that was designed on perceptual criteria. Sequences of distorted images were generated with a fine-grained, perceptually equidistant spacing of distortion levels, only 0.25\,JND (or 0.1\,JND) apart from each other. 

In a large crowdsourced field study, we collected over 1.7 million responses to triplet comparison questions. We gave a detailed analysis of this data in terms of scale reconstructions, their accuracy, convergence, detection rates, the sensitivity gain achieved by the different boosting methods, and more. Boosting methods proved to increase the discriminatory power for the fine-grained dataset, allowing to reduce the number of subjective quality comparisons while improving the accuracy of the resulting relative image impairment scales in terms of SROCC w.r.t.\ available ground truth. 

Increasing perceptual sensitivity by boosting necessarily implies that the obtained impairment scales are larger than for conventional methods such as degradation category rating. However, these larger ranges  also depend on the respective image contents and the type of distortion. In order to map the high precision boosted impairment scales back to the ranges corresponding to the original distorted images without boosting, we proposed a hybrid method that applies a monotonic numerical transformation of scale values based on a few auxiliary triplet comparisons without boosting. 

Our boosting techniques pave the way to fine-grained image quality datasets, allowing for an increased number of distortion levels, yet with high-quality subjective ground truth annotations facilitated by an amplified perceptual discrimination power.

\bibliographystyle{IEEEtran}
\bibliography{bib}

\end{document}